\documentclass[12pt]{article}
\usepackage{epsfig,amsfonts,amssymb}
\usepackage{hyperref}
\usepackage{cite}
\usepackage{cite}

\def\ZZZ{{\hbox{ Z\kern-1.6mm Z}}}
\def\RRR{{\hbox{ R\kern-2.4mm R}}}
\def\CCC{{\hbox{ C\kern-2.0mm C}}}
\def\zzz{{\hbox{z\kern-1mm z}}}

\newcommand{\qeq}{{\hbox{=\kern-2.3mm ? \kern.5mm }}}
\renewcommand{\qeq}{=}

\newcommand{\eps}{\epsilon}

\newcommand{\ve}{\varepsilon}

\newcommand{\bJ}{{\bf J}}

\newcommand{\II}{{\cal I}}

\newcommand{\CC}{{\cal C}}

\newcommand{\wt}{\widetilde}
\newcommand{\wh}{\widehat}

\newcommand{\be}{\begin{equation}}
\newcommand{\ee}{\end{equation}}
\newcommand{\ben}{\begin{eqnarray}\displaystyle}
\newcommand{\een}{\end{eqnarray}}

\newcommand{\refb}[1]{(\ref{#1})}
\newcommand{\p}{\partial}
\newcommand{\sectiono}[1]{\section{#1}\setcounter{equation}{0}}

\def\one{{\hbox{ 1\kern-.8mm l}}}
\def\zero{{\hbox{ 0\kern-1.5mm 0}}}

\newcommand{\bea}[1]{\begin{eqnarray}\label{#1} }
\newcommand{\eea}{\end{eqnarray}}

\newcommand{\eqref}{\refb}

\usepackage{bm}
\usepackage[table]{xcolor}

\def\asnotex#1{{\color{red} #1}}
\def\asnotex#1{{\color{black} #1}}

\newcommand{\f}{\frac}

\newcommand{\non}{\nonumber}

\def\fignewfiggg{

\def\JPicScale{0.7}
\ifx\JPicScale\undefined\def\JPicScale{1}\fi
\unitlength \JPicScale mm
\begin{picture}(70,70)(0,0)
\linethickness{1mm}
\multiput(30,60)(0.12,-0.12){333}{\line(1,0){0.12}}
\linethickness{1mm}
\multiput(30,20)(0.12,0.12){333}{\line(1,0){0.12}}
\linethickness{0.4mm}
\qbezier(40,50)(39.97,55.22)(42.38,57.62)
\qbezier(42.38,57.62)(44.78,60.03)(50,60)
\linethickness{0.4mm}
\qbezier(40,50)(45.22,49.97)(47.62,52.38)
\qbezier(47.62,52.38)(50.03,54.78)(50,60)
\linethickness{0.4mm}
\put(50,60){\line(0,1){10}}
\put(50,25){\makebox(0,0)[cc]{$\cdots$}}

\put(50,65){\makebox(0,0)[cc]{$g$}}

\put(45,51){\makebox(0,0)[cc]{$g$}}

\put(45,59){\makebox(0,0)[cc]{$g$}}

\end{picture}

}

\def\fignewfiggp{

\def\JPicScale{0.7}
\ifx\JPicScale\undefined\def\JPicScale{1}\fi
\unitlength \JPicScale mm
\begin{picture}(70,70)(0,0)
\linethickness{1mm}
\multiput(30,60)(0.12,-0.12){333}{\line(1,0){0.12}}
\linethickness{1mm}
\multiput(30,20)(0.12,0.12){333}{\line(1,0){0.12}}
\linethickness{0.4mm}
\qbezier(40,50)(39.97,55.22)(42.38,57.62)
\qbezier(42.38,57.62)(44.78,60.03)(50,60)
\linethickness{0.2mm}
\qbezier(40,50)(45.22,49.97)(47.62,52.38)
\qbezier(47.62,52.38)(50.03,54.78)(50,60)
\linethickness{0.2mm}
\put(50,60){\line(0,1){10}}
\put(50,25){\makebox(0,0)[cc]{$\cdots$}}

\put(50,63){\makebox(0,0)[cc]{$\gamma$}}

\put(47,52){\makebox(0,0)[cc]{$\gamma$}}

\put(42,57){\makebox(0,0)[cc]{$g$}}

\end{picture}

}

\def\fignewfigpg{

\def\JPicScale{0.7}
\ifx\JPicScale\undefined\def\JPicScale{1}\fi
\unitlength \JPicScale mm
\begin{picture}(70,70)(0,0)
\linethickness{1mm}
\multiput(30,75)(0.12,-0.12){333}{\line(1,0){0.12}}
\linethickness{1mm}
\multiput(30,35)(0.12,0.12){333}{\line(1,0){0.12}}
\linethickness{0.2mm}
\qbezier(40,65)(39.97,70.22)(42.38,72.62)
\qbezier(42.38,72.62)(44.78,75.03)(50,75)
\linethickness{0.2mm}
\qbezier(40,65)(45.22,64.97)(47.62,67.38)
\qbezier(47.62,67.38)(50.03,69.78)(50,75)
\linethickness{0.4mm}
\put(50,75){\line(0,1){10}}
\put(50,40){\makebox(0,0)[cc]{$\cdots$}}

\put(50,80){\makebox(0,0)[cc]{$g$}}

\put(45,66){\makebox(0,0)[cc]{$\gamma$}}

\put(45,74){\makebox(0,0)[cc]{$\gamma$}}

\end{picture}

}

\def\figqedaGR{

\def\JPicScale{0.5}
\ifx\JPicScale\undefined\def\JPicScale{1}\fi
\unitlength \JPicScale mm
\begin{picture}(110,85)(0,0)
\linethickness{1mm}
\multiput(30,80)(0.12,-0.12){333}{\line(1,0){0.12}}
\linethickness{1mm}
\multiput(70,40)(0.12,0.12){333}{\line(1,0){0.12}}
\linethickness{1mm}
\multiput(40,10)(0.12,0.12){250}{\line(1,0){0.12}}
\linethickness{1mm}
\multiput(70,40)(0.12,-0.12){250}{\line(1,0){0.12}}
\linethickness{0.4mm}
\put(50,60){\line(1,0){40}}
\linethickness{0.4mm}
\put(10,70){\line(1,0){30}}
\put(35,85){\makebox(0,0)[cc]{$p_a$}}

\put(20,65){\makebox(0,0)[cc]{$k$}}

\put(58,69){\makebox(0,0)[cc]{$p_a+k$}}

\put(73,65){\makebox(0,0)[cc]{$\leftarrow\ell$}}

\put(35,50){\makebox(0,0)[cc]{$p_a+k+\ell$}}

\put(110,85){\makebox(0,0)[cc]{$p_b$}}

\put(94,50){\makebox(0,0)[cc]{$p_b-\ell$}}

\put(72,20){\makebox(0,0)[cc]{$\cdots$}}

\put(72,0){\makebox(0,0)[cc]{(a)}}

\put(60,60){\makebox(0,0)[cc]{g}}

\put(25,70){\makebox(0,0)[cc]{g}}

\end{picture}

}

\def\figqedbGR{

\def\JPicScale{0.5}
\ifx\JPicScale\undefined\def\JPicScale{1}\fi
\unitlength \JPicScale mm
\begin{picture}(110,85)(0,0)
\linethickness{1mm}
\multiput(30,80)(0.12,-0.12){333}{\line(1,0){0.12}}
\linethickness{1mm}
\multiput(70,40)(0.12,0.12){333}{\line(1,0){0.12}}
\linethickness{1mm}
\multiput(40,10)(0.12,0.12){250}{\line(1,0){0.12}}
\linethickness{1mm}
\multiput(70,40)(0.12,-0.12){250}{\line(1,0){0.12}}
\linethickness{0.4mm}
\put(50,60){\line(1,0){40}}
\linethickness{0.4mm}
\put(20,60){\line(1,0){30}}
\put(35,85){\makebox(0,0)[cc]{$p_a$}}

\put(20,65){\makebox(0,0)[cc]{$k$}}

\put(73,65){\makebox(0,0)[cc]{$\leftarrow\ell$}}

\put(35,50){\makebox(0,0)[cc]{$p_a+k+\ell$}}

\put(110,85){\makebox(0,0)[cc]{$p_b$}}

\put(94,50){\makebox(0,0)[cc]{$p_b-\ell$}}

\put(72,20){\makebox(0,0)[cc]{$\cdots$}}

\put(72,0){\makebox(0,0)[cc]{(b)}}

\put(60,60){\makebox(0,0)[cc]{g}}

\put(35,60){\makebox(0,0)[cc]{g}}

\end{picture}

}

\def\figqedcGR{

\def\JPicScale{0.5}
\ifx\JPicScale\undefined\def\JPicScale{1}\fi
\unitlength \JPicScale mm
\begin{picture}(110,85)(0,0)
\linethickness{1mm}
\multiput(30,80)(0.12,-0.12){333}{\line(1,0){0.12}}
\linethickness{1mm}
\multiput(70,40)(0.12,0.12){333}{\line(1,0){0.12}}
\linethickness{1mm}
\multiput(40,10)(0.12,0.12){250}{\line(1,0){0.12}}
\linethickness{1mm}
\multiput(70,40)(0.12,-0.12){250}{\line(1,0){0.12}}
\linethickness{0.4mm}
\put(40,70){\line(1,0){60}}
\linethickness{0.4mm}
\put(25,55){\line(1,0){30}}
\put(35,85){\makebox(0,0)[cc]{$p_a$}}

\put(25,60){\makebox(0,0)[cc]{$k$}}

\put(62,65){\makebox(0,0)[cc]{$p_a+\ell$}}

\put(73,75){\makebox(0,0)[cc]{$\leftarrow\ell$}}

\put(40,45){\makebox(0,0)[cc]{$p_a+k+\ell$}}

\put(110,85){\makebox(0,0)[cc]{$p_b$}}

\put(94,50){\makebox(0,0)[cc]{$p_b-\ell$}}

\put(72,20){\makebox(0,0)[cc]{$\cdots$}}

\put(72,0){\makebox(0,0)[cc]{(c)}}

\put(90,70){\makebox(0,0)[cc]{g}}

\put(40,55){\makebox(0,0)[cc]{g}}

\end{picture}

}

\def\figselfaGR{

\def\JPicScale{0.4}
\ifx\JPicScale\undefined\def\JPicScale{1}\fi
\unitlength \JPicScale mm
\begin{picture}(80,75)(0,0)
\linethickness{1mm}
\multiput(20,70)(0.12,-0.12){250}{\line(1,0){0.12}}
\linethickness{1mm}
\multiput(50,40)(0.12,0.12){250}{\line(1,0){0.12}}
\linethickness{1mm}
\multiput(20,10)(0.12,0.12){250}{\line(1,0){0.12}}
\linethickness{1mm}
\multiput(50,40)(0.12,-0.12){250}{\line(1,0){0.12}}
\linethickness{0.4mm}
\put(10,60){\line(1,0){20}}
\linethickness{0.4mm}
\qbezier(25,65)(32.84,67.66)(35.25,65.25)
\qbezier(35.25,65.25)(37.66,62.84)(35,55)
\put(20,75){\makebox(0,0)[cc]{$p_a$}}

\put(10,55){\makebox(0,0)[cc]{$k$}}

\put(40,65){\makebox(0,0)[cc]{$\ell$}}

\put(50,15){\makebox(0,0)[cc]{$\cdots$}}

\put(20,60){\makebox(0,0)[cc]{$g$}}

\put(31,66){\makebox(0,0)[cc]{$g$}}

\end{picture}

}

\def\figselfbGR{

\ifx\JPicScale\undefined\def\JPicScale{1}\fi
\unitlength \JPicScale mm
\begin{picture}(80,75)(0,0)
\linethickness{1mm}
\multiput(15,75)(0.12,-0.12){292}{\line(1,0){0.12}}
\linethickness{1mm}
\multiput(50,40)(0.12,0.12){250}{\line(1,0){0.12}}
\linethickness{1mm}
\multiput(20,10)(0.12,0.12){250}{\line(1,0){0.12}}
\linethickness{1mm}
\multiput(50,40)(0.12,-0.12){250}{\line(1,0){0.12}}
\linethickness{0.4mm}
\put(10,60){\line(1,0){20}}
\linethickness{0.4mm}
\qbezier(20,70)(27.84,72.66)(30.25,70.25)
\qbezier(30.25,70.25)(32.66,67.84)(30,60)
\put(20,80){\makebox(0,0)[cc]{$p_a$}}

\put(10,55){\makebox(0,0)[cc]{$k$}}

\put(35,70){\makebox(0,0)[cc]{$\ell$}}

\put(50,15){\makebox(0,0)[cc]{$\cdots$}}

\put(20,60){\makebox(0,0)[cc]{$g$}}

\put(26,71){\makebox(0,0)[cc]{$g$}}

\end{picture}

}

\def\figselfcGR{

\ifx\JPicScale\undefined\def\JPicScale{1}\fi
\unitlength \JPicScale mm
\begin{picture}(80,75)(0,0)
\linethickness{1mm}
\multiput(20,70)(0.12,-0.12){250}{\line(1,0){0.12}}
\linethickness{1mm}
\multiput(50,40)(0.12,0.12){250}{\line(1,0){0.12}}
\linethickness{1mm}
\multiput(20,10)(0.12,0.12){250}{\line(1,0){0.12}}
\linethickness{1mm}
\multiput(50,40)(0.12,-0.12){250}{\line(1,0){0.12}}
\linethickness{0.4mm}
\put(10,60){\line(1,0){20}}
\linethickness{0.4mm}
\qbezier(30,60)(37.84,62.66)(40.25,60.25)
\qbezier(40.25,60.25)(42.66,57.84)(40,50)
\put(20,75){\makebox(0,0)[cc]{$p_a$}}

\put(10,55){\makebox(0,0)[cc]{$k$}}

\put(45,60){\makebox(0,0)[cc]{$\ell$}}

\put(50,15){\makebox(0,0)[cc]{$\cdots$}}

\put(20,60){\makebox(0,0)[cc]{$g$}}

\put(36,61){\makebox(0,0)[cc]{$g$}}

\end{picture}

}

\def\figselfdGR{

\def\JPicScale{0.4}
\ifx\JPicScale\undefined\def\JPicScale{1}\fi
\unitlength \JPicScale mm
\begin{picture}(80,75)(0,0)
\linethickness{1mm}
\multiput(20,70)(0.12,-0.12){250}{\line(1,0){0.12}}
\linethickness{1mm}
\multiput(50,40)(0.12,0.12){250}{\line(1,0){0.12}}
\linethickness{1mm}
\multiput(20,10)(0.12,0.12){250}{\line(1,0){0.12}}
\linethickness{1mm}
\multiput(50,40)(0.12,-0.12){250}{\line(1,0){0.12}}
\linethickness{0.4mm}
\put(10,60){\line(1,0){20}}
\linethickness{0.4mm}
\qbezier(35,55)(42.84,57.66)(45.25,55.25)
\qbezier(45.25,55.25)(47.66,52.84)(45,45)
\put(20,75){\makebox(0,0)[cc]{$p_a$}}

\put(10,55){\makebox(0,0)[cc]{$k$}}

\put(50,55){\makebox(0,0)[cc]{$\ell$}}

\put(50,15){\makebox(0,0)[cc]{$\cdots$}}

\put(20,60){\makebox(0,0)[cc]{$g$}}

\put(40,56){\makebox(0,0)[cc]{$g$}}

\end{picture}

}

\def\figselfeGR{

\def\JPicScale{0.4}
\ifx\JPicScale\undefined\def\JPicScale{1}\fi
\unitlength \JPicScale mm
\begin{picture}(80,75)(0,0)
\linethickness{1mm}
\multiput(20,70)(0.12,-0.12){250}{\line(1,0){0.12}}
\linethickness{1mm}
\multiput(50,40)(0.12,0.12){250}{\line(1,0){0.12}}
\linethickness{1mm}
\multiput(20,10)(0.12,0.12){250}{\line(1,0){0.12}}
\linethickness{1mm}
\multiput(50,40)(0.12,-0.12){250}{\line(1,0){0.12}}
\linethickness{0.4mm}
\put(10,60){\line(1,0){20}}

\put(20,75){\makebox(0,0)[cc]{$p_a$}}

\put(10,55){\makebox(0,0)[cc]{$k$}}

\put(50,55){\makebox(0,0)[cc]{$\ell$}}

\linethickness{0.4mm}
\put(43,55){\circle{10}}
\put(43,55){\circle{9}}
\put(43,55){\circle{8.5}}
\put(43,55){\circle{8}}
\put(43,55){\circle{9.5}}

\put(50,15){\makebox(0,0)[cc]{$\cdots$}}

\put(20,60){\makebox(0,0)[cc]{$g$}}

\put(41,59){\makebox(0,0)[cc]{$g$}}

\end{picture}

}

\def\figselfgGR{

\def\JPicScale{0.4}
\ifx\JPicScale\undefined\def\JPicScale{1}\fi
\unitlength \JPicScale mm
\begin{picture}(80,75)(0,0)
\linethickness{1mm}
\multiput(20,70)(0.12,-0.12){250}{\line(1,0){0.12}}
\linethickness{1mm}
\multiput(50,40)(0.12,0.12){250}{\line(1,0){0.12}}
\linethickness{1mm}
\multiput(20,10)(0.12,0.12){250}{\line(1,0){0.12}}
\linethickness{1mm}
\multiput(50,40)(0.12,-0.12){250}{\line(1,0){0.12}}
\linethickness{0.4mm}
\put(10,60){\line(1,0){20}}

\put(20,75){\makebox(0,0)[cc]{$p_a$}}

\put(10,55){\makebox(0,0)[cc]{$k$}}

\put(41,62){\makebox(0,0)[cc]{$\ell$}}

\linethickness{0.4mm}
\put(34,64){\circle{10}}
\put(34,64){\circle{9}}
\put(34,64){\circle{9.5}}
\put(34,64){\circle{8.5}}
\put(34,64){\circle{8}}

\put(50,15){\makebox(0,0)[cc]{$\cdots$}}

\put(20,60){\makebox(0,0)[cc]{$g$}}

\put(32,68){\makebox(0,0)[cc]{$g$}}

\end{picture}

}

\def\figselffGR{

\def\JPicScale{0.4}
\ifx\JPicScale\undefined\def\JPicScale{1}\fi
\unitlength \JPicScale mm
\begin{picture}(80,75)(0,0)
\linethickness{1mm}
\multiput(20,70)(0.12,-0.12){250}{\line(1,0){0.12}}
\linethickness{1mm}
\multiput(50,40)(0.12,0.12){250}{\line(1,0){0.12}}
\linethickness{1mm}
\multiput(20,10)(0.12,0.12){250}{\line(1,0){0.12}}
\linethickness{1mm}
\multiput(50,40)(0.12,-0.12){250}{\line(1,0){0.12}}
\linethickness{0.4mm}
\put(10,60){\line(1,0){20}}

\put(20,75){\makebox(0,0)[cc]{$p_a$}}

\put(10,55){\makebox(0,0)[cc]{$k$}}

\linethickness{0.4mm}

\linethickness{0.3mm}
\put(33,53){\line(1,0){8}}
\put(37,49){\line(0,1){8}}

\put(50,15){\makebox(0,0)[cc]{$\cdots$}}

\put(20,60){\makebox(0,0)[cc]{$g$}}

\end{picture}

}

\def\figcircle{

\def\JPicScale{0.6}
\ifx\JPicScale\undefined\def\JPicScale{1}\fi
\unitlength \JPicScale mm
\begin{picture}(110,70)(0,0)
\linethickness{1mm}
\multiput(10,60)(0.12,-0.12){333}{\line(1,0){0.12}}
\linethickness{1mm}
\multiput(10,20)(0.12,0.12){333}{\line(1,0){0.12}}
\linethickness{0.4mm}
\put(28.5,41.5){\line(0,1){20}}
\linethickness{0.3mm}

\linethickness{0.4mm}
\put(28,42){\circle{5}}

\linethickness{0.4mm}
\put(88.46,42){\circle{5}}

\linethickness{1mm}
\multiput(70,60)(0.12,-0.12){333}{\line(1,0){0.12}}
\linethickness{1mm}
\multiput(70,20)(0.12,0.12){333}{\line(1,0){0.12}}
\linethickness{0.4mm}
\put(80,50){\line(0,1){20}}
\linethickness{0.4mm}
\put(88.5,41.5){\line(0,1){20}}
\put(30,20){\makebox(0,0)[cc]{$\cdots$}}

\put(90,20){\makebox(0,0)[cc]{$\cdots$}}

\put(5,65){\makebox(0,0)[cc]{$p_a$}}

\put(70,65){\makebox(0,0)[cc]{$p_a$}}

\put(85,70){\makebox(0,0)[cc]{$k$}}

\put(88.5,53){\makebox(0,0)[cc]{$g$}}

\put(80,60){\makebox(0,0)[cc]{$g$}}

\put(28.5,55){\makebox(0,0)[cc]{$g$}}

\end{picture}

}

\def\figtadpoleanew{

\def\JPicScale{0.4}
\ifx\JPicScale\undefined\def\JPicScale{1}\fi
\unitlength \JPicScale mm
\begin{picture}(120,70)(0,0)
\linethickness{1mm}
\multiput(40,30)(0.32,0.12){250}{\line(1,0){0.32}}
\linethickness{1mm}
\multiput(40,60)(0.32,-0.12){250}{\line(1,0){0.32}}
\linethickness{0.4mm}
\qbezier(80,45)(74.75,52.8)(74.75,58.81)
\qbezier(74.75,58.81)(74.75,64.83)(80,70)
\linethickness{0.4mm}
\qbezier(80,45)(85.25,52.8)(85.25,58.81)
\qbezier(85.25,58.81)(85.25,64.83)(80,70)
\linethickness{0.2mm}
\put(34,50){\line(1,0){30}}

\put(80,35){\makebox(0,0)[cc]{$\cdots$}}

\put(45,50){\makebox(0,0)[cc]{$\gamma$}}

\put(75,60){\makebox(0,0)[cc]{$g$}}

\end{picture}

}

\def\figtadpolea{

\def\JPicScale{0.5}
\ifx\JPicScale\undefined\def\JPicScale{1}\fi
\unitlength \JPicScale mm
\begin{picture}(120,70)(0,0)
\linethickness{1mm}
\multiput(40,30)(0.32,0.12){250}{\line(1,0){0.32}}
\linethickness{1mm}
\multiput(40,60)(0.32,-0.12){250}{\line(1,0){0.32}}
\linethickness{0.4mm}
\qbezier(80,45)(74.75,52.8)(74.75,58.81)
\qbezier(74.75,58.81)(74.75,64.83)(80,70)
\linethickness{0.4mm}
\qbezier(80,45)(85.25,52.8)(85.25,58.81)
\qbezier(85.25,58.81)(85.25,64.83)(80,70)
\linethickness{0.4mm}
\put(34,50){\line(1,0){30}}

\put(80,35){\makebox(0,0)[cc]{$\cdots$}}

\put(45,50){\makebox(0,0)[cc]{$g$}}

\put(75,60){\makebox(0,0)[cc]{$g$}}

\end{picture}

}

\def\figtadpoleb{

\def\JPicScale{0.5}
\ifx\JPicScale\undefined\def\JPicScale{1}\fi
\unitlength \JPicScale mm
\begin{picture}(120,70)(0,0)
\linethickness{1mm}
\multiput(40,30)(0.32,0.12){250}{\line(1,0){0.32}}
\linethickness{1mm}
\multiput(40,60)(0.32,-0.12){250}{\line(1,0){0.32}}
\linethickness{0.4mm}
\qbezier(80,45)(74.75,52.8)(74.75,58.81)
\qbezier(74.75,58.81)(74.75,64.83)(80,70)
\linethickness{0.4mm}
\qbezier(80,45)(85.25,52.8)(85.25,58.81)
\qbezier(85.25,58.81)(85.25,64.83)(80,70)
\linethickness{0.4mm}
\put(50,45){\line(1,0){30}}

\put(80,35){\makebox(0,0)[cc]{$\cdots$}}

\put(60,45){\makebox(0,0)[cc]{$g$}}

\put(75,60){\makebox(0,0)[cc]{$g$}}

\end{picture}

}

\def\figtadpolec{

\def\JPicScale{0.6}
\ifx\JPicScale\undefined\def\JPicScale{1}\fi
\unitlength \JPicScale mm
\begin{picture}(120,70)(0,0)
\linethickness{1mm}
\multiput(40,30)(0.32,0.12){250}{\line(1,0){0.32}}
\linethickness{1mm}
\multiput(40,60)(0.32,-0.12){250}{\line(1,0){0.32}}
\linethickness{0.4mm}
\qbezier(80,45)(74.75,52.8)(74.75,58.81)
\qbezier(74.75,58.81)(74.75,64.83)(80,70)
\linethickness{0.4mm}
\qbezier(80,45)(85.25,52.8)(85.25,58.81)
\qbezier(85.25,58.81)(85.25,64.83)(80,70)
\linethickness{0.4mm}
\put(80,70){\line(0,1){10}}

\put(80,30){\makebox(0,0)[cc]{$\cdots$}}

\put(80,75){\makebox(0,0)[cc]{$g$}}

\put(74.7,60){\makebox(0,0)[cc]{$g$}}

\put(85.5,60){\makebox(0,0)[cc]{$g$}}

\end{picture}
}

\def\figextranew{

\def\JPicScale{0.5}
\ifx\JPicScale\undefined\def\JPicScale{1}\fi
\unitlength \JPicScale mm
\begin{picture}(95,60)(0,0)
\linethickness{1mm}
\multiput(55,60)(0.12,-0.12){333}{\line(1,0){0.12}}
\linethickness{1mm}
\multiput(55,20)(0.12,0.12){333}{\line(1,0){0.12}}
\linethickness{0.4mm}
\put(50,40){\line(1,0){25}}
\linethickness{0.4mm}
\put(65,50){\line(1,0){20}}

\put(75,25){\makebox(0,0)[cc]{$\cdots$}}

\put(75,50){\makebox(0,0)[cc]{$g$}}

\put(60,40){\makebox(0,0)[cc]{$g$}}

\end{picture}

}

\def\figgrextraa{

\def\JPicScale{0.4}
\ifx\JPicScale\undefined\def\JPicScale{1}\fi
\unitlength \JPicScale mm
\begin{picture}(90,70)(0,0)
\linethickness{1mm}
\multiput(30,70)(0.18,-0.12){333}{\line(1,0){0.18}}
\linethickness{1mm}
\multiput(30,30)(0.18,0.12){333}{\line(1,0){0.18}}
\linethickness{0.4mm}
\qbezier(45,60)(52.83,65.28)(56.44,62.88)
\qbezier(56.44,62.88)(60.05,60.47)(60,50)
\linethickness{0.4mm}
\put(23,55){\line(1,0){30}}

\put(60,35){\makebox(0,0)[cc]{$\cdots$}}

\put(30,55){\makebox(0,0)[cc]{$g$}}

\put(52,64){\makebox(0,0)[cc]{$g$}}

\end{picture}

}

\def\figgrextrab{

\def\JPicScale{0.4}
\ifx\JPicScale\undefined\def\JPicScale{1}\fi
\unitlength \JPicScale mm
\begin{picture}(90,70)(0,0)
\linethickness{1mm}
\multiput(30,70)(0.18,-0.12){333}{\line(1,0){0.18}}
\linethickness{1mm}
\multiput(30,30)(0.18,0.12){333}{\line(1,0){0.18}}
\linethickness{0.4mm}
\qbezier(45,60)(52.83,65.28)(56.44,62.88)
\qbezier(56.44,62.88)(60.05,60.47)(60,50)
\linethickness{0.4mm}
\put(15,60){\line(1,0){30}}

\put(60,35){\makebox(0,0)[cc]{$\cdots$}}

\put(30,60){\makebox(0,0)[cc]{$g$}}

\put(52,64){\makebox(0,0)[cc]{$g$}}

\end{picture}

}

\def\figgamman{

\def\JPicScale{0.5}
\ifx\JPicScale\undefined\def\JPicScale{1}\fi
\unitlength \JPicScale mm
\begin{picture}(90,70)(0,0)
\linethickness{1mm}
\multiput(30,70)(0.18,-0.12){333}{\line(1,0){0.18}}
\linethickness{1mm}
\multiput(30,30)(0.18,0.12){333}{\line(1,0){0.18}}
\linethickness{0.4mm}
\qbezier(45,60)(52.83,65.28)(56.44,62.88)
\qbezier(56.44,62.88)(60.05,60.47)(60,50)
\linethickness{0.4mm}

\put(60,35){\makebox(0,0)[cc]{$\cdots$}}

\put(55,63){\makebox(0,0)[cc]{$g$}}

\end{picture}

}

\def\figgrextrac{

\def\JPicScale{0.4}
\ifx\JPicScale\undefined\def\JPicScale{1}\fi
\unitlength \JPicScale mm
\begin{picture}(90,70)(0,0)
\linethickness{1mm}
\multiput(30,70)(0.18,-0.12){333}{\line(1,0){0.18}}
\linethickness{1mm}
\multiput(30,30)(0.18,0.12){333}{\line(1,0){0.18}}
\linethickness{0.4mm}
\qbezier(45,60)(52.83,65.28)(56.44,62.88)
\qbezier(56.44,62.88)(60.05,60.47)(60,50)
\linethickness{0.4mm}
\put(8,65){\line(1,0){30}}

\put(60,35){\makebox(0,0)[cc]{$\cdots$}}

\put(15,65){\makebox(0,0)[cc]{$g$}}

\put(52,64){\makebox(0,0)[cc]{$g$}}

\end{picture}

}

\def\figgrextrad{

\def\JPicScale{0.6}
\ifx\JPicScale\undefined\def\JPicScale{1}\fi
\unitlength \JPicScale mm
\begin{picture}(90,70)(0,0)
\linethickness{1mm}
\multiput(30,70)(0.18,-0.12){333}{\line(1,0){0.18}}
\linethickness{1mm}
\multiput(30,30)(0.18,0.12){333}{\line(1,0){0.18}}
\linethickness{0.4mm}
\qbezier(45,60)(52.83,65.28)(56.44,62.88)
\qbezier(56.44,62.88)(60.05,60.47)(60,50)
\linethickness{0.4mm}
\put(55,64){\line(0,1){15}}

\put(60,35){\makebox(0,0)[cc]{$\cdots$}}

\put(59,60){\makebox(0,0)[cc]{$g$}}

\put(55,70){\makebox(0,0)[cc]{$g$}}

\put(50,63){\makebox(0,0)[cc]{$g$}}

\end{picture}

}

\def\figgrextrae{

\def\JPicScale{0.4}
\ifx\JPicScale\undefined\def\JPicScale{1}\fi
\unitlength \JPicScale mm
\begin{picture}(90,70)(0,0)
\linethickness{1mm}
\multiput(30,70)(0.18,-0.12){333}{\line(1,0){0.18}}
\linethickness{1mm}
\multiput(30,30)(0.18,0.12){333}{\line(1,0){0.18}}
\linethickness{0.4mm}
\qbezier(45,60)(52.83,65.28)(56.44,62.88)
\qbezier(56.44,62.88)(60.05,60.47)(60,50)
\linethickness{0.4mm}
\put(27,50){\line(1,0){30}}

\put(60,35){\makebox(0,0)[cc]{$\cdots$}}

\put(40,50){\makebox(0,0)[cc]{$g$}}

\put(54,64){\makebox(0,0)[cc]{$g$}}

\end{picture}

}

\def\figgrextraf{

\def\JPicScale{0.4}
\ifx\JPicScale\undefined\def\JPicScale{1}\fi
\unitlength \JPicScale mm
\begin{picture}(90,70)(0,0)
\linethickness{1mm}
\multiput(30,70)(0.18,-0.12){333}{\line(1,0){0.18}}
\linethickness{1mm}
\multiput(30,30)(0.18,0.12){333}{\line(1,0){0.18}}
\linethickness{0.4mm}
\qbezier(45,60)(52.83,65.28)(56.44,62.88)
\qbezier(56.44,62.88)(60.05,60.47)(60,50)
\linethickness{0.4mm}
\put(70,57){\line(0,1){15}}

\put(60,35){\makebox(0,0)[cc]{$\cdots$}}

\put(70,65){\makebox(0,0)[cc]{$g$}}

\put(52,64){\makebox(0,0)[cc]{$g$}}

\end{picture}

}

\def\figemextraa{

\def\JPicScale{0.4}
\ifx\JPicScale\undefined\def\JPicScale{1}\fi
\unitlength \JPicScale mm
\begin{picture}(90,70)(0,0)
\linethickness{1mm}
\multiput(30,70)(0.18,-0.12){333}{\line(1,0){0.18}}
\linethickness{1mm}
\multiput(30,30)(0.18,0.12){333}{\line(1,0){0.18}}
\linethickness{0.4mm}
\qbezier(45,60)(52.83,65.28)(56.44,62.88)
\qbezier(56.44,62.88)(60.05,60.47)(60,50)
\linethickness{0.2mm}
\put(23,55){\line(1,0){30}}

\put(60,35){\makebox(0,0)[cc]{$\cdots$}}

\put(35,55){\makebox(0,0)[cc]{$\gamma$}}

\put(52,63){\makebox(0,0)[cc]{$g$}}

\end{picture}

}

\def\figemextrab{

\def\JPicScale{0.4}
\ifx\JPicScale\undefined\def\JPicScale{1}\fi
\unitlength \JPicScale mm
\begin{picture}(90,70)(0,0)
\linethickness{1mm}
\multiput(30,70)(0.18,-0.12){333}{\line(1,0){0.18}}
\linethickness{1mm}
\multiput(30,30)(0.18,0.12){333}{\line(1,0){0.18}}
\linethickness{0.4mm}
\qbezier(45,60)(52.83,65.28)(56.44,62.88)
\qbezier(56.44,62.88)(60.05,60.47)(60,50)
\linethickness{0.2mm}
\put(15,60){\line(1,0){30}}

\put(60,35){\makebox(0,0)[cc]{$\cdots$}}

\put(32,60){\makebox(0,0)[cc]{$\gamma$}}

\put(52,63){\makebox(0,0)[cc]{$g$}}

\end{picture}

}

\def\figemextrac{

\def\JPicScale{0.4}
\ifx\JPicScale\undefined\def\JPicScale{1}\fi
\unitlength \JPicScale mm
\begin{picture}(90,70)(0,0)
\linethickness{1mm}
\multiput(30,70)(0.18,-0.12){333}{\line(1,0){0.18}}
\linethickness{1mm}
\multiput(30,30)(0.18,0.12){333}{\line(1,0){0.18}}
\linethickness{0.4mm}
\qbezier(45,60)(52.83,65.28)(56.44,62.88)
\qbezier(56.44,62.88)(60.05,60.47)(60,50)
\linethickness{0.2mm}
\put(8,65){\line(1,0){30}}

\put(60,35){\makebox(0,0)[cc]{$\cdots$}}

\put(20,65){\makebox(0,0)[cc]{$\gamma$}}

\put(52,63){\makebox(0,0)[cc]{$g$}}

\end{picture}

}

\def\figemextrad{

\def\JPicScale{0.6}
\ifx\JPicScale\undefined\def\JPicScale{1}\fi
\unitlength \JPicScale mm
\begin{picture}(90,70)(0,0)
\linethickness{1mm}
\multiput(30,70)(0.18,-0.12){333}{\line(1,0){0.18}}
\linethickness{1mm}
\multiput(30,30)(0.18,0.12){333}{\line(1,0){0.18}}
\linethickness{0.2mm}
\qbezier(45,60)(52.83,65.28)(56.44,62.88)
\linethickness{0.4mm}
\qbezier(56.44,62.88)(60.05,60.47)(60,50)
\linethickness{0.2mm}
\put(56,63){\line(0,1){15}}

\put(60,35){\makebox(0,0)[cc]{$\cdots$}}

\put(50,63){\makebox(0,0)[cc]{$\gamma$}}

\put(56,70){\makebox(0,0)[cc]{$\gamma$}}

\put(60,57){\makebox(0,0)[cc]{$g$}}

\end{picture}

}

\def\figemextrae{

\def\JPicScale{0.4}
\ifx\JPicScale\undefined\def\JPicScale{1}\fi
\unitlength \JPicScale mm
\begin{picture}(90,70)(0,0)
\linethickness{1mm}
\multiput(30,70)(0.18,-0.12){333}{\line(1,0){0.18}}
\linethickness{1mm}
\multiput(30,30)(0.18,0.12){333}{\line(1,0){0.18}}
\linethickness{0.4mm}
\qbezier(45,60)(52.83,65.28)(56.44,62.88)
\linethickness{0.4mm}
\qbezier(56.44,62.88)(60.05,60.47)(60,50)
\linethickness{0.2mm}
\put(75,58){\line(0,1){15}}

\put(60,35){\makebox(0,0)[cc]{$\cdots$}}

\put(75,68){\makebox(0,0)[cc]{$\gamma$}}

\put(52,63){\makebox(0,0)[cc]{$g$}}

\end{picture}

}

\def\figemselfgr{

\def\JPicScale{0.6}
\ifx\JPicScale\undefined\def\JPicScale{1}\fi
\unitlength \JPicScale mm
\begin{picture}(80,75)(0,0)
\linethickness{1mm}
\multiput(40,70)(0.12,-0.12){333}{\line(1,0){0.12}}
\linethickness{1mm}
\multiput(40,30)(0.12,0.12){333}{\line(1,0){0.12}}
\put(60,35){\makebox(0,0)[cc]{$\cdots$}}

\linethickness{0.2mm}
\qbezier(45,65)(44.98,65)(46.78,65)
\qbezier(46.78,65)(48.59,65)(52.5,65)
\linethickness{0.4mm}
\qbezier(52.5,65)(56.43,65.03)(57.03,62.62)
\qbezier(57.03,62.62)(57.63,60.22)(55,55)
\linethickness{0.2mm}
\put(52,65){\line(0,1){10}}

\put(52,70){\makebox(0,0)[cc]{$\gamma$}}

\put(57,62){\makebox(0,0)[cc]{$g$}}

\put(50,65){\makebox(0,0)[cc]{$\gamma$}}

\end{picture}

}

\def\figemselfgrnew{

\def\JPicScale{0.6}
\ifx\JPicScale\undefined\def\JPicScale{1}\fi
\unitlength \JPicScale mm
\begin{picture}(80,75)(0,0)
\linethickness{1mm}
\multiput(40,70)(0.12,-0.12){333}{\line(1,0){0.12}}
\linethickness{1mm}
\multiput(40,30)(0.12,0.12){333}{\line(1,0){0.12}}
\put(60,35){\makebox(0,0)[cc]{$\cdots$}}

\linethickness{0.4mm}
\qbezier(45,65)(44.98,65)(46.78,65)
\qbezier(46.78,65)(48.59,65)(52.5,65)
\linethickness{0.2mm}
\qbezier(52.5,65)(56.43,65.03)(57.03,62.62)
\qbezier(57.03,62.62)(57.63,60.22)(55,55)
\linethickness{0.2mm}
\put(52,65){\line(0,1){10}}

\put(52,70){\makebox(0,0)[cc]{$\gamma$}}

\put(57,62){\makebox(0,0)[cc]{$\gamma$}}

\put(50,65){\makebox(0,0)[cc]{$g$}}

\end{picture}

}

\def\figgrselfem{

\def\JPicScale{0.8}
\ifx\JPicScale\undefined\def\JPicScale{1}\fi
\unitlength \JPicScale mm
\begin{picture}(80,75)(0,0)
\linethickness{1mm}
\multiput(40,70)(0.12,-0.12){333}{\line(1,0){0.12}}
\linethickness{1mm}
\multiput(40,30)(0.12,0.12){333}{\line(1,0){0.12}}
\put(60,35){\makebox(0,0)[cc]{$\cdots$}}

\linethickness{0.2mm}
\qbezier(45,65)(44.98,65)(46.78,65)
\qbezier(46.78,65)(48.59,65)(52.5,65)
\linethickness{0.2mm}
\qbezier(52.5,65)(56.43,65.03)(57.03,62.62)
\qbezier(57.03,62.62)(57.63,60.22)(55,55)
\linethickness{0.4mm}
\put(52,65){\line(0,1){10}}

\put(48,65){\makebox(0,0)[cc]{$\gamma$}}

\put(57,62){\makebox(0,0)[cc]{$\gamma$}}

\put(52,70){\makebox(0,0)[cc]{$g$}}

\end{picture}

}

\def\figgrvertex{

\def\JPicScale{0.8}
\ifx\JPicScale\undefined\def\JPicScale{1}\fi
\unitlength \JPicScale mm
\begin{picture}(190,70)(0,0)
\linethickness{1mm}
\put(10,50){\line(1,0){30}}
\linethickness{0.4mm}
\put(25,50){\line(0,1){20}}
\linethickness{1mm}
\put(50,50){\line(1,0){30}}
\linethickness{0.4mm}
\multiput(50,70)(0.12,-0.16){125}{\line(0,-1){0.16}}
\linethickness{0.4mm}
\multiput(65,50)(0.12,0.16){125}{\line(0,1){0.16}}
\linethickness{0.4mm}
\put(110,50){\line(0,1){10}}
\linethickness{0.4mm}
\multiput(100,70)(0.12,-0.12){83}{\line(1,0){0.12}}
\linethickness{0.4mm}
\multiput(110,60)(0.12,0.12){83}{\line(1,0){0.12}}
\linethickness{1mm}
\multiput(140,50)(0.12,0.12){167}{\line(1,0){0.12}}
\linethickness{1mm}
\multiput(140,70)(0.12,-0.12){167}{\line(1,0){0.12}}
\linethickness{0.4mm}
\put(140,60){\line(1,0){10}}
\linethickness{1mm}
\multiput(170,50)(0.12,0.12){167}{\line(1,0){0.12}}
\linethickness{1mm}
\multiput(170,70)(0.12,-0.12){167}{\line(1,0){0.12}}
\linethickness{0.4mm}
\linethickness{0.4mm}
\put(180,60){\line(1,0){10}}
\linethickness{0.4mm}
\put(170,60){\line(1,0){10}}
\put(150,50){\makebox(0,0)[cc]{$\cdots$}}

\put(180,50){\makebox(0,0)[cc]{$\cdots$}}

\put(185,60){\makebox(0,0)[cc]{$g$}}

\put(175,60){\makebox(0,0)[cc]{$g$}}

\put(145,60){\makebox(0,0)[cc]{$g$}}

\put(110,55){\makebox(0,0)[cc]{$g$}}

\put(115,65){\makebox(0,0)[cc]{$g$}}

\put(105,65){\makebox(0,0)[cc]{$g$}}

\put(76,65){\makebox(0,0)[cc]{$g$}}

\put(54,65){\makebox(0,0)[cc]{$g$}}

\put(25,65){\makebox(0,0)[cc]{$g$}}

\end{picture}

}

\def\figphgr{

\def\JPicScale{0.5}
\ifx\JPicScale\undefined\def\JPicScale{1}\fi
\unitlength \JPicScale mm
\begin{picture}(100,75)(0,0)
\linethickness{1mm}
\multiput(40,110)(0.12,-0.12){500}{\line(1,0){0.12}}
\linethickness{1mm}
\multiput(40,50)(0.12,0.12){500}{\line(1,0){0.12}}
\linethickness{0.2mm}
\put(50,100){\line(1,0){20}}
\linethickness{0.2mm}
\put(70,100){\line(1,0){20}}
\linethickness{0.4mm}
\put(70,100){\line(0,1){15}}
\put(70,55){\makebox(0,0)[cc]{$\cdots$}}

\put(60,100){\makebox(0,0)[cc]{$\gamma$}}

\put(80,100){\makebox(0,0)[cc]{$\gamma$}}

\put(70,110){\makebox(0,0)[cc]{$g$}}

\end{picture}

}

\def\figintgr{

\def\JPicScale{0.5}
\ifx\JPicScale\undefined\def\JPicScale{1}\fi
\unitlength \JPicScale mm
\begin{picture}(100,75)(0,0)
\linethickness{1mm}
\multiput(40,70)(0.12,-0.12){500}{\line(1,0){0.12}}
\linethickness{1mm}
\multiput(40,10)(0.12,0.12){500}{\line(1,0){0.12}}
\linethickness{0.2mm}
\put(50,60){\line(1,0){20}}
\linethickness{0.4mm}
\put(70,60){\line(1,0){20}}
\linethickness{0.2mm}
\put(70,60){\line(0,1){15}}
\put(70,15){\makebox(0,0)[cc]{$\cdots$}}

\put(60,60){\makebox(0,0)[cc]{$\gamma$}}

\put(80,60){\makebox(0,0)[cc]{$\gamma$}}

\put(70,70){\makebox(0,0)[cc]{$g$}}

\end{picture}

}

\def\figintgr{

\def\JPicScale{0.5}
\ifx\JPicScale\undefined\def\JPicScale{1}\fi
\unitlength \JPicScale mm
\begin{picture}(100,75)(0,0)
\linethickness{1mm}
\multiput(40,90)(0.12,-0.12){500}{\line(1,0){0.12}}
\linethickness{1mm}
\multiput(40,30)(0.12,0.12){500}{\line(1,0){0.12}}
\linethickness{0.2mm}
\put(50,80){\line(1,0){20}}
\linethickness{0.4mm}
\put(70,80){\line(1,0){20}}
\linethickness{0.2mm}
\put(70,80){\line(0,1){15}}
\put(70,35){\makebox(0,0)[cc]{$\cdots$}}

\put(60,80){\makebox(0,0)[cc]{$\gamma$}}

\put(80,80){\makebox(0,0)[cc]{$g$}}

\put(70,90){\makebox(0,0)[cc]{$\gamma$}}

\end{picture}

}

\def\figKnon{

\def\JPicScale{0.5}
\ifx\JPicScale\undefined\def\JPicScale{1}\fi
\unitlength \JPicScale mm
\begin{picture}(135,70)(0,0)
\linethickness{1mm}
\put(20,50){\line(1,0){30}}
\linethickness{0.4mm}
\put(35,50){\line(0,1){20}}
\linethickness{0.4mm}
\put(30,30){\line(0,1){20}}
\linethickness{1mm}
\put(60,50){\line(1,0){30}}
\linethickness{0.4mm}
\put(75,50){\line(0,1){20}}
\linethickness{0.4mm}
\put(80,30){\line(0,1){20}}
\linethickness{1mm}
\put(100,50){\line(1,0){30}}
\linethickness{0.4mm}
\put(115,50){\line(0,1){20}}
\linethickness{0.4mm}
\put(115,30){\line(0,1){20}}
\put(55,50){\makebox(0,0)[cc]{+}}

\put(95,50){\makebox(0,0)[cc]{+}}

\put(145,50){\makebox(0,0)[cc]{$\neq$}}

\put(30,40){\makebox(0,0)[cc]{K}}

\put(80,40){\makebox(0,0)[cc]{K}}

\put(115,40){\makebox(0,0)[cc]{K}}

\end{picture}

}

\def\figKgauge{

\def\JPicScale{0.5}
\ifx\JPicScale\undefined\def\JPicScale{1}\fi
\unitlength \JPicScale mm
\begin{picture}(135,70)(0,0)
\linethickness{1mm}
\put(20,50){\line(1,0){30}}
\linethickness{0.2mm}
\put(35,50){\line(0,1){20}}
\linethickness{0.2mm}
\put(30,30){\line(0,1){20}}
\linethickness{1mm}
\put(60,50){\line(1,0){30}}
\linethickness{0.2mm}
\put(75,50){\line(0,1){20}}
\linethickness{0.2mm}
\put(80,30){\line(0,1){20}}
\linethickness{1mm}
\put(100,50){\line(1,0){30}}
\linethickness{0.2mm}
\put(115,50){\line(0,1){20}}
\linethickness{0.2mm}
\put(115,30){\line(0,1){20}}
\put(55,50){\makebox(0,0)[cc]{+}}

\put(95,50){\makebox(0,0)[cc]{+}}

\put(150,50){\makebox(0,0)[cc]{=}}

\put(30,40){\makebox(0,0)[cc]{K}}

\put(80,40){\makebox(0,0)[cc]{K}}

\put(115,40){\makebox(0,0)[cc]{K}}

\end{picture}

}

\def\figKgaugenew{

\def\JPicScale{0.5}
\ifx\JPicScale\undefined\def\JPicScale{1}\fi
\unitlength \JPicScale mm
\begin{picture}(135,70)(0,0)
\linethickness{1mm}
\put(20,50){\line(1,0){30}}
\linethickness{0.2mm}
\put(33,50){\line(0,1){20}}
\linethickness{0.2mm}
\put(30,30){\line(0,1){20}}
\linethickness{1mm}
\put(60,50){\line(1,0){30}}
\linethickness{0.2mm}
\put(75,50){\line(0,1){20}}
\linethickness{0.2mm}
\put(78,30){\line(0,1){20}}
\linethickness{1mm}
\put(100,50){\line(1,0){30}}
\linethickness{0.2mm}
\put(115,50){\line(0,1){20}}
\linethickness{0.2mm}
\put(115,30){\line(0,1){20}}
\put(55,50){\makebox(0,0)[cc]{+}}

\put(95,50){\makebox(0,0)[cc]{+}}

\put(12,50){\makebox(0,0)[cc]{$-$}}

\put(28,65){\makebox(0,0)[cc]{$k$}}

\put(25,35){\makebox(0,0)[cc]{$\ell$}}

\put(150,50){\makebox(0,0)[cc]{= \quad 0}}

\put(30,50){\circle{5}}

\put(78,50){\circle{5}}

\put(115,46){\makebox(0,0)[cc]{$\Uparrow$}}

\put(30,40){\makebox(0,0)[cc]{$\gamma$}}

\put(33,60){\makebox(0,0)[cc]{$\gamma$}}

\put(78,40){\makebox(0,0)[cc]{$\gamma$}}

\put(75,60){\makebox(0,0)[cc]{$\gamma$}}

\put(115,35){\makebox(0,0)[cc]{$\gamma$}}

\put(115,60){\makebox(0,0)[cc]{$\gamma$}}

\end{picture}

}

\def\figKgaugenewqedgr{

\def\JPicScale{0.5}
\ifx\JPicScale\undefined\def\JPicScale{1}\fi
\unitlength \JPicScale mm
\begin{picture}(135,70)(0,0)
\linethickness{1mm}
\put(20,50){\line(1,0){30}}
\linethickness{0.4mm}
\put(33,50){\line(0,1){20}}
\linethickness{0.2mm}
\put(30,30){\line(0,1){20}}
\linethickness{1mm}
\put(60,50){\line(1,0){30}}
\linethickness{0.4mm}
\put(75,50){\line(0,1){20}}
\linethickness{0.2mm}
\put(78,30){\line(0,1){20}}
\linethickness{1mm}
\put(100,50){\line(1,0){30}}
\linethickness{0.4mm}
\put(115,50){\line(0,1){20}}
\linethickness{0.2mm}
\put(115,30){\line(0,1){20}}
\put(55,50){\makebox(0,0)[cc]{+}}

\put(95,50){\makebox(0,0)[cc]{+}}

\put(12,50){\makebox(0,0)[cc]{$-$}}

\put(28,65){\makebox(0,0)[cc]{$k$}}

\put(25,35){\makebox(0,0)[cc]{$\ell$}}

\put(150,50){\makebox(0,0)[cc]{= \quad 0}}

\put(30,50){\circle{5}}

\put(78,50){\circle{5}}

\put(115,46){\makebox(0,0)[cc]{$\Uparrow$}}

\put(78,35){\makebox(0,0)[cc]{$\gamma$}}

\put(115,35){\makebox(0,0)[cc]{$\gamma$}}

\put(30,35){\makebox(0,0)[cc]{$\gamma$}}

\put(75,60){\makebox(0,0)[cc]{$g$}}

\put(115,60){\makebox(0,0)[cc]{$g$}}

\put(33,60){\makebox(0,0)[cc]{$g$}}

\end{picture}

}

\def\figmiddle{

\def\JPicScale{0.5}
\ifx\JPicScale\undefined\def\JPicScale{1}\fi
\unitlength \JPicScale mm
\begin{picture}(135,80)(0,0)

\linethickness{1mm}
\put(-30,50){\line(1,0){30}}
\linethickness{0.2mm}
\put(-15,20){\line(0,1){30}}

\put(-15,47){\makebox(0,0)[cc]{$\Uparrow$}}

\put(10,50){\makebox(0,0)[cc]{=}}

\linethickness{1mm}
\put(30,50){\line(1,0){40}}
\linethickness{0.2mm}
\linethickness{0.2mm}
\multiput(30,50)(0.12,-0.18){167}{\line(0,-1){0.18}}
\linethickness{0.2mm}
\put(30,50){\circle{5}}

\linethickness{1mm}
\put(90,50){\line(1,0){40}}
\linethickness{0.2mm}
\linethickness{0.3mm}
\multiput(110,20)(0.12,0.18){167}{\line(0,1){0.18}}
\linethickness{0.3mm}
\put(130,50){\circle{5}}

\put(-25,55){\makebox(0,0)[cc]{$p_c$}}

\put(-20,20){\makebox(0,0)[cc]{$\ell$}}

\put(80,50){\makebox(0,0)[cc]{$-$}}

\put(140,50){\makebox(0,0)[cc]{,}}

\put(116,30){\makebox(0,0)[cc]{$\gamma$}}

\put(43,30){\makebox(0,0)[cc]{$\gamma$}}

\put(-15,35){\makebox(0,0)[cc]{$\gamma$}}

\end{picture}

}

\def\figKgaugenewgr{

\def\JPicScale{0.7}
\ifx\JPicScale\undefined\def\JPicScale{1}\fi
\unitlength \JPicScale mm
\begin{picture}(135,70)(0,0)
\linethickness{1mm}
\put(20,50){\line(1,0){30}}
\linethickness{0.4mm}
\put(33,50){\line(0,1){20}}
\linethickness{0.4mm}
\put(30,30){\line(0,1){20}}
\linethickness{1mm}
\put(60,50){\line(1,0){30}}
\linethickness{0.4mm}
\put(75,50){\line(0,1){20}}
\linethickness{0.4mm}
\put(78,30){\line(0,1){20}}
\linethickness{1mm}
\put(100,50){\line(1,0){30}}
\linethickness{0.4mm}
\put(115,50){\line(0,1){20}}
\linethickness{0.4mm}
\put(115,30){\line(0,1){20}}
\put(55,50){\makebox(0,0)[cc]{+}}

\put(95,50){\makebox(0,0)[cc]{+}}

\put(12,50){\makebox(0,0)[cc]{$-$}}

\put(20,55){\makebox(0,0)[cc]{$p_c\rightarrow $}}

\put(28,65){\makebox(0,0)[cc]{$\ell\downarrow$}}

\put(25,35){\makebox(0,0)[cc]{$k\uparrow$}}

\put(170,50){\makebox(0,0)[cc]{$=$ \quad $A(p_c,k,\ell,\xi,\zeta)$}}

\put(30,50){\circle{5}}

\put(78,50){\circle{5}}

\put(115,47){\makebox(0,0)[cc]{$\Uparrow$}}

\put(115,40){\makebox(0,0)[cc]{$g$}}

\put(115,60){\makebox(0,0)[cc]{$g$}}

\put(78,40){\makebox(0,0)[cc]{$g$}}

\put(75,60){\makebox(0,0)[cc]{$g$}}

\put(30,40){\makebox(0,0)[cc]{$g$}}

\put(33,60){\makebox(0,0)[cc]{$g$}}

\end{picture}

}

\def\figKgaugenewgraa{

\def\JPicScale{0.5}
\ifx\JPicScale\undefined\def\JPicScale{1}\fi
\unitlength \JPicScale mm
\begin{picture}(135,70)(0,0)
\linethickness{1mm}
\put(20,50){\line(1,0){30}}
\linethickness{0.2mm}
\put(33,50){\line(0,1){20}}
\linethickness{0.4mm}
\put(30,30){\line(0,1){20}}
\linethickness{1mm}
\put(60,50){\line(1,0){30}}
\linethickness{0.2mm}
\put(75,50){\line(0,1){20}}
\linethickness{0.4mm}
\put(78,30){\line(0,1){20}}
\linethickness{1mm}
\put(100,50){\line(1,0){30}}
\linethickness{0.2mm}
\put(115,50){\line(0,1){20}}
\linethickness{0.4mm}
\put(115,30){\line(0,1){20}}
\put(55,50){\makebox(0,0)[cc]{+}}

\put(95,50){\makebox(0,0)[cc]{+}}

\put(12,50){\makebox(0,0)[cc]{$-$}}

\put(28,65){\makebox(0,0)[cc]{$k$}}

\put(25,35){\makebox(0,0)[cc]{$\ell$}}

\put(220,50){\makebox(0,0)[cc]{$=-2\, i\, q_c\, \left\{ \xi.k\, \eps.(2p+k+\ell) + \xi.\eps \, \ell.(2p+k+\ell)
\right\} $}}

\put(30,50){\circle{5}}

\put(78,50){\circle{5}}

\put(115,46){\makebox(0,0)[cc]{$\Uparrow$}}

\put(78,35){\makebox(0,0)[cc]{$g$}}

\put(115,35){\makebox(0,0)[cc]{$g$}}

\put(30,35){\makebox(0,0)[cc]{$g$}}

\put(75,60){\makebox(0,0)[cc]{$\gamma$}}

\put(115,60){\makebox(0,0)[cc]{$\gamma$}}

\put(33,60){\makebox(0,0)[cc]{$\gamma$}}

\end{picture}

}

\def\figmiddlegr{

\def\JPicScale{0.5}
\ifx\JPicScale\undefined\def\JPicScale{1}\fi
\unitlength \JPicScale mm
\begin{picture}(135,80)(0,0)

\linethickness{1mm}
\put(-30,50){\line(1,0){30}}
\linethickness{0.4mm}
\put(-15,20){\line(0,1){30}}

\put(-15,47){\makebox(0,0)[cc]{$\Uparrow$}}

\put(10,50){\makebox(0,0)[cc]{=}}

\linethickness{1mm}
\put(30,50){\line(1,0){40}}
\linethickness{0.4mm}
\linethickness{0.4mm}
\multiput(30,50)(0.12,-0.18){167}{\line(0,-1){0.18}}
\linethickness{0.3mm}
\put(30,50){\circle{5}}

\linethickness{1mm}
\put(90,50){\line(1,0){40}}
\linethickness{0.2mm}
\linethickness{0.4mm}
\multiput(110,20)(0.12,0.18){167}{\line(0,1){0.18}}
\linethickness{0.3mm}
\put(130,50){\circle{5}}

\put(-25,55){\makebox(0,0)[cc]{$p_c\rightarrow$}}

\put(-20,30){\makebox(0,0)[cc]{$\uparrow$}}

\put(-20,20){\makebox(0,0)[cc]{$k$}}

\put(80,50){\makebox(0,0)[cc]{$-$}}

\put(140,50){\makebox(0,0)[cc]{,}}

\put(-15,30){\makebox(0,0)[cc]{$g$}}

\put(-15,30){\makebox(0,0)[cc]{$g$}}

\put(43,30){\makebox(0,0)[cc]{$g$}}

\put(117,30){\makebox(0,0)[cc]{$g$}}

\end{picture}

}

\def\figward{

\def\JPicScale{0.5}
\ifx\JPicScale\undefined\def\JPicScale{1}\fi
\unitlength \JPicScale mm
\begin{picture}(135,80)(0,0)
\linethickness{1mm}
\put(30,50){\line(1,0){40}}
\linethickness{0.2mm}
\put(50,50){\line(0,1){30}}
\linethickness{0.2mm}
\multiput(30,50)(0.12,-0.18){167}{\line(0,-1){0.18}}
\linethickness{1mm}
\put(90,50){\line(1,0){40}}
\linethickness{0.2mm}
\put(110,50){\line(0,1){30}}
\linethickness{0.3mm}
\multiput(110,20)(0.12,0.18){167}{\line(0,1){0.18}}
\linethickness{0.2mm}
\put(30,50){\circle{10}}

\linethickness{0.3mm}
\put(130,50){\circle{10}}

\put(80,50){\makebox(0,0)[cc]{$-$}}

\put(38,40){\makebox(0,0)[cc]{K}}

\put(122,40){\makebox(0,0)[cc]{K}}

\end{picture}

}

\def\figgrward{

\def\JPicScale{0.5}
\ifx\JPicScale\undefined\def\JPicScale{1}\fi
\unitlength \JPicScale mm
\begin{picture}(135,80)(0,0)
\linethickness{1mm}
\put(30,50){\line(1,0){40}}
\linethickness{0.4mm}
\put(50,50){\line(0,1){30}}
\linethickness{0.4mm}
\multiput(30,50)(0.12,-0.18){167}{\line(0,-1){0.18}}
\linethickness{1mm}
\put(90,50){\line(1,0){40}}
\linethickness{0.4mm}
\put(110,50){\line(0,1){30}}
\linethickness{0.4mm}
\multiput(110,20)(0.12,0.18){167}{\line(0,1){0.18}}
\linethickness{0.3mm}
\put(30,50){\circle{10}}

\linethickness{0.3mm}
\put(130,50){\circle{10}}

\put(80,50){\makebox(0,0)[cc]{$-$}}

\put(38,40){\makebox(0,0)[cc]{K}}

\put(122,40){\makebox(0,0)[cc]{K}}

\end{picture}

}

\def\figemward{

\def\JPicScale{0.5}
\ifx\JPicScale\undefined\def\JPicScale{1}\fi
\unitlength \JPicScale mm
\begin{picture}(135,80)(0,0)
\linethickness{1mm}
\put(30,50){\line(1,0){40}}
\linethickness{0.2mm}
\put(50,50){\line(0,1){30}}
\linethickness{0.4mm}
\multiput(30,50)(0.12,-0.18){167}{\line(0,-1){0.18}}
\linethickness{1mm}
\put(90,50){\line(1,0){40}}
\linethickness{0.2mm}
\put(110,50){\line(0,1){30}}
\linethickness{0.4mm}
\multiput(110,20)(0.12,0.18){167}{\line(0,1){0.18}}
\linethickness{0.3mm}
\put(30,50){\circle{10}}

\linethickness{0.3mm}
\put(130,50){\circle{10}}

\put(80,50){\makebox(0,0)[cc]{$-$}}

\put(38,40){\makebox(0,0)[cc]{K}}

\put(122,40){\makebox(0,0)[cc]{K}}

\end{picture}

}

\def\fignonph{

\def\JPicScale{0.5}
\ifx\JPicScale\undefined\def\JPicScale{1}\fi
\unitlength \JPicScale mm
\begin{picture}(135,70)(0,0)
\linethickness{1mm}
\put(20,50){\line(1,0){30}}
\linethickness{0.2mm}
\put(35,50){\line(0,1){20}}
\linethickness{0.4mm}
\put(30,30){\line(0,1){20}}
\linethickness{1mm}
\put(60,50){\line(1,0){30}}
\linethickness{0.2mm}
\put(75,50){\line(0,1){20}}
\linethickness{0.4mm}
\put(80,30){\line(0,1){20}}
\linethickness{1mm}
\put(100,50){\line(1,0){30}}
\linethickness{0.2mm}
\put(115,50){\line(0,1){20}}
\linethickness{0.4mm}
\put(115,30){\line(0,1){20}}
\put(55,50){\makebox(0,0)[cc]{+}}

\put(95,50){\makebox(0,0)[cc]{+}}

\put(145,50){\makebox(0,0)[cc]{$\neq$}}

\put(30,40){\makebox(0,0)[cc]{K}}

\put(80,40){\makebox(0,0)[cc]{K}}

\put(115,40){\makebox(0,0)[cc]{K}}

\end{picture}

}

\def\figgravitya{

\def\JPicScale{0.6}
\ifx\JPicScale\undefined\def\JPicScale{1}\fi
\unitlength \JPicScale mm
\begin{picture}(80,80)(0,0)
\linethickness{1mm}
\multiput(20,70)(0.14,-0.12){417}{\line(1,0){0.14}}
\linethickness{1mm}
\multiput(20,20)(0.14,0.12){417}{\line(1,0){0.14}}
\linethickness{0.4mm}
\put(32,60){\line(1,0){35}}
\linethickness{0.4mm}
\put(50,60){\line(0,1){20}}
\put(20,75){\makebox(0,0)[cc]{$p_a$}}

\put(80,75){\makebox(0,0)[cc]{$p_b$}}

\put(50,25){\makebox(0,0)[cc]{$\cdots$}}

\put(55,80){\makebox(0,0)[cc]{$k$}}

\put(40,65){\makebox(0,0)[cc]{$\ell$}}

\put(60,65){\makebox(0,0)[cc]{$k-\ell$}}

\put(30,50){\makebox(0,0)[cc]{$p_a+\ell$}}

\put(72,50){\makebox(0,0)[cc]{$p_b+k-\ell$}}

\put(45,60){\makebox(0,0)[cc]{$g$}}

\put(60,60){\makebox(0,0)[cc]{$g$}}

\put(50,70){\makebox(0,0)[cc]{$g$}}

\end{picture}

}

\def\figgravityb{

\def\JPicScale{0.6}
\ifx\JPicScale\undefined\def\JPicScale{1}\fi
\unitlength \JPicScale mm
\begin{picture}(80,80)(0,0)
\linethickness{1mm}
\multiput(20,70)(0.14,-0.12){417}{\line(1,0){0.14}}
\linethickness{1mm}
\multiput(20,20)(0.14,0.12){417}{\line(1,0){0.14}}
\put(20,75){\makebox(0,0)[cc]{$p_a$}}

\put(80,75){\makebox(0,0)[cc]{$p_b$}}

\put(50,25){\makebox(0,0)[cc]{$\cdots$}}

\linethickness{0.4mm}
\qbezier(25,65)(38.06,67.67)(42.88,64.06)
\qbezier(42.88,64.06)(47.69,60.45)(45,50)
\linethickness{0.4mm}
\put(40,65){\line(0,1){15}}

\put(45,75){\makebox(0,0)[cc]{$k$}}

\put(30,70){\makebox(0,0)[cc]{$\ell$}}

\put(55,60){\makebox(0,0)[cc]{$k-\ell$}}

\put(22,55){\makebox(0,0)[cc]{$p_a+\ell$}}

\put(45,62){\makebox(0,0)[cc]{$g$}}

\put(40,70){\makebox(0,0)[cc]{$g$}}

\put(34,66){\makebox(0,0)[cc]{$g$}}

\end{picture}

}

\def\figqeda{

\def\JPicScale{0.5}
\ifx\JPicScale\undefined\def\JPicScale{1}\fi
\unitlength \JPicScale mm
\begin{picture}(110,85)(0,0)
\linethickness{1mm}
\multiput(30,80)(0.12,-0.12){333}{\line(1,0){0.12}}
\linethickness{1mm}
\multiput(70,40)(0.12,0.12){333}{\line(1,0){0.12}}
\linethickness{1mm}
\multiput(40,10)(0.12,0.12){250}{\line(1,0){0.12}}
\linethickness{1mm}
\multiput(70,40)(0.12,-0.12){250}{\line(1,0){0.12}}
\linethickness{0.2mm}
\put(50,60){\line(1,0){40}}
\linethickness{0.2mm}
\put(10,70){\line(1,0){30}}
\put(35,85){\makebox(0,0)[cc]{$p_a$}}

\put(20,65){\makebox(0,0)[cc]{$k$}}

\put(58,69){\makebox(0,0)[cc]{$p_a+k$}}

\put(79,65){\makebox(0,0)[cc]{$\ell\rightarrow$}}

\put(35,50){\makebox(0,0)[cc]{$p_a+k-\ell$}}

\put(110,85){\makebox(0,0)[cc]{$p_b$}}

\put(94,50){\makebox(0,0)[cc]{$p_b+\ell$}}

\put(72,20){\makebox(0,0)[cc]{$\cdots$}}

\put(72,0){\makebox(0,0)[cc]{(a)}}

\put(68,60){\makebox(0,0)[cc]{$\gamma$}}

\put(30,70){\makebox(0,0)[cc]{$\gamma$}}

\put(148,60){\makebox(0,0)[cc]{$\gamma$}}

\put(180,60){\makebox(0,0)[cc]{$\gamma$}}

\put(275,55){\makebox(0,0)[cc]{$\gamma$}}

\put(290,70){\makebox(0,0)[cc]{$\gamma$}}

\end{picture}

}

\def\figqedb{

\def\JPicScale{0.5}
\ifx\JPicScale\undefined\def\JPicScale{1}\fi
\unitlength \JPicScale mm
\begin{picture}(110,85)(0,0)
\linethickness{1mm}
\multiput(30,80)(0.12,-0.12){333}{\line(1,0){0.12}}
\linethickness{1mm}
\multiput(70,40)(0.12,0.12){333}{\line(1,0){0.12}}
\linethickness{1mm}
\multiput(40,10)(0.12,0.12){250}{\line(1,0){0.12}}
\linethickness{1mm}
\multiput(70,40)(0.12,-0.12){250}{\line(1,0){0.12}}
\linethickness{0.2mm}
\put(50,60){\line(1,0){40}}
\linethickness{0.2mm}
\put(20,60){\line(1,0){30}}
\put(35,85){\makebox(0,0)[cc]{$p_a$}}

\put(20,65){\makebox(0,0)[cc]{$k$}}

\put(79,65){\makebox(0,0)[cc]{$\ell\rightarrow$}}

\put(35,50){\makebox(0,0)[cc]{$p_a+k-\ell$}}

\put(110,85){\makebox(0,0)[cc]{$p_b$}}

\put(94,50){\makebox(0,0)[cc]{$p_b+\ell$}}

\put(72,20){\makebox(0,0)[cc]{$\cdots$}}

\put(72,0){\makebox(0,0)[cc]{(b)}}

\end{picture}

}

\def\figqedc{

\def\JPicScale{0.5}
\ifx\JPicScale\undefined\def\JPicScale{1}\fi
\unitlength \JPicScale mm
\begin{picture}(110,85)(0,0)
\linethickness{1mm}
\multiput(30,80)(0.12,-0.12){333}{\line(1,0){0.12}}
\linethickness{1mm}
\multiput(70,40)(0.12,0.12){333}{\line(1,0){0.12}}
\linethickness{1mm}
\multiput(40,10)(0.12,0.12){250}{\line(1,0){0.12}}
\linethickness{1mm}
\multiput(70,40)(0.12,-0.12){250}{\line(1,0){0.12}}
\linethickness{0.2mm}
\put(40,70){\line(1,0){60}}
\linethickness{0.2mm}
\put(25,55){\line(1,0){30}}
\put(35,85){\makebox(0,0)[cc]{$p_a$}}

\put(25,60){\makebox(0,0)[cc]{$k$}}

\put(62,65){\makebox(0,0)[cc]{$p_a-\ell$}}

\put(79,75){\makebox(0,0)[cc]{$\ell\rightarrow$}}

\put(40,45){\makebox(0,0)[cc]{$p_a+k-\ell$}}

\put(110,85){\makebox(0,0)[cc]{$p_b$}}

\put(94,50){\makebox(0,0)[cc]{$p_b+\ell$}}

\put(72,20){\makebox(0,0)[cc]{$\cdots$}}

\put(72,0){\makebox(0,0)[cc]{(c)}}

\end{picture}

}

\def\figqedafin{

\def\JPicScale{0.4}
\ifx\JPicScale\undefined\def\JPicScale{1}\fi
\unitlength \JPicScale mm
\begin{picture}(110,85)(0,0)
\linethickness{1mm}
\multiput(30,80)(0.12,-0.12){333}{\line(1,0){0.12}}
\linethickness{1mm}
\multiput(70,40)(0.12,0.12){333}{\line(1,0){0.12}}
\linethickness{1mm}
\multiput(40,10)(0.12,0.12){250}{\line(1,0){0.12}}
\linethickness{1mm}
\multiput(70,40)(0.12,-0.12){250}{\line(1,0){0.12}}
\linethickness{0.4mm}
\put(50,60){\line(1,0){40}}
\linethickness{0.2mm}
\put(10,70){\line(1,0){30}}
\put(35,85){\makebox(0,0)[cc]{$p_a$}}

\put(20,65){\makebox(0,0)[cc]{$k$}}

\put(110,85){\makebox(0,0)[cc]{$p_b$}}

\put(72,20){\makebox(0,0)[cc]{$\cdots$}}

\put(72,0){\makebox(0,0)[cc]{(a)}}

\put(70,60){\makebox(0,0)[cc]{$g$}}

\put(27,70){\makebox(0,0)[cc]{$\gamma$}}

\end{picture}

}

\def\figqedbfin{

\def\JPicScale{0.4}
\ifx\JPicScale\undefined\def\JPicScale{1}\fi
\unitlength \JPicScale mm
\begin{picture}(110,85)(0,0)
\linethickness{1mm}
\multiput(30,80)(0.12,-0.12){333}{\line(1,0){0.12}}
\linethickness{1mm}
\multiput(70,40)(0.12,0.12){333}{\line(1,0){0.12}}
\linethickness{1mm}
\multiput(40,10)(0.12,0.12){250}{\line(1,0){0.12}}
\linethickness{1mm}
\multiput(70,40)(0.12,-0.12){250}{\line(1,0){0.12}}
\linethickness{0.4mm}
\put(50,60){\line(1,0){40}}
\linethickness{0.2mm}
\put(20,60){\line(1,0){30}}
\put(35,85){\makebox(0,0)[cc]{$p_a$}}

\put(20,65){\makebox(0,0)[cc]{$k$}}

\put(110,85){\makebox(0,0)[cc]{$p_b$}}

\put(72,20){\makebox(0,0)[cc]{$\cdots$}}

\put(72,0){\makebox(0,0)[cc]{(b)}}

\put(70,60){\makebox(0,0)[cc]{$g$}}

\put(35,60){\makebox(0,0)[cc]{$\gamma$}}

\end{picture}

}

\def\figqedcfin{

\def\JPicScale{0.4}
\ifx\JPicScale\undefined\def\JPicScale{1}\fi
\unitlength \JPicScale mm
\begin{picture}(110,85)(0,0)
\linethickness{1mm}
\multiput(30,80)(0.12,-0.12){333}{\line(1,0){0.12}}
\linethickness{1mm}
\multiput(70,40)(0.12,0.12){333}{\line(1,0){0.12}}
\linethickness{1mm}
\multiput(40,10)(0.12,0.12){250}{\line(1,0){0.12}}
\linethickness{1mm}
\multiput(70,40)(0.12,-0.12){250}{\line(1,0){0.12}}
\linethickness{0.4mm}
\put(40,70){\line(1,0){60}}
\linethickness{0.2mm}
\put(25,55){\line(1,0){30}}
\put(35,85){\makebox(0,0)[cc]{$p_a$}}

\put(25,60){\makebox(0,0)[cc]{$k$}}

\put(110,85){\makebox(0,0)[cc]{$p_b$}}

\put(72,20){\makebox(0,0)[cc]{$\cdots$}}

\put(72,0){\makebox(0,0)[cc]{(c)}}

\put(70,70){\makebox(0,0)[cc]{$g$}}

\put(40,55){\makebox(0,0)[cc]{$\gamma$}}

\end{picture}

}

\def\figqedaqgr{

\def\JPicScale{0.4}
\ifx\JPicScale\undefined\def\JPicScale{1}\fi
\unitlength \JPicScale mm
\begin{picture}(110,85)(0,0)
\linethickness{1mm}
\multiput(30,80)(0.12,-0.12){333}{\line(1,0){0.12}}
\linethickness{1mm}
\multiput(70,40)(0.12,0.12){333}{\line(1,0){0.12}}
\linethickness{1mm}
\multiput(40,10)(0.12,0.12){250}{\line(1,0){0.12}}
\linethickness{1mm}
\multiput(70,40)(0.12,-0.12){250}{\line(1,0){0.12}}
\linethickness{0.2mm}
\put(50,60){\line(1,0){40}}
\linethickness{0.4mm}
\put(10,70){\line(1,0){30}}
\put(35,85){\makebox(0,0)[cc]{$p_a$}}

\put(20,65){\makebox(0,0)[cc]{$k$}}

\put(110,85){\makebox(0,0)[cc]{$p_b$}}

\put(72,20){\makebox(0,0)[cc]{$\cdots$}}

\put(72,0){\makebox(0,0)[cc]{(a)}}

\put(30,70){\makebox(0,0)[cc]{$g$}}

\put(70,60){\makebox(0,0)[cc]{$\gamma$}}

\end{picture}

}

\def\figqedbqgr{

\def\JPicScale{0.4}
\ifx\JPicScale\undefined\def\JPicScale{1}\fi
\unitlength \JPicScale mm
\begin{picture}(110,85)(0,0)
\linethickness{1mm}
\multiput(30,80)(0.12,-0.12){333}{\line(1,0){0.12}}
\linethickness{1mm}
\multiput(70,40)(0.12,0.12){333}{\line(1,0){0.12}}
\linethickness{1mm}
\multiput(40,10)(0.12,0.12){250}{\line(1,0){0.12}}
\linethickness{1mm}
\multiput(70,40)(0.12,-0.12){250}{\line(1,0){0.12}}
\linethickness{0.2mm}
\put(50,60){\line(1,0){40}}
\linethickness{0.4mm}
\put(20,60){\line(1,0){30}}
\put(35,85){\makebox(0,0)[cc]{$p_a$}}

\put(20,65){\makebox(0,0)[cc]{$k$}}

\put(110,85){\makebox(0,0)[cc]{$p_b$}}

\put(72,20){\makebox(0,0)[cc]{$\cdots$}}

\put(72,0){\makebox(0,0)[cc]{(b)}}

\put(35,60){\makebox(0,0)[cc]{$g$}}

\put(70,60){\makebox(0,0)[cc]{$\gamma$}}

\end{picture}

}

\def\figqedcqgr{

\def\JPicScale{0.4}
\ifx\JPicScale\undefined\def\JPicScale{1}\fi
\unitlength \JPicScale mm
\begin{picture}(110,85)(0,0)
\linethickness{1mm}
\multiput(30,80)(0.12,-0.12){333}{\line(1,0){0.12}}
\linethickness{1mm}
\multiput(70,40)(0.12,0.12){333}{\line(1,0){0.12}}
\linethickness{1mm}
\multiput(40,10)(0.12,0.12){250}{\line(1,0){0.12}}
\linethickness{1mm}
\multiput(70,40)(0.12,-0.12){250}{\line(1,0){0.12}}
\linethickness{0.2mm}
\put(40,70){\line(1,0){60}}
\linethickness{0.4mm}
\put(25,55){\line(1,0){30}}
\put(35,85){\makebox(0,0)[cc]{$p_a$}}

\put(25,60){\makebox(0,0)[cc]{$k$}}

\put(110,85){\makebox(0,0)[cc]{$p_b$}}

\put(72,20){\makebox(0,0)[cc]{$\cdots$}}

\put(72,0){\makebox(0,0)[cc]{(c)}}

\put(70,70){\makebox(0,0)[cc]{$\gamma$}}

\put(35,55){\makebox(0,0)[cc]{$g$}}

\end{picture}

}

\def\figqedtwo{

\def\JPicScale{0.5}
\ifx\JPicScale\undefined\def\JPicScale{1}\fi
\unitlength \JPicScale mm
\begin{picture}(110,85)(0,0)
\linethickness{1mm}
\multiput(30,80)(0.12,-0.12){333}{\line(1,0){0.12}}
\linethickness{1mm}
\multiput(70,40)(0.12,0.12){333}{\line(1,0){0.12}}
\linethickness{1mm}
\multiput(40,10)(0.12,0.12){250}{\line(1,0){0.12}}
\linethickness{1mm}
\multiput(70,40)(0.12,-0.12){250}{\line(1,0){0.12}}
\linethickness{0.2mm}
\put(50,60){\line(1,0){40}}
\linethickness{0.2mm}
\put(35,85){\makebox(0,0)[cc]{$p_a$}}

\put(79,65){\makebox(0,0)[cc]{$\ell\rightarrow$}}

\put(45,45){\makebox(0,0)[cc]{$p_a-\ell$}}

\put(110,85){\makebox(0,0)[cc]{$p_b$}}

\put(94,50){\makebox(0,0)[cc]{$p_b+\ell$}}

\put(72,20){\makebox(0,0)[cc]{$\cdots$}}

\put(60,60){\makebox(0,0)[cc]{$\gamma$}}

\end{picture}

}

\def\figqedtwoGR{

\def\JPicScale{0.5}
\ifx\JPicScale\undefined\def\JPicScale{1}\fi
\unitlength \JPicScale mm
\begin{picture}(110,85)(0,0)
\linethickness{1mm}
\multiput(30,80)(0.12,-0.12){333}{\line(1,0){0.12}}
\linethickness{1mm}
\multiput(70,40)(0.12,0.12){333}{\line(1,0){0.12}}
\linethickness{1mm}
\multiput(40,10)(0.12,0.12){250}{\line(1,0){0.12}}
\linethickness{1mm}
\multiput(70,40)(0.12,-0.12){250}{\line(1,0){0.12}}
\linethickness{0.4mm}
\put(50,60){\line(1,0){40}}
\linethickness{0.4mm}
\put(35,85){\makebox(0,0)[cc]{$p_a$}}

\put(75,65){\makebox(0,0)[cc]{$\leftarrow\ell$}}

\put(45,45){\makebox(0,0)[cc]{$p_a+\ell$}}

\put(110,85){\makebox(0,0)[cc]{$p_b$}}

\put(94,50){\makebox(0,0)[cc]{$p_b-\ell$}}

\put(72,20){\makebox(0,0)[cc]{$\cdots$}}

\put(65,60){\makebox(0,0)[cc]{$g$}}

\end{picture}

}

\def\figselfsum{

\def\JPicScale{0.5}
\ifx\JPicScale\undefined\def\JPicScale{1}\fi
\unitlength \JPicScale mm
\begin{picture}(80,75)(0,0)
\linethickness{1mm}
\multiput(15,75)(0.12,-0.12){292}{\line(1,0){0.12}}
\linethickness{1mm}
\multiput(50,40)(0.12,0.12){250}{\line(1,0){0.12}}
\linethickness{1mm}
\multiput(20,10)(0.12,0.12){250}{\line(1,0){0.12}}
\linethickness{1mm}
\multiput(50,40)(0.12,-0.12){250}{\line(1,0){0.12}}
\linethickness{0.2mm}
\put(12,58){\line(1,0){20}}
\linethickness{0.2mm}
\qbezier(20,70)(27.84,72.66)(30.25,70.25)
\qbezier(30.25,70.25)(32.66,67.84)(30,60)
\put(20,80){\makebox(0,0)[cc]{$p_a$}}

\put(12,53){\makebox(0,0)[cc]{$k$}}

\put(35,70){\makebox(0,0)[cc]{$\ell$}}

\put(32,58){\circle{5}}

\put(0,40){\makebox(0,0)[cc]{$-$}}

\put(50,15){\makebox(0,0)[cc]{$\cdots$}}

\put(30,70){\makebox(0,0)[cc]{$\gamma$}}

\put(20,58){\makebox(0,0)[cc]{$\gamma$}}

\end{picture}

}

\def\figselfa{

\def\JPicScale{0.4}
\ifx\JPicScale\undefined\def\JPicScale{1}\fi
\unitlength \JPicScale mm
\begin{picture}(80,75)(0,0)
\linethickness{1mm}
\multiput(20,70)(0.12,-0.12){250}{\line(1,0){0.12}}
\linethickness{1mm}
\multiput(50,40)(0.12,0.12){250}{\line(1,0){0.12}}
\linethickness{1mm}
\multiput(20,10)(0.12,0.12){250}{\line(1,0){0.12}}
\linethickness{1mm}
\multiput(50,40)(0.12,-0.12){250}{\line(1,0){0.12}}
\linethickness{0.2mm}
\put(10,60){\line(1,0){20}}
\linethickness{0.2mm}
\qbezier(25,65)(32.84,67.66)(35.25,65.25)
\qbezier(35.25,65.25)(37.66,62.84)(35,55)
\put(20,75){\makebox(0,0)[cc]{$p_a$}}

\put(10,55){\makebox(0,0)[cc]{$k$}}

\put(40,65){\makebox(0,0)[cc]{$\ell$}}

\put(50,15){\makebox(0,0)[cc]{$\cdots$}}

\put(20,60){\makebox(0,0)[cc]{$\gamma$}}

\put(31,66){\makebox(0,0)[cc]{$\gamma$}}

\end{picture}

}

\def\figselfb{

\ifx\JPicScale\undefined\def\JPicScale{1}\fi
\unitlength \JPicScale mm
\begin{picture}(80,75)(0,0)
\linethickness{1mm}
\multiput(15,75)(0.12,-0.12){292}{\line(1,0){0.12}}
\linethickness{1mm}
\multiput(50,40)(0.12,0.12){250}{\line(1,0){0.12}}
\linethickness{1mm}
\multiput(20,10)(0.12,0.12){250}{\line(1,0){0.12}}
\linethickness{1mm}
\multiput(50,40)(0.12,-0.12){250}{\line(1,0){0.12}}
\linethickness{0.2mm}
\put(10,60){\line(1,0){20}}
\linethickness{0.2mm}
\qbezier(20,70)(27.84,72.66)(30.25,70.25)
\qbezier(30.25,70.25)(32.66,67.84)(30,60)
\put(20,80){\makebox(0,0)[cc]{$p_a$}}

\put(10,55){\makebox(0,0)[cc]{$k$}}

\put(35,70){\makebox(0,0)[cc]{$\ell$}}

\put(50,15){\makebox(0,0)[cc]{$\cdots$}}

\put(20,60){\makebox(0,0)[cc]{$\gamma$}}

\put(27,70){\makebox(0,0)[cc]{$\gamma$}}

\end{picture}

}

\def\figselfc{

\ifx\JPicScale\undefined\def\JPicScale{1}\fi
\unitlength \JPicScale mm
\begin{picture}(80,75)(0,0)
\linethickness{1mm}
\multiput(20,70)(0.12,-0.12){250}{\line(1,0){0.12}}
\linethickness{1mm}
\multiput(50,40)(0.12,0.12){250}{\line(1,0){0.12}}
\linethickness{1mm}
\multiput(20,10)(0.12,0.12){250}{\line(1,0){0.12}}
\linethickness{1mm}
\multiput(50,40)(0.12,-0.12){250}{\line(1,0){0.12}}
\linethickness{0.2mm}
\put(10,60){\line(1,0){20}}
\linethickness{0.2mm}
\qbezier(30,60)(37.84,62.66)(40.25,60.25)
\qbezier(40.25,60.25)(42.66,57.84)(40,50)
\put(20,75){\makebox(0,0)[cc]{$p_a$}}

\put(10,55){\makebox(0,0)[cc]{$k$}}

\put(45,60){\makebox(0,0)[cc]{$\ell$}}

\put(50,15){\makebox(0,0)[cc]{$\cdots$}}

\put(20,60){\makebox(0,0)[cc]{$\gamma$}}

\put(36,61){\makebox(0,0)[cc]{$\gamma$}}

\end{picture}

}

\def\figselfd{

\def\JPicScale{0.4}
\ifx\JPicScale\undefined\def\JPicScale{1}\fi
\unitlength \JPicScale mm
\begin{picture}(80,75)(0,0)
\linethickness{1mm}
\multiput(20,70)(0.12,-0.12){250}{\line(1,0){0.12}}
\linethickness{1mm}
\multiput(50,40)(0.12,0.12){250}{\line(1,0){0.12}}
\linethickness{1mm}
\multiput(20,10)(0.12,0.12){250}{\line(1,0){0.12}}
\linethickness{1mm}
\multiput(50,40)(0.12,-0.12){250}{\line(1,0){0.12}}
\linethickness{0.2mm}
\put(10,60){\line(1,0){20}}
\linethickness{0.2mm}
\qbezier(35,55)(42.84,57.66)(45.25,55.25)
\qbezier(45.25,55.25)(47.66,52.84)(45,45)
\put(20,75){\makebox(0,0)[cc]{$p_a$}}

\put(10,55){\makebox(0,0)[cc]{$k$}}

\put(50,55){\makebox(0,0)[cc]{$\ell$}}

\put(50,15){\makebox(0,0)[cc]{$\cdots$}}

\put(20,60){\makebox(0,0)[cc]{$\gamma$}}

\put(41,55){\makebox(0,0)[cc]{$\gamma$}}

\end{picture}

}

\def\figselfe{

\def\JPicScale{0.4}
\ifx\JPicScale\undefined\def\JPicScale{1}\fi
\unitlength \JPicScale mm
\begin{picture}(80,75)(0,0)
\linethickness{1mm}
\multiput(20,70)(0.12,-0.12){250}{\line(1,0){0.12}}
\linethickness{1mm}
\multiput(50,40)(0.12,0.12){250}{\line(1,0){0.12}}
\linethickness{1mm}
\multiput(20,10)(0.12,0.12){250}{\line(1,0){0.12}}
\linethickness{1mm}
\multiput(50,40)(0.12,-0.12){250}{\line(1,0){0.12}}
\linethickness{0.2mm}
\put(10,60){\line(1,0){20}}

\put(20,75){\makebox(0,0)[cc]{$p_a$}}

\put(10,55){\makebox(0,0)[cc]{$k$}}

\put(50,55){\makebox(0,0)[cc]{$\ell$}}

\linethickness{0.2mm}
\put(43,55){\circle{10}}

\put(50,15){\makebox(0,0)[cc]{$\cdots$}}

\put(20,60){\makebox(0,0)[cc]{$\gamma$}}

\put(42,59){\makebox(0,0)[cc]{$\gamma$}}

\end{picture}

}

\def\figselff{

\def\JPicScale{0.4}
\ifx\JPicScale\undefined\def\JPicScale{1}\fi
\unitlength \JPicScale mm
\begin{picture}(80,75)(0,0)
\linethickness{1mm}
\multiput(20,70)(0.12,-0.12){250}{\line(1,0){0.12}}
\linethickness{1mm}
\multiput(50,40)(0.12,0.12){250}{\line(1,0){0.12}}
\linethickness{1mm}
\multiput(20,10)(0.12,0.12){250}{\line(1,0){0.12}}
\linethickness{1mm}
\multiput(50,40)(0.12,-0.12){250}{\line(1,0){0.12}}
\linethickness{0.2mm}
\put(10,60){\line(1,0){20}}

\put(20,75){\makebox(0,0)[cc]{$p_a$}}

\put(10,55){\makebox(0,0)[cc]{$k$}}

\linethickness{0.2mm}

\linethickness{0.3mm}
\put(33,53){\line(1,0){8}}
\put(37,49){\line(0,1){8}}

\put(50,15){\makebox(0,0)[cc]{$\cdots$}}

\put(20,60){\makebox(0,0)[cc]{$\gamma$}}

\end{picture}

}

\def\figselfafin{

\def\JPicScale{0.4}
\ifx\JPicScale\undefined\def\JPicScale{1}\fi
\unitlength \JPicScale mm
\begin{picture}(80,75)(0,0)
\linethickness{1mm}
\multiput(20,70)(0.12,-0.12){250}{\line(1,0){0.12}}
\linethickness{1mm}
\multiput(50,40)(0.12,0.12){250}{\line(1,0){0.12}}
\linethickness{1mm}
\multiput(20,10)(0.12,0.12){250}{\line(1,0){0.12}}
\linethickness{1mm}
\multiput(50,40)(0.12,-0.12){250}{\line(1,0){0.12}}
\linethickness{0.2mm}
\put(10,60){\line(1,0){20}}
\linethickness{0.4mm}
\qbezier(25,65)(32.84,67.66)(35.25,65.25)
\qbezier(35.25,65.25)(37.66,62.84)(35,55)
\put(20,75){\makebox(0,0)[cc]{$p_a$}}

\put(10,55){\makebox(0,0)[cc]{$k$}}

\put(40,65){\makebox(0,0)[cc]{$\ell$}}

\put(50,15){\makebox(0,0)[cc]{$\cdots$}}

\put(20,60){\makebox(0,0)[cc]{$\gamma$}}

\put(30,67){\makebox(0,0)[cc]{${\small g}$}}
 
\end{picture}

}

\def\figselfbfin{

\ifx\JPicScale\undefined\def\JPicScale{1}\fi
\unitlength \JPicScale mm
\begin{picture}(80,75)(0,0)
\linethickness{1mm}
\multiput(15,75)(0.12,-0.12){292}{\line(1,0){0.12}}
\linethickness{1mm}
\multiput(50,40)(0.12,0.12){250}{\line(1,0){0.12}}
\linethickness{1mm}
\multiput(20,10)(0.12,0.12){250}{\line(1,0){0.12}}
\linethickness{1mm}
\multiput(50,40)(0.12,-0.12){250}{\line(1,0){0.12}}
\linethickness{0.2mm}
\put(10,60){\line(1,0){20}}
\linethickness{0.4mm}
\qbezier(20,70)(27.84,72.66)(30.25,70.25)
\qbezier(30.25,70.25)(32.66,67.84)(30,60)
\put(20,80){\makebox(0,0)[cc]{$p_a$}}

\put(10,55){\makebox(0,0)[cc]{$k$}}

\put(35,70){\makebox(0,0)[cc]{$\ell$}}

\put(50,15){\makebox(0,0)[cc]{$\cdots$}}

\put(20,60){\makebox(0,0)[cc]{$\gamma$}}

\put(26,71){\makebox(0,0)[cc]{${\small g}$}}

\end{picture}

}

\def\figselfcfin{

\ifx\JPicScale\undefined\def\JPicScale{1}\fi
\unitlength \JPicScale mm
\begin{picture}(80,75)(0,0)
\linethickness{1mm}
\multiput(20,70)(0.12,-0.12){250}{\line(1,0){0.12}}
\linethickness{1mm}
\multiput(50,40)(0.12,0.12){250}{\line(1,0){0.12}}
\linethickness{1mm}
\multiput(20,10)(0.12,0.12){250}{\line(1,0){0.12}}
\linethickness{1mm}
\multiput(50,40)(0.12,-0.12){250}{\line(1,0){0.12}}
\linethickness{0.2mm}
\put(10,60){\line(1,0){20}}
\linethickness{0.4mm}
\qbezier(30,60)(37.84,62.66)(40.25,60.25)
\qbezier(40.25,60.25)(42.66,57.84)(40,50)
\put(20,75){\makebox(0,0)[cc]{$p_a$}}

\put(10,55){\makebox(0,0)[cc]{$k$}}

\put(45,60){\makebox(0,0)[cc]{$\ell$}}

\put(50,15){\makebox(0,0)[cc]{$\cdots$}}

\put(20,60){\makebox(0,0)[cc]{$\gamma$}}

\put(37,61){\makebox(0,0)[cc]{${\small g}$}}

\end{picture}

}

\def\figselfdfin{

\def\JPicScale{0.4}
\ifx\JPicScale\undefined\def\JPicScale{1}\fi
\unitlength \JPicScale mm
\begin{picture}(80,75)(0,0)
\linethickness{1mm}
\multiput(20,70)(0.12,-0.12){250}{\line(1,0){0.12}}
\linethickness{1mm}
\multiput(50,40)(0.12,0.12){250}{\line(1,0){0.12}}
\linethickness{1mm}
\multiput(20,10)(0.12,0.12){250}{\line(1,0){0.12}}
\linethickness{1mm}
\multiput(50,40)(0.12,-0.12){250}{\line(1,0){0.12}}
\linethickness{0.2mm}
\put(10,60){\line(1,0){20}}
\linethickness{0.4mm}
\qbezier(35,55)(42.84,57.66)(45.25,55.25)
\qbezier(45.25,55.25)(47.66,52.84)(45,45)
\put(20,75){\makebox(0,0)[cc]{$p_a$}}

\put(10,55){\makebox(0,0)[cc]{$k$}}

\put(50,55){\makebox(0,0)[cc]{$\ell$}}

\put(50,15){\makebox(0,0)[cc]{$\cdots$}}

\put(20,60){\makebox(0,0)[cc]{$\gamma$}}

\put(41,56){\makebox(0,0)[cc]{${\small g}$}}

\end{picture}

}

\def\figselfefin{

\def\JPicScale{0.4}
\ifx\JPicScale\undefined\def\JPicScale{1}\fi
\unitlength \JPicScale mm
\begin{picture}(80,75)(0,0)
\linethickness{1mm}
\multiput(20,70)(0.12,-0.12){250}{\line(1,0){0.12}}
\linethickness{1mm}
\multiput(50,40)(0.12,0.12){250}{\line(1,0){0.12}}
\linethickness{1mm}
\multiput(20,10)(0.12,0.12){250}{\line(1,0){0.12}}
\linethickness{1mm}
\multiput(50,40)(0.12,-0.12){250}{\line(1,0){0.12}}
\linethickness{0.2mm}
\put(10,60){\line(1,0){20}}

\put(20,75){\makebox(0,0)[cc]{$p_a$}}

\put(10,55){\makebox(0,0)[cc]{$k$}}

\put(50,55){\makebox(0,0)[cc]{$\ell$}}

\linethickness{0.4mm}
\put(43,55){\circle{10}}
\put(43,55){\circle{9.5}}
\put(43,55){\circle{9}}
\put(43,55){\circle{8.5}}
\put(43,55){\circle{8}}

\put(50,15){\makebox(0,0)[cc]{$\cdots$}}

\put(20,60){\makebox(0,0)[cc]{$\gamma$}}

\put(43,59){\makebox(0,0)[cc]{${\small g}$}}

\end{picture}

}

\def\figselfffin{

\def\JPicScale{0.4}
\ifx\JPicScale\undefined\def\JPicScale{1}\fi
\unitlength \JPicScale mm
\begin{picture}(80,75)(0,0)
\linethickness{1mm}
\multiput(20,70)(0.12,-0.12){250}{\line(1,0){0.12}}
\linethickness{1mm}
\multiput(50,40)(0.12,0.12){250}{\line(1,0){0.12}}
\linethickness{1mm}
\multiput(20,10)(0.12,0.12){250}{\line(1,0){0.12}}
\linethickness{1mm}
\multiput(50,40)(0.12,-0.12){250}{\line(1,0){0.12}}
\linethickness{0.2mm}
\put(10,60){\line(1,0){20}}

\put(20,75){\makebox(0,0)[cc]{$p_a$}}

\put(10,55){\makebox(0,0)[cc]{$k$}}

\linethickness{0.2mm}

\linethickness{0.3mm}
\put(33,53){\line(1,0){8}}
\put(37,49){\line(0,1){8}}

\put(50,15){\makebox(0,0)[cc]{$\cdots$}}

\put(20,60){\makebox(0,0)[cc]{$\gamma$}}

\end{picture}

}

\def\figselfaqgr{

\def\JPicScale{0.4}
\ifx\JPicScale\undefined\def\JPicScale{1}\fi
\unitlength \JPicScale mm
\begin{picture}(80,75)(0,0)
\linethickness{1mm}
\multiput(20,70)(0.12,-0.12){250}{\line(1,0){0.12}}
\linethickness{1mm}
\multiput(50,40)(0.12,0.12){250}{\line(1,0){0.12}}
\linethickness{1mm}
\multiput(20,10)(0.12,0.12){250}{\line(1,0){0.12}}
\linethickness{1mm}
\multiput(50,40)(0.12,-0.12){250}{\line(1,0){0.12}}
\linethickness{0.4mm}
\put(10,60){\line(1,0){20}}
\linethickness{0.2mm}
\qbezier(25,65)(32.84,67.66)(35.25,65.25)
\qbezier(35.25,65.25)(37.66,62.84)(35,55)
\put(20,75){\makebox(0,0)[cc]{$p_a$}}

\put(10,55){\makebox(0,0)[cc]{$k$}}

\put(40,65){\makebox(0,0)[cc]{$\ell$}}

\put(50,15){\makebox(0,0)[cc]{$\cdots$}}

\put(20,60){\makebox(0,0)[cc]{$g$}}

\put(30,65){\makebox(0,0)[cc]{$\gamma$}}

\end{picture}

}

\def\figselfbqgr{

\ifx\JPicScale\undefined\def\JPicScale{1}\fi
\unitlength \JPicScale mm
\begin{picture}(80,75)(0,0)
\linethickness{1mm}
\multiput(15,75)(0.12,-0.12){292}{\line(1,0){0.12}}
\linethickness{1mm}
\multiput(50,40)(0.12,0.12){250}{\line(1,0){0.12}}
\linethickness{1mm}
\multiput(20,10)(0.12,0.12){250}{\line(1,0){0.12}}
\linethickness{1mm}
\multiput(50,40)(0.12,-0.12){250}{\line(1,0){0.12}}
\linethickness{0.4mm}
\put(10,60){\line(1,0){20}}
\linethickness{0.2mm}
\qbezier(20,70)(27.84,72.66)(30.25,70.25)
\qbezier(30.25,70.25)(32.66,67.84)(30,60)
\put(20,80){\makebox(0,0)[cc]{$p_a$}}

\put(10,55){\makebox(0,0)[cc]{$k$}}

\put(35,70){\makebox(0,0)[cc]{$\ell$}}

\put(50,15){\makebox(0,0)[cc]{$\cdots$}}

\put(20,60){\makebox(0,0)[cc]{$g$}}

\put(25,70){\makebox(0,0)[cc]{$\gamma$}}

\end{picture}

}

\def\figselfcqgr{

\ifx\JPicScale\undefined\def\JPicScale{1}\fi
\unitlength \JPicScale mm
\begin{picture}(80,75)(0,0)
\linethickness{1mm}
\multiput(20,70)(0.12,-0.12){250}{\line(1,0){0.12}}
\linethickness{1mm}
\multiput(50,40)(0.12,0.12){250}{\line(1,0){0.12}}
\linethickness{1mm}
\multiput(20,10)(0.12,0.12){250}{\line(1,0){0.12}}
\linethickness{1mm}
\multiput(50,40)(0.12,-0.12){250}{\line(1,0){0.12}}
\linethickness{0.4mm}
\put(10,60){\line(1,0){20}}
\linethickness{0.2mm}
\qbezier(30,60)(37.84,62.66)(40.25,60.25)
\qbezier(40.25,60.25)(42.66,57.84)(40,50)
\put(20,75){\makebox(0,0)[cc]{$p_a$}}

\put(10,55){\makebox(0,0)[cc]{$k$}}

\put(45,60){\makebox(0,0)[cc]{$\ell$}}

\put(50,15){\makebox(0,0)[cc]{$\cdots$}}

\put(20,60){\makebox(0,0)[cc]{$g$}}

\put(38,60){\makebox(0,0)[cc]{$\gamma$}}

\end{picture}

}

\def\figselfdqgr{

\def\JPicScale{0.4}
\ifx\JPicScale\undefined\def\JPicScale{1}\fi
\unitlength \JPicScale mm
\begin{picture}(80,75)(0,0)
\linethickness{1mm}
\multiput(20,70)(0.12,-0.12){250}{\line(1,0){0.12}}
\linethickness{1mm}
\multiput(50,40)(0.12,0.12){250}{\line(1,0){0.12}}
\linethickness{1mm}
\multiput(20,10)(0.12,0.12){250}{\line(1,0){0.12}}
\linethickness{1mm}
\multiput(50,40)(0.12,-0.12){250}{\line(1,0){0.12}}
\linethickness{0.4mm}
\put(10,60){\line(1,0){20}}
\linethickness{0.2mm}
\qbezier(35,55)(42.84,57.66)(45.25,55.25)
\qbezier(45.25,55.25)(47.66,52.84)(45,45)
\put(20,75){\makebox(0,0)[cc]{$p_a$}}

\put(10,55){\makebox(0,0)[cc]{$k$}}

\put(50,55){\makebox(0,0)[cc]{$\ell$}}

\put(50,15){\makebox(0,0)[cc]{$\cdots$}}

\put(20,60){\makebox(0,0)[cc]{$g$}}

\put(41,56){\makebox(0,0)[cc]{$\gamma$}}

\end{picture}

}

\def\figselfeqgr{

\def\JPicScale{0.4}
\ifx\JPicScale\undefined\def\JPicScale{1}\fi
\unitlength \JPicScale mm
\begin{picture}(80,75)(0,0)
\linethickness{1mm}
\multiput(20,70)(0.12,-0.12){250}{\line(1,0){0.12}}
\linethickness{1mm}
\multiput(50,40)(0.12,0.12){250}{\line(1,0){0.12}}
\linethickness{1mm}
\multiput(20,10)(0.12,0.12){250}{\line(1,0){0.12}}
\linethickness{1mm}
\multiput(50,40)(0.12,-0.12){250}{\line(1,0){0.12}}
\linethickness{0.4mm}
\put(10,60){\line(1,0){20}}

\put(20,75){\makebox(0,0)[cc]{$p_a$}}

\put(10,55){\makebox(0,0)[cc]{$k$}}

\put(50,55){\makebox(0,0)[cc]{$\ell$}}

\linethickness{0.2mm}
\put(43,55){\circle{10}}

\put(50,15){\makebox(0,0)[cc]{$\cdots$}}

\put(20,60){\makebox(0,0)[cc]{$g$}}

\put(42,58){\makebox(0,0)[cc]{$\gamma$}}

\end{picture}

}

\def\figselfgqgr{

\def\JPicScale{0.4}
\ifx\JPicScale\undefined\def\JPicScale{1}\fi
\unitlength \JPicScale mm
\begin{picture}(80,75)(0,0)
\linethickness{1mm}
\multiput(20,70)(0.12,-0.12){250}{\line(1,0){0.12}}
\linethickness{1mm}
\multiput(50,40)(0.12,0.12){250}{\line(1,0){0.12}}
\linethickness{1mm}
\multiput(20,10)(0.12,0.12){250}{\line(1,0){0.12}}
\linethickness{1mm}
\multiput(50,40)(0.12,-0.12){250}{\line(1,0){0.12}}
\linethickness{0.4mm}
\put(20,50){\line(1,0){20}}

\put(20,75){\makebox(0,0)[cc]{$p_a$}}

\put(10,55){\makebox(0,0)[cc]{$k$}}

\put(50,55){\makebox(0,0)[cc]{$\ell$}}

\linethickness{0.2mm}
\put(43,55){\circle{10}}

\put(50,15){\makebox(0,0)[cc]{$\cdots$}}

\put(30,50){\makebox(0,0)[cc]{$g$}}

\put(42,58){\makebox(0,0)[cc]{$\gamma$}}

\end{picture}

}

\def\figselfggrav{

\def\JPicScale{0.4}
\ifx\JPicScale\undefined\def\JPicScale{1}\fi
\unitlength \JPicScale mm
\begin{picture}(80,75)(0,0)
\linethickness{1mm}
\multiput(20,70)(0.12,-0.12){250}{\line(1,0){0.12}}
\linethickness{1mm}
\multiput(50,40)(0.12,0.12){250}{\line(1,0){0.12}}
\linethickness{1mm}
\multiput(20,10)(0.12,0.12){250}{\line(1,0){0.12}}
\linethickness{1mm}
\multiput(50,40)(0.12,-0.12){250}{\line(1,0){0.12}}
\linethickness{0.2mm}
\put(20,50){\line(1,0){20}}

\put(20,75){\makebox(0,0)[cc]{$p_a$}}

\put(10,55){\makebox(0,0)[cc]{$k$}}

\put(50,55){\makebox(0,0)[cc]{$\ell$}}

\linethickness{0.2mm}
\put(43,55){\circle{10}}
\put(43,55){\circle{9.5}}
\put(43,55){\circle{9}}
\put(43,55){\circle{8.5}}
\put(43,55){\circle{8}}

\put(50,15){\makebox(0,0)[cc]{$\cdots$}}

\put(30,50){\makebox(0,0)[cc]{$\gamma$}}

\put(43,59){\makebox(0,0)[cc]{${\small g}$}}

\end{picture}

}

\def\figselffqgr{

\def\JPicScale{0.4}
\ifx\JPicScale\undefined\def\JPicScale{1}\fi
\unitlength \JPicScale mm
\begin{picture}(80,75)(0,0)
\linethickness{1mm}
\multiput(20,70)(0.12,-0.12){250}{\line(1,0){0.12}}
\linethickness{1mm}
\multiput(50,40)(0.12,0.12){250}{\line(1,0){0.12}}
\linethickness{1mm}
\multiput(20,10)(0.12,0.12){250}{\line(1,0){0.12}}
\linethickness{1mm}
\multiput(50,40)(0.12,-0.12){250}{\line(1,0){0.12}}
\linethickness{0.4mm}
\put(10,60){\line(1,0){20}}

\put(20,75){\makebox(0,0)[cc]{$p_a$}}

\put(10,55){\makebox(0,0)[cc]{$k$}}

\linethickness{0.2mm}

\linethickness{0.3mm}
\put(33,53){\line(1,0){8}}
\put(37,49){\line(0,1){8}}

\put(50,15){\makebox(0,0)[cc]{$\cdots$}}

\put(20,60){\makebox(0,0)[cc]{$g$}}

\end{picture}

}

\begin{document}

\baselineskip 24pt

\baselineskip 24pt

\begin{center}

{\Large \bf Classical and Quantum Results on Logarithmic Terms in the Soft 
Theorem in Four Dimensions}


\end{center}

\vskip .6cm
\medskip

\vspace*{4.0ex}

\baselineskip=18pt

\centerline{\large \rm Biswajit Sahoo and Ashoke Sen}

\vspace*{4.0ex}


\centerline{\large \it Harish-Chandra Research Institute, HBNI}
\centerline{\large \it  Chhatnag Road, Jhusi,
Allahabad 211019, India}


\vspace*{1.0ex}
\centerline{\small E-mail:  biswajitsahoo@hri.res.in, sen@hri.res.in}

\vspace*{5.0ex}

\centerline{\bf Abstract} \bigskip

We explore the logarithmic terms in the soft theorem 
in four dimensions by analyzing classical scattering with
generic incoming and outgoing states and
one loop quantum scattering amplitudes. The classical and quantum results are consistent with each other.
Although most of our analysis in quantum theory is carried out for one loop amplitudes in a theory of 
(charged) scalars interacting
via gravitational and electromagnetic interactions, we expect the results to be valid more generally.



\vfill \eject

\baselineskip 18pt

\tableofcontents

\sectiono{Introduction} \label{sintro}

Soft theorems\cite{Gell-Mann,low,saito,burnett,bell,duca,
weinberg1,weinberg2,jackiw1,jackiw2,ademollo,shapiro} 
have been analyzed recently from different 
perspectives, both
using asymptotic symmetries\cite{1312.2229,
1401.7026,1408.2228,
1411.5745,1502.02318,1502.06120,1502.07644,1505.05346,
1506.05789,1509.01406,1605.09094,1605.09677,1608.00685,1612.08294,1701.00496,
1612.05886,1703.01351,1703.05448,1707.08016,
1709.03850,1711.04371,1712.01204,1712.09591,1803.03023}
and also by direct analysis of amplitudes in field theory 
and string theory\cite{1103.2981,1404.4091,1404.5551,1404.7749,
1405.1015,1405.1410,1405.2346,
1405.3413,1405.3533,1406.4172,1406.5155,1406.6574,1406.6987,
1406.7184,1407.5936,
1407.5982,1408.4179,1410.6406,1411.6661,1412.3699,1502.05258,
1503.04816,1504.01364,1504.05558,1504.05559,
1505.05854,1507.00938,1507.08829,1507.08882,
1509.07840,1511.04921,
1512.00803,1601.03457,1604.00650,1604.03355,1604.02834,
1604.03893,1607.02700,1610.03481,1611.02172,1611.07534,1611.03137,
1702.02350,1702.03934,1703.00024,1705.06175,1706.00759,1707.06803,1801.05528,1802.03148,1805.11079}.
There are general arguments
establishing their validity in any space-time dimensions in any theory as long as one maintains the relevant 
gauge symmetries -- general coordinate invariance for soft graviton theorem and U(1) gauge invariance for
soft photon theorem\cite{1703.00024,1706.00759,1707.06803}. 
However these arguments break down in four space-time dimensions due to infrared
divergences\cite{1405.1015} where more care may be needed\cite{1405.3413,1701.00496}. 
Indeed, since the S-matrix itself is infrared divergent, it is not {\it a priori} clear how to
interpret a relation whose both sides are divergent.

Although soft theorem is a relation between quantum scattering amplitudes -- amplitudes with soft photon or
graviton to amplitudes without soft photon or graviton -- one can also relate soft theorem to classical scattering
amplitudes. In four space-time dimensions this can be done via asymptotic 
symmetries\cite{1411.5745,1707.08016,1712.01204}. 
Ref.\cite{1801.07719} produced a more direct relation between 
soft theorem and classical scattering in generic space-time
dimensions by directly taking the classical limit of a quantum scattering amplitude. 
This relates various
terms in soft theorem to appropriate terms in the radiative part of the 
electromagnetic and gravitational fields in classical scattering
in generic space-time dimensions. Reversing the logic, one can use the classical scattering data 
to give an 
alternative definition of the soft factors.

Since classical scattering is well defined even in four space-time dimensions, one 
can hope to use 
the classical definition of soft factors to understand soft theorem in four dimensions.
Since in higher dimensions the soft theorem expresses the low frequency 
radiative part of the
electromagnetic and gravitational fields in terms of momenta and
angular momenta of incoming and outgoing finite energy
particles, the naive guess will be that the same
formula will continue to hold in four dimensions.
However in carrying out this procedure
we encounter an obstacle\cite{1804.09193,1806.01872}. 
The subleading terms in the soft theorem contain a factor of angular momentum
$j^{\mu\nu}$ of the individual particles involved in the scattering, with the orbital contribution to the angular momentum
given by $x^\mu p^\nu - x^\nu p^\mu$,  where $x^\mu(\tau)$ and $p^\mu(\tau)$ label the asymptotic coordinates and
momenta of the particle as a function of the proper time. In dimensions larger than four, $p^\mu$ approaches a constant
and $x^\mu$ approaches the form $c^\mu + \alpha \, p^\mu \, \tau$ with constant $c^\mu$ and $\alpha$ 
as $\tau\to\infty$. Therefore $j^{\mu\nu}$ is
independent of $\tau$ as $\tau\to\infty$ and we can use the asymptotic value of $j^{\mu\nu}$ computed this way to
evaluate the soft factor. However in four space-time dimensions the long range gravitational and / or electromagnetic 
forces acting on the particles produce an additional term of the form $b^\mu \ln\tau$ in the expression for $x^\mu$.
This gives a logarithmically divergent term of the form $(b^\mu p^\nu - b^\nu p^\mu)\ln\tau$ in the expression for
$j^{\mu\nu}$, making the subleading soft factor divergent.

Since we do not expect the radiative component of the metric or gauge fields to diverge in classical scattering in four
space-time dimensions, this suggests that the divergence in the subleading soft factor is a breakdown of the power
series expansion in the  energy $\omega$ of the soft particle. Therefore the soft factor must contain 
non-analytic terms in $\omega$. The natural guess is that the soft factor at the subleading order is given by replacing
the factors of $\ln\tau$  in the  naive expression by $\ln\omega^{-1}$. This has been 
tested in \cite{1804.09193} by
considering several examples of classical scattering in four space-time 
dimensions.\footnote{The existence of various logarithmic terms in classical scattering has been
known earlier\cite{peter,blanchet,0912.4254,1211.6095}.
Soft theorem provides a systematic procedure for computing the
coefficient of the logarithmic term in the subleading soft factor
without detailed knowledge of the forces responsible for
the scattering.}

The purpose of this paper is two fold. In all the examples considered in \cite{1804.09193} 
the scattering process considered had
one heavy center producing the long range Coulomb or gravitational field, and other particles
carrying smaller masses were taken to be moving under the influence of the long range fields produced by
the heavy center. In this paper we relax this assumption and consider a general scattering process where all particles
involved in the scattering have masses of the same order, and then determine the logarithmic terms in the
classical soft factor using the $\ln\tau \to \ln\omega^{-1}$ replacement rule.  We also analyze directly the quantum 
subleading soft
factor by considering one loop scattering of charged scalar fields in the presence of gravitational and electromagnetic
interaction.  The difference with the previous analysis, {\it e.g.} in 
\cite{1405.1015}, is that we do not insist on a power series expansion
in $\omega$ and calculating the coefficients of the power series expansion. Instead we allow for possible 
non-analytic terms of order $\ln\omega^{-1}$ in the soft expansion. This analysis yields results consistent with the 
classical results, although the quantum results contain additional real part which we interpret as the result of
back reaction of the radiation on the motion of the particles.

Since \cite{1703.00024,1706.00759,1707.06803} gave a general derivation of 
soft theorem including loop corrections as long as 1PI vertices do not
generate soft factor in the denominator, one could ask to what extent we could derive the results of the current
paper using the result of \cite{1703.00024,1706.00759,1707.06803}. 
To this end we note that there are two distinct sources of logarithmic terms in the
soft theorem. The first is the region of integration in which the loop momentum is large compared to the energy of the
external soft particle. In this region we expect the arguments of \cite{1703.00024,1706.00759,1707.06803} 
to be valid, and we find that the contribution
from this region can indeed be obtained by applying the usual soft operator on the amplitude without the soft
graviton. The other source is the region of integration in which the loop momentum is small compared to the external
soft momentum. The contribution from this region cannot be derived using the usual soft theorem, and need to be
computed explicitly.

The rest of the paper is organized as follows. \S\ref{ssummary} contains a summary and a discussion of our results
where we also discuss various special cases of our classical result. \S\ref{sclassical} describes the analysis
of the logarithmic terms in the soft expansion for general classical scattering. 
\S\ref{smom} describes some general strategy for dealing with the infrared divergent part of the S-matrix and 
extracting the quantum soft factor by making use of momentum conservation.
\S\ref{s3} describes one loop quantum
computation of the logarithmic terms in the soft photon theorem in scalar quantum electrodynamics (scaler QED).
\S\ref{s5} describes a similar computation in the soft graviton theorem in a theory of charge neutral scalars interacting
with the gravitational field. In \S\ref{s6} we consider  charged scalars interacting via both gravitational and
electromagnetic interaction, and determine the one loop contribution to the quantum soft graviton factor due to
electromagnetic interaction and one loop contribution to the quantum soft photon factor due to
gravitational interaction.

Classical gravitational radiation during a high energy scattering process has been analyzed in  \cite{1409.4555,1512.00281}.
We have been informed by G.~Veneziano that for massless particle scattering, results related to the ones
described here were found in 
\cite{1901.10986}, and also that the logarithmic terms in the classical scattering have been derived in \cite{1812.08137} by
taking the soft limit of the results of \cite{1409.4555,1512.00281}.

\asnotex{Note added: The papers quoted above have now appeared in the arXiv\cite{1812.08137,1901.10986}. 
In particular the results of \cite{1812.08137}
can be shown to be in perfect agreement with our results for scattering of massless particles.}

\sectiono{Summary and analysis of the results} \label{ssummary}

In this section we shall first summarize our results and then discuss various aspects of the results. Finally we shall
consider some special limits and compare with known results. We shall use $\hbar=c=8\pi
G=1$
units. 

\subsection{Summary of the results}

In order to give a uniform treatment of the classical soft photon and soft graviton theorem,
we shall denote by $\phi(\vec x, t)$ the radiative part of the metric or 
electromagnetic field at a point $\vec x$ at time $t$
for a scattering event around the origin. For electromagnetic field, $\phi$ can be directly
identified with the gauge field.
For the gravitational field we define
\be 
h_{\mu\nu}= (g_{\mu\nu}-\eta_{\mu\nu})/2, \qquad 
e_{\mu\nu} = h_{\mu\nu} -{1\over 2}\eta_{\mu\nu} \, h_\rho^{~\rho}\, ,
\ee
and take $\phi$  to be $e_{\mu\nu}$.
For both electromagnetism and gravity we define classical soft factor $S(\ve, k)$ 
in $D$ space-time dimensions via the relation:
\ben\label{e100}
\int dt\, e^{i\omega t} \, \ve . \phi(\vec x,t)  &=& e^{i\omega R} \, 
\left({\omega\over 2\pi i R}\right)^{(D-2)/2}
\, {1\over 2\omega}  \, S(\ve, k) \non\\ &=&
-{i\over 4\pi R}\, e^{i\omega R} \, S(\ve, k) \quad \hbox{for \quad $D=4$}\, ,
\een
where  $\ve$ is the polarization tensor of the soft particle so that $\ve.\phi=\ve^\mu A_\mu$ for gauge fields
and $\ve^{\mu\nu} e_{\mu\nu}$ for gravity,
and 
\be
k = -\omega (1, \hat n), \quad \hat n\equiv \vec x / |\vec x|, \quad R=|\vec x|\, .
\ee
On the other hand the quantum soft factor $S(\ve, k)$ is the ratio of an amplitude with an outgoing 
soft photon or graviton with 
momentum $k$ and polarization $\ve$ and an amplitude without such a soft particle. 
It was shown in \cite{1801.07719} 
that in the classical limit the quantum soft factor reduces to the
classical soft factor for $D>4$. Our interest will be in analyzing the
situation in $D=4$.

We consider the scattering of $n$ particles carrying electric charges $\{q_a\}$ and momenta $\{p_a\}$ for
$a=1,\cdots n$. In our convention 
the momenta / charges carry extra minus sign if they are outgoing. The particles are taken to
interact via electromagnetic and gravitational interactions besides other short range 
interactions whose nature we need not know. 
The symbol $\eta_a$ takes value $+1$ ($-1$) if the
$a$-th particle is ingoing (outgoing). Then the classical result for the soft photon factor
$S_{\rm em}(\ve, k)$, containing terms of order $\omega^{-1}$ and $\ln\,\omega^{-1}$, 
is\footnote{In this and subsequent expressions $R$ arises as an infrared cut-off. For the
classical result the $\ln R$ terms arise due to long range gravitational force on the 
soft photon or graviton during its journey from the scattering center to the detector
over a distance $R$. For the
quantum part, the natural infrared cut-off is provided by the resolution of the detector. For
a detector placed at a distance $R$ from the scattering center, the best energy resolution
possible is of order $1/R$. Therefore it is again natural to take $R$ as the infrared
regulator. \label{foo1}}
\ben\label{e1.16int}
S_{\rm em} &=& \sum_a {\ve_\mu p_a^\mu \over p_a.k} \, q_a - i \, \ln \omega^{-1}
\sum_a {q_a \, \ve_\mu k_\rho \over p_a.k}
 \sum_{b\ne a\atop \eta_a\eta_b=1} \, {q_a q_b\over 4\pi} \,
{m_a^2 m_b^2 \, \{p_{b}^{\rho} p_a^\mu - p_b^\mu p_a^\rho\} \over 
\{ (p_b.p_a)^2 -m_a^2 m_b^2\}^{3/2}} \nonumber \\
&& + {i\over 4\pi}\, (\ln\omega^{-1}+\ln R^{-1})
\, \, \sum_{b\atop \eta_b=-1} k.p_b \, \sum_a {\ve_\mu p_a^\mu \over p_a.k} \, q_a 
\nonumber \\ && \hskip -.3in
+ {i\over 8\pi} 
\, \ln \omega^{-1}
\sum_a {q_a \, \ve_\mu k_\rho \over p_a.k}
 \sum_{b\ne a\atop \eta_a\eta_b=1} 
 \,  {p_b.p_a\over 
\{ (p_b.p_a)^2 -m_a^2 m_b^2\}^{3/2}}\, (p_b^\rho p_a^\mu - p_b^\mu p_a^\rho) 
\left\{2(p_b.p_a)^2  - 3 m_a^2 m_b^2 \right\} 
\, . \nonumber \\
\een
Since for real polarization the subleading contribution is purely imaginary, it does not affect the flux to this order.
However the flux for circular polarization and / or the wave-form of the electromagnetic field do receive subleading
contribution. An identical situation prevails for gravity.

The  quantum result for $S_{\rm em}$ has additional terms:\footnote{Note however that when we express the
results in terms of the frequency / wavelength of the soft photon / graviton and momenta of the finite energy
particles, neither the classical nor the quantum result has any power of $\hbar$. We shall discuss later the conditions
under which we expect the quantum results to be small compared to the classical results.} 
\ben \label{e3.23Gint}
&& 
\Delta S_{\rm em} = 
{1\over 16\pi^2} \, \ln\omega^{-1} \,  
\sum_{a} q_a\, {\ve_\mu k_\nu \over p_a.k} \left\{p_a^\mu {\p\over \p p_{a\nu}} - p_a^\nu {\p\over \p p_{a\mu}}
\right\} \nonumber \\ && \hskip .3in \sum_{b\ne a} \left[
\f{\left\{ 2\, q_a q_b p_a.p_b + 2\, (p_a.p_b)^2 - p_a^2 p_b^2\right\}
}{\sqrt{(p_{a}.p_{b})^{2}-p_{a}^{2}p_{b}^{2}}} \, \ln\left({p_a.p_b +  \sqrt{(p_a.p_b)^2 - p_a^2 p_b^2}\over
p_a.p_b -  \sqrt{(p_a.p_b)^2 - p_a^2 p_b^2}}
\right)
\right]
\nonumber \\
&& +\f{1}{8\pi^{2}}\,  (\ln\omega^{-1}+\ln R^{-1})\,
\sum_{a}\ \f{q_{a}\ve_\mu p_{a}^\mu}{p_{a}.k}\ \sum_{b}\ (p_{b}.k)\ 
\ln\Bigg(\f{m_b^2}{(p_{b}.\hat{k})^{2}}\Bigg)\, .
\een
The classical results are universal, independent of the theory and the nature of external particles. We expect that
the quantum results are also universal, but we have derived them by working with one loop amplitudes in scalar
QED coupled to gravity. It is easy to check that \refb{e1.16int}, 
\refb{e3.23Gint} are invariant under gauge transformation 
$\ve_\mu\to \ve_\mu + \xi\, k_\mu$ for any constant $\xi$. 

As will be discussed in \S\ref{sdiss}, the quantum correction \refb{e3.23Gint} should not be directly added to
\refb{e1.16int} and substituted into \refb{e100} to compute the radiative component of the 
classical electromagnetic field.
Rather, when the contribution \refb{e3.23Gint} is small compared to \refb{e1.16int}, we can substitute \refb{e1.16int} 
into \refb{e100} to compute
the classical electromagnetic field produced by a scattering event.

As discussed in \S\ref{smom}, the quantum results are ambiguous and are defined 
up to addition of a term to $S_{\rm em}$
of the form $\ln R^{-1}\, k.U \, S^{(0)}_{\rm em}$ where $S^{(0)}_{\rm em}$ is the leading soft factor given by
the first term on the right hand side of \refb{e1.16int} 
and $U$ is a vector constructed out of the $p_a$'s. By choosing
$U=(8\pi^2)^{-1} \sum_b p_b \, \ln (m_b^2/\mu^2)$,
we can replace the $\ln m_b^2$ term in the coefficient of $\ln R^{-1}$ in the last line of \refb{e3.23Gint} by $\ln\mu^2$
for any mass parameter $\mu$. This makes manifest the fact that the coefficient is not divergent in the $m_b\to 0$
limit. The coefficient of $\ln\omega^{-1}$ cannot be changed this way, but in this case the finiteness of $m_b\to 0$
limit follows as a consequence of cancellation between the second and third line of \refb{e3.23Gint} and momentum
conservation.

If we want to consider the situation where we ignore the effect of gravity, then we need to set the terms proportional
to $\ln\omega^{-1}$ that are linear in $q_c$'s to zero. On the other hand if we want to consider the situation where
we ignore the effect of electromagnetic interaction between the particles during scattering (but still use electromagnetic
interaction to compute soft photon emission process), we have to set the terms proportional
to $\ln\omega^{-1}$ that are cubic in the $q_c$'s to zero.

The classical result for soft graviton factor takes the form
\ben\label{e1.17int}
S_{\rm gr} &=& \sum_a {\ve_{\mu\nu} p_a^\mu p_a^\nu \over p_a.k}  - i  \, \ln \omega^{-1}
\sum_a {\ve_{\mu\nu} p_a^\nu k_\rho 
 \over p_a.k}
 \sum_{b\ne a\atop \eta_a\eta_b=1} \, {q_a q_b\over 4\pi} \,
{m_a^2 m_b^2 \, \{p_{b}^{\rho} p_a^\mu - p_b^\mu p_a^\rho\} \over 
\{  (p_b.p_a)^2 -m_a^2 m_b^2\}^{3/2}}
\nonumber \\ &&
+  {i\over 4\pi}
\, (\ln\omega^{-1} +\ln R^{-1})\, \, \sum_{b\atop \eta_b=-1} k.p_b \,  \sum_a {\ve_{\mu\nu} p_a^\mu p_a^\nu \over p_a.k}  
\nonumber \\ && \hskip -.3in
+ {i\over 8\pi}  
\, \ln \omega^{-1}
\sum_a {\ve_{\mu\nu} p_a^\nu k_\rho 
 \over p_a.k}
 \sum_{b\ne a\atop \eta_a\eta_b=1} 
 \,  { p_b.p_a\over 
\{ (p_b.p_a)^2 -m_a^2 m_b^2\}^{3/2}}\, (p_b^\rho p_a^\mu - p_b^\mu p_a^\rho) 
\left\{2(p_b.p_a)^2  - 3 m_a^2 m_b^2\right\} 
\, .\nonumber \\
\een
The quantum result has additional terms
\ben \label{eqgrsoft}
\Delta S_{\rm gr} &=&  {1\over 16\pi^2} \, \ln\omega^{-1} \,  
 \sum_{a} {\ve_{\mu\rho} p_a^\rho k_\nu \over p_a.k} 
\left\{p_a^\mu {\p\over \p p_{a\nu}} - p_a^\nu {\p\over \p p_{a\mu}}
\right\}\nonumber \\ && \hskip 1in \sum_{b\ne a} \left[
\f{\left\{ 2\, q_a q_b p_a.p_b + 2\, (p_a.p_b)^2 - p_a^2 p_b^2\right\}
}{\sqrt{(p_{a}.p_{b})^{2}-p_{a}^{2}p_{b}^{2}}} \, \ln\left({p_a.p_b +  \sqrt{(p_a.p_b)^2 - p_a^2 p_b^2}\over
p_a.p_b -  \sqrt{(p_a.p_b)^2 - p_a^2 p_b^2}}
\right)
\right] \nonumber \\
&& + {1\over 8\pi^2} \, (\ln\omega^{-1} +\ln R^{-1})\, 
\sum_a {\ve_{\mu\nu} p_a^\mu p_a^\nu \over p_a.k} \sum_b  p_b.k\, \ln{m_b^2 \over (p_b.\hat k)^2}
\, ,
\een
where $\hat k =-k/\omega =(1, \hat n)$. Again the classical results are valid universally. The
quantum results are obtained from one loop calculation in scalar QED coupled
to gravity, but we expect them
to be universal. As in the case of \refb{e3.23Gint}, the $\ln m_b^2$ term in the coefficient of $\ln R^{-1}$ in
the last line of \refb{eqgrsoft} can be replaced by $\ln\mu^2$ by exploiting the ambiguity in the definition
of the soft factor discussed in \S\ref{smom}. One can check that \refb{e1.17int}, \refb{eqgrsoft} are
invariant under gauge transformations $\ve_{\mu\nu}\to
\ve_{\mu\nu}+\xi_\mu k_\nu + \xi_\nu k_\mu$ for any
constant vector $\xi_\mu$.

If we want to consider the situation where we ignore the effect of electromagnetic interactions, 
then we need to set the terms proportional
to $\ln\omega^{-1}$ that are quadratic in $q_c$'s to zero. On the other hand if we want to consider the 
situation where
we ignore the effect of gravitational interaction between the particles during scattering 
(but still use gravitational
interaction to compute soft graviton emission process), we have to set the $q_c$ independent terms
in the coefficient of $\ln\omega^{-1}$  to zero.

\subsection{Discussion of results} \label{sdiss}

First we shall briefly outline how these results are derived; more details can be found in later sections. The classical
results \refb{e1.16int} and \refb{e1.17int} are the result of direct application of classical soft theorem to subleading order.
As described in \cite{1804.09193}, 
the soft factor involves orbital angular momenta of initial and final particles and these diverge
logarithmically in the elapsed time $\tau$ in four dimensions due to the long range gravitational / electromagnetic force
on the incoming and outgoing particles that generates a term proportional to $\ln|\tau|$ in the trajectory. We follow the
prescription of \cite{1804.09193} 
of replacing $\ln|\tau|$ by $\ln\omega^{-1}$ to arrive at the first and third lines of the classical results 
\refb{e1.16int}, \refb{e1.17int}. The second lines of \refb{e1.16int} and \refb{e1.17int} arise from additional phases
that are not directly determined by soft theorem. They represent the effect of long 
range gravitational force on the outgoing soft 
photon or graviton which causes the soft particle to slow down and also
backscatter.

Quantum results are the result of direct one loop 
computation in a field theory of multiple charged scalars, coupled 
to electromagnetic and gravitational fields. We simply evaluate the order $\omega^{-1}$ and 
$\ln\omega^{-1}$ terms in the scattering amplitude of multiple finite energy scalars and an outgoing  soft photon or 
graviton
of energy $\omega$, and express this as the product of the amplitude without the soft photon or graviton and 
a multiplicative factor that we call the soft factor. The latter is given by the sum of \refb{e1.16int} and \refb{e3.23Gint} for
soft photon and the sum of \refb{e1.17int} and \refb{eqgrsoft} for the soft graviton. 
Even though the S-matrix elements with and without the soft particle are infrared
divergent, much of this cancels when we take the ratio of the two. The remaining infrared divergent part is 
regulated by the infra-red length cut-off $R$ and is responsible for the terms proportional to $\ln R$ in these
expressions. This is related to the quantity $\sigma_n'$ introduced in \cite{1405.1015}.

The different terms proportional to $\ln\omega^{-1}$ in \refb{e1.16int},
\refb{e3.23Gint} and in
\refb{e1.17int}, \refb{eqgrsoft} have different origin. We shall explain them in the context of the soft graviton factor,
but the case of soft photon factor is very similar.
\begin{enumerate}
\item We begin with the classical result \refb{e1.17int}. The term proportional to $q_aq_b$ in the first line represents
the effect of late time gravitational radiation due to the late time acceleration of the particles via long range
electromagnetic
interaction. The term in the last line of \refb{e1.17int} represents the effect of late time 
gravitational radiation due to the late time acceleration of the particles via long range gravitational
interaction. 
\asnotex{We expect the scale of these logarithms to be set by the largest length scale involved in the classical
scattering process, {\it e.g.} the typical distance of closest approach between the  particles involved
in the scattering. This is taken to be
larger than or of the order of the Schwarzschild radii of the particles
and much larger than the
Compton wave-lengths of the particles involved in the scattering.}
In the quantum one loop computation both these terms arise from the region of loop momentum integration
where the loop momentum is large compared to 
$\omega$  but small compared to the energies of the other particles. \asnotex{In this case the scale
of these logarithms is again set by the largest length scale involved in the quantum scattering which is
the inverse of the typical energy
carried by the finite energy external states. For one loop result to be reliable, 
this needs to be taken to be large compared to the Schwarzschild
radii of these particles.}
\item The term in the second line of \refb{e1.17int}
proportional to $(\ln\omega^{-1}+\ln R^{-1})$ represents the effect of
gravitational drag on the soft graviton due to the other finite energy particles in the final state. This has the 
effect of causing a time delay, represented by the $\ln R^{-1}$ term, for the soft graviton to travel to a distance
$R$. This also has the effect of inducing backscattering of the soft graviton, represented by the $\ln\omega^{-1}$
term. In the quantum computation these terms arise from region of loop momentum integration where the loop
momentum is smaller than $\omega$ and larger than the infrared cut-off $R^{-1}$. This term has appeared {\it e.g.}
in \cite{peter,0912.4254,1211.6095}. \asnotex{As mentioned in footnote \ref{foo1},
the scale of these logarithms is set by the effective infrared cut-off, {\it e.g.}
the distance $R$ to the detector for the classical scattering and the resolution of the detector for the quantum 
scattering. The latter in turn has a lower limit set by $R^{-1}$ since we cannot measure the energy of the outgoing
particle with an accuracy better than $R^{-1}$ if the detector is placed at a distance $R$ from the scattering center.}
\item We emphasize that the classical results are obtained by replacing in the classical soft
theorem the logarithmically divergent terms by $\ln\omega^{-1}$ and not by direct calculation
of electromagnetic and 
gravitational radiation during classical scattering. In special cases the equivalence of these
two procedures was tested in \cite{1804.09193} 
by direct classical computation. In principle similar tests  can
be done for the general formulae\ \refb{e1.16int} and \refb{e1.17int}, but we have not done
this.
\item We now turn to the additional terms \refb{eqgrsoft} that arise in the quantum computation. First note that both these
terms are real for real polarizations unlike the classical result where the coefficients of  $\ln\omega^{-1}$ terms are
imaginary for real polarizations. The terms in the first two lines come from regions of loop momentum integration where
the loop momentum is large compared to $\omega$ but 
small compared to the energies of the other particles, while the 
term in the third line arise from  region of loop momentum integration where the loop
momentum is small compared to $\omega$ and large compared to 
the infrared cut-off $R^{-1}$.
\item In the quantum computation the terms that arise from loop momenta large
compared to $\omega$, namely the terms
in the first and third line of \refb{e1.17int} and the first two lines of \refb{eqgrsoft}, can be generated using a simple
algorithm. As discussed earlier, the amplitude without the soft graviton has an infrared divergent factor multiplying
it. Let us call this the IR factor.
If in the integration over loop momenta of this IR factor we restrict the loop momentum integration to be
large compared to 
$\omega$ and apply the usual subleading soft differential operator that arises in higher dimensions to
this IR factor, we recover precisely the results given in the first and third line of 
\refb{e1.17int} and the first two lines of \refb{eqgrsoft}. The rest of the contribution that arises from integration
region where the loop momentum is small compared to $\omega$ cannot be recovered this way. This indicates
that the general argument of \cite{1703.00024,1706.00759}, 
based on general coordinate invariance of 1PI effective action and power
counting assuming that loops do not generate inverse power of soft momentum, remain valid in four dimensions
as well as long as the loop momentum is large
compared to the external soft momentum.

\end{enumerate}

Since the real infrared divergent 
part of the amplitude reflects the effect of real graviton emission, our interpretation of the extra
contributions \refb{eqgrsoft} in the quantum theory 
is that they reflect the effect of backreaction of soft radiation on the classical
trajectories. To this end note that the validity of the classical limit described in 
\cite{1801.07719} requires that the total energy
carried by soft radiation should remain small compared to the energies of the finite energy objects taking part in
the scattering. Here `soft radiation' represents those particles which are not included in the sum over $a$ in
\refb{e1.17int}. 
Therefore we should expect that the extra terms arising in the quantum theory should be small in the
limit when the total energy carried by the soft radiation is small.

In order to test this hypothesis we need to consider a scattering where the energy carried away by soft radiation 
remains small compared to the energies of finite energy objects. One way to achieve this is to consider scattering
at large impact parameter so that each incoming particle gets deflected by a small amount and the energy radiated
during this process remains small. In this case the momenta $\{p_a\}$ come in approximately equal and opposite
pairs -- the incoming and the corresponding outgoing particle. Now in eq.\refb{eqgrsoft} the last term changes sign under 
$p_b\to -p_b$ and also under $p_a\to -p_a$. This shows that it is small for small deflection scattering. The first
term on the right hand side of \refb{eqgrsoft} changes sign 
under $(p_b, q_b)\to -(p_b, q_b)$ and also under $(p_a, q_a)\to -(p_a, q_a)$, due to the argument of the log
getting inverted under each of these operations. This
shows that the terms approximately cancel making the result small. There is one exception to this that arises 
when $q_b=-q_a$, $p_b\simeq -p_a$, i.e.\ the pairs $(a,b)$ represent the incoming and the corresponding outgoing
particle. In this case there is no other term that cancels this since the sum does not include the $b=a$ term, 
and we need to explicitly evaluate this
and show that it vanishes. This can be checked explicitly by first evaluating the derivatives in the second line of
\refb{eqgrsoft}, then setting $p_b=-p_a+\eps$ and then carefully 
evaluating the result in the $\eps\to 0$ limit. Even though individual terms diverge in the $\eps\to 0$ limit, a careful
analysis shows that the result vanishes. This
confirms that quantum corrections are small in this limit.

Another situation discussed in \cite{1801.07719}, where the radiated energy remains small compared to the energies
of the hard particles, is the probe limit in which one of the particles has a large mass $M$ and
the other particles are lighter carrying energy small compared to $M$. We shall now verify
that in this case too the quantum corrections \refb{eqgrsoft} are small compared to the classical result \refb{e1.17int}.
For this we shall work in a frame in which the heavy particle is initially at rest, and using gauge invariance choose
the polarization tensor $\ve$ to have only spatial components.  After the scattering the heavy particle acquires a
momentum but it is small compared to $M$. In this case the dominant contribution to \refb{e1.17int}, of order $M$,
comes from choosing $a$ to be one of the light particles and $b$ to be the heavy particle in the second and third
line of \refb{e1.17int}. However in the quantum correction \refb{eqgrsoft} similar contribution cancels between the choice
of $b$ as the initial state heavy particle and the final state heavy particle, and we do not get any contribution
proportional to $M$. This again shows that quantum corrections are small compared to the classical result 
in this limit.

We must emphasize however that the quantum analysis
is carried out
for single soft graviton emission. If we want to relate the quantum result to the radiative component of the
classical gravitational field as in \cite{1801.07719}, then we need to first consider multiple
soft graviton emission and then take the classical limit. The analysis of \cite{1801.07719} 
relied on the fact that the soft factors associated with different
bins in the phase space are independent of each other, i.e.\ the probability of emitting certain number of soft
particles in one bin does not depend on how many soft particles are emitted in the other bin. This 
independence breaks
down when the total energy carried by the soft particles becomes comparable to the energies of the hard
particles -- precisely when the quantum correction \refb{eqgrsoft} becomes comparable to the classical
result \refb{e1.17int}. 
Therefore we should not use \refb{eqgrsoft}  to modify the classical result \refb{e1.17int}.
Instead we should use the smallness of 
\refb{eqgrsoft} as a  test of when the classical result \refb{e1.17int} is valid. 
An identical discussion holds for
electromagnetism.

\subsection{Special cases}

As a special case we can consider the situation described in 
\cite{1806.01872} where a neutral
massive object of mass $M$ at rest
decays into a heavy
object of mass $M_0\simeq M$ and a set of neutral 
light objects carrying mass $m_a<< M$ and momentum $p_a=-e_a(1,\vec \beta_a)$ with $e_a<<M$ 
for $a=1,\cdots N$. Our goal will be to write down the classical soft graviton factor for this
case. We shall take the polarization tensor of the soft graviton to have components only along the spatial direction,
since the result for the other components may be found by using invariance under the gauge transformation
$\ve_{\mu\nu}\to \ve_{\mu\nu} + \xi_\mu k_\nu + \xi_\nu k_\mu$ for any vector $\xi$. If we denote the momentum 
carried by final state
heavy object of mass $M_0$ by $p_{N+1}$, then we have $p^0_{N+1}\simeq -M_0$ and $|p^i_{N+1}|<< M_0$.
Examining \refb{e1.17int} with $q_a=q_b=0$ we see that dominant term proportional to $\ln\omega^{-1}$ comes
from the terms where we choose $b=N+1$ and $a$ labels any of the $N$ finite energy states. Using the relation
$e_a^2 = m_a^2 / (1-\vec\beta_a^2)$, the net contribution
takes the form:
\ben\label{e101}
&& {i\over 4\pi} \ln\omega^{-1}\, M_0\, \sum_{a=1}^N  e_a\,{\ve^{ij} \beta_{ai} \beta_{aj}\over 1-\hat n.\vec \beta_a}
+{i\over 8\pi} \ln\omega^{-1}\, M_0\, \sum_{a=1}^N e_a\,{\ve^{ij} \beta_{ai} \beta_{aj}\over 1-\hat n.\vec \beta_a}
{ (-e_a) (2 e_a^2 - 3 m_a^2) \over (e_a^2 - m_a^2)^{3/2}} \nonumber\\
&=& {i\over 8\pi} \ln\omega^{-1}\, M_0\,  \sum_{a=1}^N e_a\,{\ve^{ij} \beta_{ai} \beta_{aj}\over 1-\hat n.\vec \beta_a}
\, {2 \vec \beta_a^3+1 - 3\vec\beta_a^2
\over |\vec \beta_a|^3}+\cdots\, ,
\een
where $\cdots$ contain terms without a factor of $M_0$ and are therefore smaller in the limit of large $M_0$.
This agrees with the results of \cite{1806.01872}. 
As discussed in \cite{1806.01872}, this produces a late time tail in the gravitational
wave-form that falls off as inverse power of time.

Note that when all the final state light particles are massless, so that $|\vec\beta_a|=1$ for $1\le a\le N$, the expression
\refb{e101} vanishes. This would be the situation during binary black hole merger when the final state particles are
only gravitons. However since in such processes the radiation carries away an appreciable fraction of the mass of the
parent system, the $\cdots$ terms in \refb{e101} could be significant even though their contribution will be
suppressed by the ratio of the total energy carried away by radiation to the mass of the parent system. We shall now evaluate the result  without making any approximation. In this case
in the sum over $a$ and $b$ in \refb{e1.17int}, either $a$ or $b$ (or both) represents a
massless particle. 
Recalling that
when $p_a$ and $p_b$ are both outgoing then $p_a.p_b$ is negative, we can express the terms in \refb{e1.17int} 
proportional to $\ln\omega^{-1}$ as
\be 
{i\over 4\pi} \ln\omega^{-1} \sum_{a=1}^{N+1} \, \ve^{ij} p_{ai} p_{aj} 
+ {i\over 4\pi} \ln\omega^{-1} \sum_{a=1}^{N+1} \, \ve^{ij} p_{ai}  \sum_{b=1\atop b\ne a}^{N+1} p_{bj}=0\, ,
\ee
where in the last step we have used conservation of spatial momentum $\sum_{b=1}^{N+1} p_{bj}=0$. Therefore
we see that even without making any approximation, 
the coefficient of the $\ln\omega^{-1}$ term in the classical soft graviton
factor continues to vanish.

Another special case we can consider is when a charge neutral 
object of mass $M$ at rest breaks apart into two charge neutral objects of
masses $m_1$ and $m_2$, spatial momenta $\vec p$ and $-\vec p$ and energies 
$e_1=\sqrt{m_1^2+\vec p^2}$ and $e_2=\sqrt{\vec p^2+m_2^2}$. In this case if we
take the polarization tensor of the soft graviton to have components only along the spatial direction, then the
contribution from the initial state to \refb{e1.17int} vanishes and we need to only compute the contribution
from a pair of final states. This can be easily evaluated and the terms proportional to $\ln\omega^{-1}$
take the form 
\ben
&& 
{i\over 8\pi} \,  \ln\omega^{-1} \, \ve_{ij} p^i p^j \, (e_1+e_2)
\, \left\{ {1\over e_1 - \hat n.\vec p}
+ {1\over e_2 + \hat n.\vec p}
\right\} \nonumber \\ && \hskip .5in \times 
\left[{e_1 e_2 + \vec p^2\over \{(e_1e_2+\vec p^2)^2 - m_1^2 m_2^2\}^{3/2}}
\left\{ 2(e_1 e_2+\vec p^2)^2 - 3 m_1^2 m_2^2\right\}  -2\right]\, .
\een

Next special case we shall analyze is that of scattering of massless particles, again focussing on the classical result
\refb{e1.17int}. Defining
\be \label{edefP}
P\equiv \sum_{\eta_a=1} p_a =- \sum_{\eta_a=-1} p_a\, ,
\ee
and the fact that $p_a.p_b$ is negative for $\eta_a\eta_b=1$, we can express the term proportional to
$\ln\omega^{-1}$ in \refb{e1.17int} for massless particles as
\be \label{emassless}
-{i\over 2\pi} \, \ln\omega^{-1}\, k.P\, \sum_{a\atop \eta_a=1} 
{\ve_{\mu\nu} p_a^\mu p_a^\nu\over
p_a.k} +{i\over 2\pi} \, \ln\omega^{-1}\, \ve_{\mu\nu} P^\mu P^\nu\, .
\ee
Note that this involves only the momenta of the initial state particles and is insensitive to the momenta of the
final state particles. This asymmetry is related to the fact that in our analysis we are considering soft particle
only in the final state and not in the initial state.

More generally one can show that for a general scattering process involving both
massive and massless particles,  the terms proportional to $\ln\omega^{-1}$
in the classical formula \refb{e1.17int} is not sensitive to the details of the
final state massless particles except through overall momentum conservation. 
To see this let us first consider terms that could involve a final state massless particle momenta and the initial
state momenta. These come from choosing $a$ to be an initial state and $b$ to be a final state massless
state in the term in the second line of \refb{e1.17int}. The net contribution from such terms is given by
\be
{i\over 4\pi} \ln\omega^{-1} 
\sum_{b \, \rm massless \atop \eta_b=-1} k.p_b \, \sum_{a\atop \eta_a=1} {\ve_{\mu\nu} p_a^\mu p_a^\nu\over
k.p_a} = -{i\over 4\pi} \, \ln\omega^{-1}  \, \,
k. (P-P_{\rm massive}) \, \sum_{a\atop \eta_a=1} {\ve_{\mu\nu} p_a^\mu p_a^\nu\over
k.p_a} \, ,
\ee
where $-P$ denotes total outgoing momentum as defined in \refb{edefP} and $-P_{\rm massive}$ denotes the
total outgoing momentum carried by the massive particles. Therefore this does not depend explicitly on the momenta
of the outgoing massless states except through momentum conservation.

Next we consider terms that involve a pair of final state momenta at least one of which is massless. This term receives
contribution from all three lines on the right hand side of \refb{e1.17int} with the restriction $\eta_a=1$, $\eta_b=1$,
and either $m_a$ or $m_b$ or both zero.
Therefore  the term proportional to $q_aq_b$ vanishes. Also the coefficient of
$\ln\omega^{-1}$ in the summand in
the last two lines simplifies to
\be \label{esim1}
{i\over 4\pi} {\ve_{\mu\nu} p_a^\mu p_a^\nu\over p_a.k} \, p_b.k
-{i\over 4\pi} {\ve_{\mu\nu} p_a^\mu p_a^\nu\over p_a.k} \, p_b.k + {i\over 4\pi} \, \ve_{\mu\nu}\, p_a^\mu p_b^\nu  \, .
\ee
In the first term the sum over $a$ and $b$ includes the term where $b=a$, but in the second and the third term
the sum excludes the $b=a$ term. Therefore the first two terms
almost cancel, leaving behind a contribution where we set $b=a$.  This left over contribution
${i\over 4\pi} \, \ve_{\mu\nu} p_a^\mu p_a^\nu$ can now be added to the last term to include in the sum over $a$ or
$b$ also the contribution where $b=a$. The net contribution from the terms where either $a$ or $b$ or both
represent massless state is then
\be
{i\over 4\pi} \ln\omega^{-1}  \sum_{a,b; \eta_a=\eta_b=-1\atop \rm either\, a\, or\, b\, massless} 
\ve_{\mu\nu}\, p_a^\mu p_b^\nu\, .
\ee
This can be rewritten as
\be
{i\over 4\pi} \ln\omega^{-1}  \ve_{\mu\nu}\left(\sum_{a,b; \eta_a=\eta_b=-1} 
 p_a^\mu p_b^\nu -  \sum_{a,b; \eta_a=\eta_b=-1\atop \rm  a\, and \, b\, massive} 
 p_a^\mu p_b^\nu\right)
= {i\over 4\pi} \ln\omega^{-1} \ve_{\mu\nu} \left(P^\mu P^\nu - P_{\rm massive}^\mu P_{\rm massive}^\nu
\right)\, .
\ee
This also does not depend on
the details of the momenta of massless final state particles except for the total momentum carried by these
particles.

\sectiono{Classical analysis} \label{sclassical}

The goal of this section will be to calculate the logarithmic terms in the soft factors in four space-time
dimensions by examining them in the classical limit. 

In dimensions larger than 4, the
soft factors for photons and gravitons are given respectively by
\be \label{eem}
S_{\rm em} = \sum_a {\ve_\mu p_a^\mu \over p_a.k} \, q_a + i \sum_a q_a \,  {\ve_\mu k_\rho \bJ_a^{\rho\mu}\over p_a.k}\, ,
\ee
and 
\be \label{egr}
S_{\rm gr} = \sum_a {\ve_{\mu\nu} p_a^\mu p_a^\nu \over p_a.k}  + i \sum_a {\ve_{\mu\nu} p_a^\nu 
k_\rho \bJ_a^{\rho\mu}\over p_a.k}\, .
\ee
Here the sum over $a$ runs over all the incoming and outgoing particles, and $q_a$, $p_a$ and
$\bJ_a$ denote the charge, momentum and angular momentum of the $a$-th particle, counted with 
positive sign for an ingoing particle and negative sign for an outgoing particle. 
$S_{\rm em}$ may also contain a non-universal term at the subleading order.
For S-matrix elements in quantum theory, $\bJ_a$ is a differential operator involving derivatives
with respect to the external momenta. However in the classical limit in which the external
finite energy states are macroscopic, $\bJ_a$ represents the classical angular momenta carried by
the external particles. In this limit the soft factors 
describe the radiative part of the low frequency electromagnetic and gravitational 
fields during a classical scattering\cite{1801.07719} as described in \refb{e100}.

In applying \refb{eem}, \refb{egr} to four dimensional theories, the complication arises from
the contribution to $\bJ_a^{\mu\nu}$ from the orbital angular momentum.
They are computed from the form of the
asymptotic trajectories:
\be \label{etraj}
r_a^\mu(\sigma) =\eta_a\, {1\over m_a} p_a^\mu \, \sigma + c_a^\mu \, \ln |\sigma| + \cdots \, ,
\ee
where $\eta_a$ is positive for incoming particles and negative for outgoing particles, $m_a$ is the
mass of the $a$-th particle and the 
proper time $\sigma$ is
large and negative for incoming particles and large and positive for outgoing particles. The term 
proportional to $\ln|\sigma|$ represents the effect of long range electromagnetic and/or gravitational
interaction between the particles.
This gives, for large $|\sigma|$,
\be \label{elog}
\bJ_a^{\mu\nu} \simeq r_a^\mu(\sigma) p_a^\nu - r_a^\nu(\sigma) p_a^\mu +\hbox{spin}
= (c_a^\mu p_a^\nu- c_a^\nu p_a^\mu) \, \ln |\sigma| +\cdots \, .
\ee
Here and in the following we shall use the convention that when a variable is followed
by an argument $(\sigma)$ it denotes the value of the variable at proper time $\sigma$,
but when a variable is written without an argument, we take it to be its $\sigma$ independent
asymptotic value. Therefore in \refb{etraj}, 
\refb{elog} the $p_a^\mu$'s denote the asymptotic values
of $p_a^\mu$, reflecting the fact that the difference between $p_a^\mu(\sigma)=
m_a \eta_a dr_a^\mu/d\sigma$ and $p_a^\mu$ approaches
zero asymptotically.

Analysis of \cite{1804.09193 } indicates that if we substitute \refb{elog} 
into \refb{eem} and \refb{egr} and
replace $\ln|\sigma|$ by $\ln\omega^{-1}$ -- where $\omega=k_0$ is the frequency of the outgoing 
soft radiation -- we can recover the logarithmic terms in the soft factors up to overall phases. This gives, up to
overall phases:
\be \label{eem1}
S_{\rm em} = \sum_a {\ve_\mu p_a^\mu \over p_a.k} \, q_a + i \, \ln \omega^{-1}
\sum_a q_a \,  {\ve_\mu k_\rho (c_a^\rho p_a^\mu - c_a^\mu p_a^\rho) \over p_a.k}\, ,
\ee
and
\be \label{egr1}
S_{\rm gr} = \sum_a {\ve_{\mu\nu} p_a^\mu p_a^\nu \over p_a.k}  + i \, \ln \omega^{-1}
\sum_a {\ve_{\mu\nu} p_a^\nu 
k_\rho (c_a^\rho p_a^\mu - c_a^\mu p_a^\rho)  \over p_a.k}\, .
\ee
Note that although $S_{\rm em}$ may contain a non-universal term at the subleading order,
the term proportional to $\ln\omega^{-1}$ comes from orbital angular momentum and is
universal.

Irrespective of what forces are operative during the scattering, the coefficient $c_a^\mu$
are determined only by the long range forces acting on the incoming and the outgoing
particles. These will be taken to be electromagnetic and / or gravitational interaction.
We shall now compute $c_a^\mu$ due to 
electromagnetic and gravitational interactions. We know from
explicit comparison with known results that in the case of scattering via electromagnetic interactions there are no
additional phases in the soft factor, but in the 
case of gravitational long range interaction there is an additional phase reflecting the
effect of backscattering of the soft photon or soft graviton in the background 
gravitational field\cite{peter,0912.4254,1211.6095}. This phase will also be
determined below.

\subsection{Effect of electromagnetic interactions}

We shall first study the effect of logarithmic correction to the trajectory due to 
long range electromagnetic interaction. For this we need to compute the gauge potential 
$A_\mu^{(b)}(x)$ at space-time 
point $x$ due to particle $b$. We have
\be \label{e1.7}
A_\mu^{(b)}(x) = {1\over 2\pi} \int d\sigma \, \eta_b\, \, q_b 
\, V_{b\mu}(\sigma) \, \delta_+( - (x-r_b(\sigma))^2), \quad
V_b^\mu(\sigma) \equiv {dr_{b\mu}(\sigma)
\over d\sigma}\simeq \eta_b \, {p_b^\mu\over m_b}\, ,
\ee
where $\delta_+$ denotes the usual Dirac delta function with the understanding 
that we have to choose the zero of the argument for which $x^0>r_b^0(\sigma)$.
$V_b$ 
denotes the asymptotic four velocity of the $b$-th particle. In evaluating \refb{e1.7}
we shall ignore the logarithmic corrections to the trajectory and take
$r_b(\sigma) \simeq V_b \, \sigma $. This gives, using $V_b^2=-1$,
\be
\delta_+(- (x-r_b(\sigma))^2) = \delta_+(-x^2 + 2\, V_b.x \, \sigma +\sigma^2 + \cdots)
\simeq {1\over 2|V_b.x+\sigma|} \, \delta(\sigma + V_b.x + \sqrt{(V_b.x)^2 + x^2})\, ,
\ee
where the sign in front of the square root has been chosen to ensure that $x^0>x_b^0(\sigma)$ at the
solution. Substituting this into \refb{e1.7} we get
\be \label{eco3}
A_\mu^{(b)}(x) \simeq {1\over 4\pi} \, {\eta_b\, q_b V_{b\mu} \over \sqrt{(V_b.x)^2 + x^2}}\, .
\ee
From this we calculate
\be 
F^{(b)}_{\mu\nu}(x) = \p_\mu A^{(b)}_\nu (x) - \p_\nu A^{(b)}_\mu(x) \simeq -{\eta_b\, 
q_b\over 4\pi} \,  {x_{\mu} V_{b\nu}- x_{\nu} V_{b\mu}\over 
\{ (V_b.x)^2 + x^2\}^{3/2}}\, .
\ee
At the location $r_a= V_a\sigma=-V_a |\sigma| \eta_a$ of the 
$a$-th particle we get, using $V_a^2=-1$
\be
F^{(b)}_{\mu\nu}(r_a(\sigma)) \simeq \eta_a \, \eta_b\, {q_b\over 4\pi\, \sigma^2} \,  {V_{a\mu} V_{b\nu}- V_{a\nu} V_{b\mu}\over 
\{ (V_b.V_a)^2 -1\}^{3/2}}\, .
\ee
Now the $a$-th particle will feel the field produced by 
the $b$-th particle if either both $a$-th and the $b$-th particle are
outgoing or if both particles are ingoing. 
Therefore the equation of motion for the $a$-th particle takes the form
\be \label{eco1}
{d p_{a\mu}(\sigma)\over d\sigma} = q_a \, \sum_{b\ne a\atop \eta_a\eta_b=1} 
F^{(b)}_{\mu\nu}(r_a(\sigma)) \, V_a^\nu(\sigma)
\simeq {1\over \sigma^2}  \, \sum_{b\ne a\atop \eta_a\eta_b=1} \eta_a \, \eta_b\, {q_a q_b\over 4\pi} \, {V_a.V_b V_{a\mu} + V_{b\mu}\over 
\{ (V_b.V_a)^2 -1\}^{3/2}}\, .
\ee
On the other hand we have
\be\label{eco2}
{d p_{a\mu}(\sigma)\over d\sigma} = {m_a\over \eta_a} {d^2 r_{a\mu}\over d\sigma^2} = - {m_a\over \eta_a} {c_{a\mu} \over \sigma^2}\, ,
\ee
where in the last step we used \refb{etraj}.
Comparing \refb{eco1}, \refb{eco2} we get
\be
c_a^\mu = -{1\over m_a} \sum_{b\ne a\atop \eta_a\eta_b=1} \, \eta_b
\, {q_a q_b\over 4\pi} \, {V_a.V_b V_{a}^{\mu} + V_{b}^{\mu}\over 
\{ (V_b.V_a)^2 -1\}^{3/2}}
= -\sum_{b\ne a\atop \eta_a\eta_b=1} \, {q_a q_b\over 4\pi} \, 
{m_b^2\, p_a.p_b \, p_{a}^{\mu} + m_a^2 m_b^2 \, p_{b}^{\mu}\over 
\{ (p_b.p_a)^2 -m_a^2 m_b^2\}^{3/2}}\, ,
\ee
and
\be
c_a^\mu p_a^\nu - c_a^\nu p_a^\mu 
= -\sum_{b\ne a\atop \eta_a\eta_b=1} \, {q_a q_b\over 4\pi} \,
{m_a^2 m_b^2 \, \{p_{b}^{\mu} p_a^\nu - p_b^\nu p_a^\mu\} \over 
\{  (p_b.p_a)^2 -m_a^2 m_b^2\}^{3/2}}\, .
\ee
Eqs.\refb{eem1} and \refb{egr1} now give\footnote{Note that even if we assume that the 
logarithmic corrections to the trajectories are generated 
predominantly by electromagnetic interaction, the resulting acceleration 
can generate logarithmic corrections to the gravitational
radiation during the scattering.}
\be\label{e1.16}
S_{\rm em} = \sum_a {\ve_\mu p_a^\mu \over p_a.k} \, q_a - i \, \ln \omega^{-1}
\sum_a {q_a \, \ve_\mu k_\rho \over p_a.k}
 \sum_{b\ne a\atop \eta_a\eta_b=1} \, {q_a q_b\over 4\pi} \,
{m_a^2 m_b^2 \, \{p_{b}^{\rho} p_a^\mu - p_b^\mu p_a^\rho\} \over 
\{ (p_b.p_a)^2 -m_a^2 m_b^2\}^{3/2}}\, ,
\ee
and
\be\label{e1.17}
S_{\rm gr} = \sum_a {\ve_{\mu\nu} p_a^\mu p_a^\nu \over p_a.k}  - i  \, \ln \omega^{-1}
\sum_a {\ve_{\mu\nu} p_a^\nu k_\rho 
 \over p_a.k}
 \sum_{b\ne a\atop \eta_a\eta_b=1} \, {q_a q_b\over 4\pi} \,
{m_a^2 m_b^2 \, \{p_{b}^{\rho} p_a^\mu - p_b^\mu p_a^\rho\} \over 
\{  (p_b.p_a)^2 -m_a^2 m_b^2\}^{3/2}}\, .
\ee

\subsection{Effect of gravitational interactions}
Let us now suppose that the logarithmic correction to the trajectories arise
due to gravitational interaction. We introduce the
graviton field $h_{\mu\nu}$ and its trace reversed version $e_{\mu\nu}$ via the equations
\be
h_{\mu\nu} \equiv (g_{\mu\nu} -\eta_{\mu\nu})/2, \qquad e_{\mu\nu} = h_{\mu\nu} 
- {1\over 2} \eta_{\mu\nu} \,
h_\rho^{~\rho}\, .
\ee
Then the
analog of \refb{e1.7} for the gravitational field produced at $x$ due to the $b$-th particle is
\be \label{e2.7}
e_{\mu\nu}^{(b)}(x) = {1\over 2\pi} \int d\sigma \, m_b \, V_{b\mu}(\sigma) \, V_{b\nu}(\sigma)
\, \delta_+( - (x-r_b(\sigma))^2) \, .
\ee
Using $r_b(\sigma)=V_b\, \sigma +\cdots$ we get the analog of \refb{eco3}
\be \label{eco4}
e_{\mu\nu}^{(b)}(x) \simeq {1\over 4\pi} \, {m_b \, V_{b\mu} \, V_{b\nu}
\over \sqrt{(V_b.x)^2 + x^2}}\, .
\ee
The associated Christoffel symbol is given by, in the weak field approximation,
\ben\label{echri}
\Gamma^{(b)\alpha}_{\rho\tau}(x) &=& -{m_b\over 4\pi} \,  {1\over 
\{ (V_b.x)^2 + x^2\}^{3/2}}\, \eta^{\alpha\mu} \, 
\left[\left\{V_{b\mu} V_{b\tau} + {1\over 2} \eta_{\mu\tau} \right\} \left\{x_\rho + V_b.x\, V_{b\rho}\right\}
\right . \nonumber \\ && \hskip -.5in \left.
+ \left\{V_{b\mu} V_{b\rho} + {1\over 2} \eta_{\mu\rho} \right\} \left\{x_\tau + V_b.x\, V_{b\tau}\right\}
- \left\{V_{b\rho} V_{b\tau} + {1\over 2} \eta_{\rho\tau} \right\} \left\{x_\mu + V_b.x\, V_{b\mu}\right\}
\right]\,.
\een
From this we can write down the
equation of motion of the $a$-th particle
\ben
{d^2 r_{a}^\alpha(\sigma)\over d\sigma^2} &=& - \sum_{b\ne a\atop \eta_a\eta_b=1} \Gamma^{(b)\alpha}_{\rho\tau}(r_a(\sigma)) \, V_a^\rho(\sigma) \, V_a^\tau(\sigma) 
 \\ &\simeq& -\eta_a\, {1\over 4\pi \sigma^2}  \, \sum_{b\ne a\atop \eta_a\eta_b=1} \, m_b \, {1\over 
\{ (V_b.V_a)^2 -1\}^{3/2}}\, \left[- {1\over 2} V_a^\alpha + {1\over 2} V_b^\alpha \left\{2(V_b.V_a)^3  
- 3 V_b.V_a\right\}\right]\, . \nonumber
\een
On the other hand using \refb{etraj} the left hand side is given by $-c_a^\alpha /\sigma^2$.
This gives
\be 
c_a^\alpha = \eta_a\,  {1\over 4\pi}  \, \sum_{b\ne a\atop \eta_a\eta_b=1} \, m_b \, {1\over 
\{ (V_b.V_a)^2 -1\}^{3/2}}\, \left\{- {1\over 2} V_a^\alpha + {1\over 2} V_b^\alpha \left(2(V_b.V_a)^3  
- 3 V_b.V_a\right)\right\} \, ,
\ee
and
\ben
c_a^\rho p_a^\mu - c_a^\mu p_a^\rho &=&
 {1\over 8\pi \sigma^2}  \, \sum_{b\ne a\atop \eta_a\eta_b=1} \,  m_a \, m_b \, {1\over 
\{ (V_b.V_a)^2 -1\}^{3/2}}\, (V_b^\rho V_a^\mu - V_b^\mu V_a^\rho) \, 
\left\{2(V_b.V_a)^3  
- 3 V_b.V_a\right\}\nonumber \\
&=&
 {1\over 8\pi}  \, \sum_{b\ne a\atop \eta_a\eta_b=1} \,  {p_b.p_a\over 
\{ (p_b.p_a)^2 -m_a^2 m_b^2\}^{3/2}}\, (p_b^\rho p_a^\mu - p_b^\mu p_a^\rho) \, 
\left\{2(p_b.p_a)^2  - 3 m_a^2 m_b^2 \right\} \, .\nonumber \\
\een
Substituting this into \refb{eem1} and \refb{egr1} we get,\footnote{Even if the logarithmic
correction to the trajectory is generated by
gravitational interaction, the particles can emit electromagnetic waves. This happens for example
if we have a scattering of a charged particle and a neutral particle.} up to overall phases:
\ben\label{eem2}
S_{\rm em} &=& \sum_a {\ve_\mu p_a^\mu \over p_a.k} \, q_a + {i\over 8\pi} 
\, \ln \omega^{-1}
\sum_a {q_a \, \ve_\mu k_\rho \over p_a.k}
 \sum_{b\ne a\atop \eta_a\eta_b=1} 
 \,  {p_b.p_a\over 
\{ (p_b.p_a)^2 -m_a^2 m_b^2\}^{3/2}}\, (p_b^\rho p_a^\mu - p_b^\mu p_a^\rho) 
\nonumber \\ && \hskip 1in \times
\left\{2(p_b.p_a)^2  - 3 m_a^2 m_b^2 \right\} 
\, ,
\een
and
\ben\label{egr2}
S_{\rm gr} &=& \sum_a {\ve_{\mu\nu} p_a^\mu p_a^\nu \over p_a.k}  + {i\over 8\pi}  
\, \ln \omega^{-1}
\sum_a {\ve_{\mu\nu} p_a^\nu k_\rho 
 \over p_a.k}
 \sum_{b\ne a\atop \eta_a\eta_b=1} 
 \,  { p_b.p_a\over 
\{ (p_b.p_a)^2 -m_a^2 m_b^2\}^{3/2}}\, (p_b^\rho p_a^\mu - p_b^\mu p_a^\rho) 
\nonumber \\ && \hskip 1in \times
\left\{2(p_b.p_a)^2  - 3 m_a^2 m_b^2\right\} 
\, .
\een

In this case we expect the wave-form of the gauge field / metric 
to also have an additional phase factor 
reflecting the effect of the gravitational drag on the soft particle due to the other particles.
For this let us characterize the asymptotic trajectory of the soft particle as
\be\label{enulltraj}
x^\mu(\tau) = n^\mu \, \tau+ m^\mu \, \ln|\tau|\, ,
\ee
where $\tau$ is the affine parameter associated with the trajectory, $n=(1,\hat n)$ is a null
vector along the asymptotic direction of motion of the soft particle and $m^\mu$  is a
four vector to be determined. Now substituting \refb{enulltraj} into 
the equation of motion 
\be
{d^2 x^\mu\over d\tau^2} = - \Gamma^\mu_{\nu\rho} \, {dx^\nu \over d\tau}\, {dx^\rho\over
d\tau}\, ,
\ee
and using the form \refb{echri} of $\Gamma^\mu_{\nu\rho}$, we get the following
expression for $m^\mu$ by comparing the $1/\tau^2$ terms on the two sides of the
equations of motion:
\be\label{emmm}
m^\alpha= -{1\over 4\pi} \sum_{b\atop \eta_b=-1} {m_b\over |n.V_b|^3} \, V_b^\alpha \, 
(V_b.n)^3 = {1\over 4\pi} \sum_{b\atop \eta_b=-1}  m_b\, V_b^\alpha =
- {1\over 4\pi} \sum_{b\atop \eta_b=-1}  p_b^\alpha\, .
\ee
Now eliminating $\tau$ in terms of $t\equiv x^0$ using \refb{enulltraj}, we can express
\refb{enulltraj} as
\be
x^i = n^i t + (m^i - n^i m^0) \ln |t| + \hbox{finite}\, .
\ee
Therefore if we denote by $k=(k^0,k)=-\omega(1, \hat n)$ the four momentum of the soft particle, the overall $-$ sign reflecting the fact that it is an outgoing particle, the
wave-function of the particle will be proportional to
\be \label{eabove}
\exp\left[-i \vec k. \left\{\vec x - \hat n t - (\vec m - \hat n \, m^0)\ln|t|\right\}\right]
= \exp[-i\omega t + i \omega \hat n.\vec x] \, \exp[i (\vec k.\vec m + \omega\, m^0)\ln |t|]\,.
\ee
The second factor can be regarded as an additional 
infrared divergent contribution to the soft factor.
Using $|t|\sim R$ where $R$ is the distance of the soft particle from the scattering center,
 and eq.\refb{emmm}, we can express
the second factor in \refb{eabove} as
\be\label{eirphase}
\exp[i k.m\, \ln R] = 
\exp\left[-{i\over 4\pi} \ln R\, \sum_{b\atop \eta_b=-1} k.p_b\right]\, .
\ee
Since this is a pure phase it does not affect the flux. However it does produce observable effect on the
electromagnetic / gravitational wave-form\cite{1806.01872}.

It follows from the analysis of  \cite{peter,0912.4254,1211.6095} that 
the effect of gravitational backscattering of the soft photon / graviton actually
converts $\ln R$ in \refb{eirphase} to $\ln(R\, \omega)$. 
This has been
reviewed in
\cite{1804.09193}.
It is natural to absorb this multiplicative factor in the wave-form into the definition of the soft factors.
Expanding the exponential in a power series, picking up the term of order 
$\omega\ln(\omega R)$ in the
expansion, and multiplying this by the leading soft factor, we get additional contributions to the
soft photon and soft graviton factor at the subleading order
\be
{i\over 4\pi}\, \left(\ln\omega^{-1}+ \ln R^{-1}\right)\, 
S^{(0)}_{\rm em}\, \sum_{b\atop \eta_b=-1} k.p_b, \qquad
\hbox{and} \qquad {i\over 4\pi}\, \left(\ln\omega^{-1}+ \ln R^{-1}\right)\,  S^{(0)}_{\rm gr} \, \sum_{b\atop \eta_b=-1} k.p_b\, .
\ee
Adding these to \refb{eem2} and \refb{egr2} we get the net soft factors to be
\ben\label{eem2tot}
S_{\rm em} &=& \sum_a {\ve_\mu p_a^\mu \over p_a.k} \, q_a 
+ {i\over 4\pi}\, \left(\ln\omega^{-1}+ \ln R^{-1}\right)\,  \, \sum_{b\atop \eta_b=-1} k.p_b \, \sum_a {\ve_\mu p_a^\mu \over p_a.k} \, q_a 
\nonumber \\ && \hskip -.3in
+ {i\over 8\pi} 
\, \ln \omega^{-1}
\sum_a {q_a \, \ve_\mu k_\rho \over p_a.k}
 \sum_{b\ne a\atop \eta_a\eta_b=1} 
 \,  {p_b.p_a\over 
\{ (p_b.p_a)^2 -m_a^2 m_b^2\}^{3/2}}\, (p_b^\rho p_a^\mu - p_b^\mu p_a^\rho) 
\left\{2(p_b.p_a)^2  - 3 m_a^2 m_b^2 \right\} 
\, ,\nonumber \\
\een
and
\ben\label{egr2tot}
S_{\rm gr}&=& \sum_a {\ve_{\mu\nu} p_a^\mu p_a^\nu \over p_a.k}  
+  {i\over 4\pi}
\, \left(\ln\omega^{-1}+ \ln R^{-1}\right)\,  \, \sum_{b\atop \eta_b=-1} k.p_b \,  \sum_a {\ve_{\mu\nu} p_a^\mu p_a^\nu \over p_a.k}  
\nonumber \\ && \hskip -.3in
+ {i\over 8\pi}  
\, \ln \omega^{-1}
\sum_a {\ve_{\mu\nu} p_a^\nu k_\rho 
 \over p_a.k}
 \sum_{b\ne a\atop \eta_a\eta_b=1} 
 \,  { p_b.p_a\over 
\{ (p_b.p_a)^2 -m_a^2 m_b^2\}^{3/2}}\, (p_b^\rho p_a^\mu - p_b^\mu p_a^\rho) 
\left\{2(p_b.p_a)^2  - 3 m_a^2 m_b^2\right\} 
\, .\nonumber \\
\een

\subsection{Effect of electromagnetic and gravitational interactions}

We now combine the results of last two subsections
to write down the
general expression for the soft factor when both gravitational interaction and electromagnetic
interactions are responsible for the logarithmic corrections to the trajectory. 
The logarithmic terms get added up, yielding the
results:
\ben \label{eem3}
S_{\rm em} &=&  \sum_a {\ve_\mu p_a^\mu \over p_a.k} \, q_a
+ {i\over 4\pi}\, \left(\ln\omega^{-1}+ \ln R^{-1}\right)\,  \, \sum_{b\atop \eta_b=-1} k.p_b \, \sum_a {\ve_\mu p_a^\mu \over p_a.k} \, q_a 
\nonumber \\ &&
- i \, \ln \omega^{-1}
\sum_a {q_a \, \ve_\mu k_\rho \over p_a.k}
 \sum_{b\ne a\atop \eta_a\eta_b=1} \, {q_a q_b\over 4\pi} \,
{m_a^2 m_b^2 \, \{p_{b}^{\rho} p_a^\mu - p_b^\mu p_a^\rho\} \over 
\{ (p_b.p_a)^2 -m_a^2 m_b^2\}^{3/2}} \nonumber \\ && \hskip -.3in
 + {i\over 8\pi} 
\, \ln \omega^{-1}
\sum_a {q_a \, \ve_\mu k_\rho \over p_a.k}
 \sum_{b\ne a\atop \eta_a\eta_b=1} 
 \,  {p_b.p_a\over 
\{ (p_b.p_a)^2 -m_a^2 m_b^2\}^{3/2}}\, (p_b^\rho p_a^\mu - p_b^\mu p_a^\rho) 
\left\{2(p_b.p_a)^2  - 3 m_a^2 m_b^2 \right\} 
\, ,\nonumber \\
\een
and
\ben\label{egr3}
S_{\rm gr} &=& \sum_a {\ve_{\mu\nu} p_a^\mu p_a^\nu \over p_a.k} +  {i\over 4\pi}
\, \left(\ln\omega^{-1}+ 
\ln R^{-1}\right)\,  \, \sum_{b\atop \eta_b=-1} k.p_b \,  \sum_a {\ve_{\mu\nu} p_a^\mu p_a^\nu \over p_a.k}  
\nonumber \\ &&
- i \, \ln \omega^{-1}
\sum_a {\ve_{\mu\nu} \, p_a^\nu\, k_\rho \over p_a.k}
 \sum_{b\ne a\atop \eta_a\eta_b=1} \, {q_a q_b\over 4\pi} \,
{m_a^2 m_b^2 \, \{p_{b}^{\rho} p_a^\mu - p_b^\mu p_a^\rho\} \over 
\{ (p_b.p_a)^2 -m_a^2 m_b^2\}^{3/2}} \nonumber \\ && \hskip -.3in
 + {i\over 8\pi}  
\, \ln \omega^{-1}
\sum_a {\ve_{\mu\nu} p_a^\nu k_\rho 
 \over p_a.k}
 \sum_{b\ne a\atop \eta_a\eta_b=1} 
 \,  { p_b.p_a\over 
\{ (p_b.p_a)^2 -m_a^2 m_b^2\}^{3/2}}\, (p_b^\rho p_a^\mu - p_b^\mu p_a^\rho) 
\left\{2(p_b.p_a)^2  - 3 m_a^2 m_b^2\right\} 
\, .\nonumber \\
\een
These reproduce \refb{e1.16int} and \refb{e1.17int} respectively.

Note that the soft factors given in \refb{eem3} and \refb{egr3} depend only 
on the charges and momenta
carried by the external states. Therefore these can be reinterpreted as 
multiplicative soft factors in the full quantum
theory -- since there is no angular momentum there is no derivative with respect to the external momenta.
In the next few sections we shall carry out some explicit quantum computations to
examine to what extent this holds.

\sectiono{How to treat momentum conservation and infrared divergences} \label{smom}

In quantum theory, 
single soft theorem is expected to relate an amplitude $\Gamma^{(n,1)}$
with $n$ finite energy external states carrying momenta 
$p_1,\cdots p_n$ and one soft particle of momentum $k$ to an amplitude 
$\Gamma^{(n)}$ with just 
$n$ finite energy external states carrying momenta 
$p_1,\cdots p_n$. 
This relation takes the form
\be \label{eqsoft}
\Gamma^{(n,1)}(p_1,\cdots p_n, k) \simeq S(\ve, k; \{p_a\}) \, \Gamma^{(n)}(p_1,\cdots p_n)\, ,
\ee
where $S(\ve, k; \{p_a\})$ is the soft factor $S_{\rm em}$ or $S_{\rm gr}$.
There is however a potential problem. While the amplitude $\Gamma^{(n,1)}$ has
momentum conservation $\sum_a p_a+k=0$, the amplitude $\Gamma^{(n)}$
has momentum conservation
$\sum_a p_a=0$. Therefore we cannot keep the $p_a$'s  and 
$k$ as independent
variables in \refb{eqsoft}. 
Usually this problem is
overcome by including the momentum conserving delta-functions 
in the definition of the amplitudes $\Gamma^{(n,1)}$ and $\Gamma^{(n)}$ and treating \refb{eqsoft} as a relation between 
distributions. 
The soft factor $S(\ve, k; \{p_a\})$ appearing in \refb{eqsoft} is treated as a differential operator that also acts 
on the delta function and generates the Taylor series expansion of $\delta\left(\sum_a p_a + k\right)$ 
in power series
of the momentum $k$ of the soft particle. The subleading term in this expansion, given by
$k^\mu \{\p/\p p_b^\mu\}\delta\left(\sum_a p_a\right)$ for any $b$, is included in
the full subleading soft theorem in dimensions $D>4$. However since in $D=4$ we 
only analyze subleading terms containing $\ln\omega^{-1}$ factors, the term proportional to derivative of
the delta function will not appear in  our analysis.

In four space-time dimensions there are additional issues due to infrared divergence. Both the amplitudes
$\Gamma^{(n,1)}$ and $\Gamma^{(n)}$ have infrared divergences which can be represented as overall multiplicative
factors multiplying infrared finite amplitudes. For electromagnetic interactions these factors are common and
can be factored out of the amplitudes but for gravity there is a residual infrared divergent factor in $\Gamma^{(n,1)}$
besides the ones that appear in $\Gamma^{(n)}$. In any case we shall denote by $\exp[K]$ the infrared divergent 
factor of $\Gamma^{(n)}$ and define regulated amplitudes via the relation:
\be \label{e4.2rep}
\Gamma^{(n)}=\exp[K] \, \Gamma^{(n)}_{\rm reg}, \quad \Gamma^{(n,1)}=\exp[K] \, \Gamma^{(n,1)}_{\rm reg}\, .
\ee
$K$ is in general a function of the momenta $p_a$ of the finite energy particles. 
This makes $\Gamma^{(n)}_{\rm reg}$ free from infrared divergences, but $\Gamma^{(n,1)}_{\rm reg}$ still
contains some residual infrared divergences for gravitational interaction.
Eq.\refb{eqsoft} is now replaced by\footnote{The 
situation
here is somewhat different from the one in \cite{1405.1015}. 
Since the logarithmic term in $S(\ve, k; \{p_a\})$ that we are after is being represented as 
a multiplicative
factor instead of a differential operator, the infrared divergent 
factor on the right hand side can be moved 
past $S$ to the extreme left.}  
\be
\label{eqsoftnew}
\Gamma^{(n,1)}_{\rm reg}(p_1,\cdots p_n, k) \simeq S(\ve, k; \{p_a\}) \, \Gamma^{(n)}_{\rm reg}(p_1,\cdots p_n)\, .
\ee
The residual infrared divergences in $\Gamma^{(n,1)}_{\rm reg}$ will be reflected in the infrared divergent
contributions to $S(\ve, k; \{p_a\})$.

There is however a potential ambiguity in the definition of $\Gamma^{(n,1)}_{\rm reg}$ and hence of
$S(\ve, k; \{p_a\})$. 
This is due to the fact that
in the definition of $K$ we can add a term of the form $Q.\sum_{a} p_a$ for any vector $Q$ (which could be a
function of the $p_a$'s) since by the momentum conserving delta function in $\Gamma^{(n)}$, $\sum_a p_a$ vanishes.
However addition of such a term changes the definition of $\Gamma^{(n,1)}_{\rm reg}$ in \refb{e4.2rep}
by a multiplicative factor of
$\exp[k.Q]$ since the momentum conserving delta function in $\Gamma^{(n,1)}$ gives $k+\sum_a p_a=0$. 
This has the effect of multiplying $S(\ve, k; \{p_a\})$ by $\exp[k.Q]$.
Expanding
$\exp(k.Q)$ as $(1+k.Q)$ we see that the the additional contribution appears at the  subleading order, and
has the form of $k.Q$ multiplying the leading soft factor.
It does not affect the $\ln\omega^{-1}$ terms that we are after since the leading soft factor has
no $\ln\omega^{-1}$ term and $Q$ is $\omega$ independent. However this can affect
the genuine infrared divergent terms proportional to $\ln R$ in the expression for $\Gamma^{(n,1)}_{\rm reg}$, 
since in the definition of $Q$ we can include
terms proportional to $\ln R$. Choosing $Q=-U \, \ln R$ for some vector $U$ constructed from the $p_a$'s
amounts to having an additive contribution to $S^{(1)}$ of the form
\be \label{eambig}
- \ln R\, k.U \, S^{(0)}(\ve, k; \{p_a\})\, .
\ee

\sectiono{Soft photon theorem in scalar QED} \label{s3}

\begin{figure}
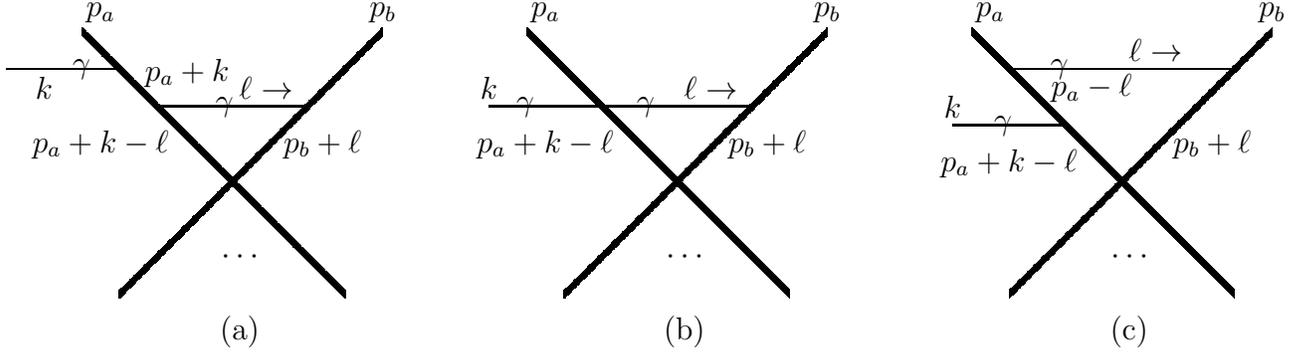


\begin{center}

\hbox{\figqeda \figqedb \figqedc}

\end{center}

\caption{One loop contribution to $\Gamma^{(n,1)}$ involving internal photon line
connecting two different legs. The thick lines represent scalar particles and \asnotex{the thin
lines carrying the symbol $\gamma$ represent photons.} There are other diagrams related to this
by permutations of the external scalar particles.
\label{figqed}}

\end{figure}

\begin{figure}
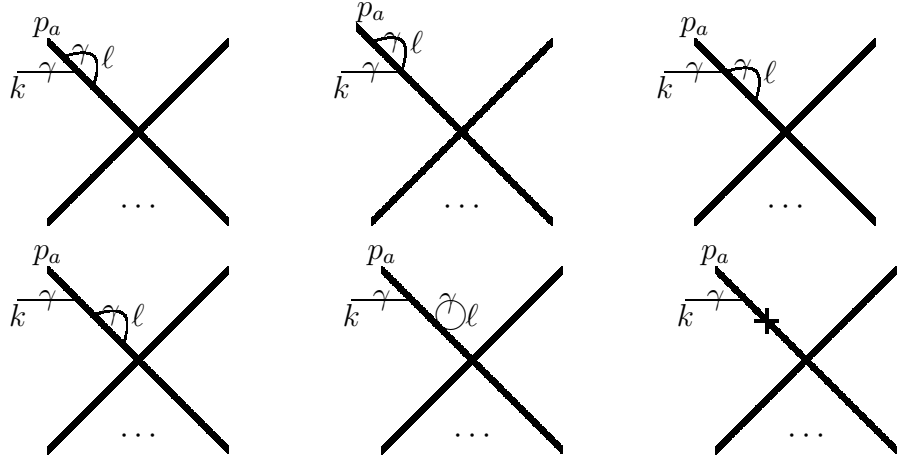


\begin{center}

\hbox{\hskip 1in \figselfa \qquad \figselfb \qquad \figselfc} 
\hbox{\hskip 1in \figselfd \qquad \figselfe \qquad \figselff}

\end{center}

\caption{One loop contribution to $\Gamma^{(n,1)}$ involving internal photon line
connecting two different points on the same leg. There are other diagrams related to this
by permutations of the external scalar particles. In the  last term the + on the scalar line
represents a counterterm associated with mass renormalization that has to be adjusted
to cancel the net contribution proportional to $1/(p_a.k)^2$.
\label{figself}}

\end{figure}

\begin{figure}
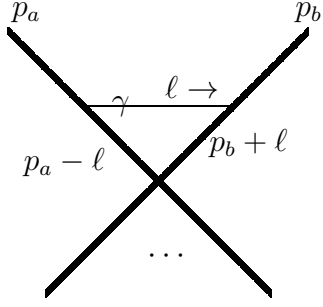


\begin{center}

\figqedtwo

\end{center}

\caption{One loop contribution to $\Gamma^{(n)}$. There are other diagrams related to this
by permutations of the external scalar particles.
\label{figqedtwo}}

\end{figure}

Consider a theory containing a U(1) gauge field $A_\mu$ and
$n$ scalars $\phi_1,\cdots \phi_n$ of  masses
$m_1,\cdots m_n$ and carrying U(1) charges $q_1,\cdots q_n$, satisfying
$\sum_{a=1}^n q_a=0$. We further assume that there is a non-derivative contact 
interaction between the $n$-scalars. Then the relevant part of the action takes
the form
\ben
&& \int d^4x \Bigg[ -{1\over 4} F_{\mu\nu} F^{\mu\nu} 
-\sum_{a=1}^n \left\{ (\p_\mu \phi_a^* + i q_a A_\mu \phi_a^*) 
(\p^\mu \phi_a - i q_a A^\mu \phi_a) + m_a^2 \phi_a^*\phi_a\right\}  \nonumber \\
&& \hskip 1in
+ \lambda \, \phi_1\cdots \phi_n + \lambda 
\, \phi_1^*\cdots
\phi_n^*
\Bigg]\, .
\een
We consider in this theory an amplitude with one external outgoing photon of momentum $k$
and $n$ external states corresponding to the fields $\phi_1,\cdots \phi_n$, carrying momenta
$p_1,\cdots p_n$. All momenta are counted as positive if ingoing so that if the $a$-th particle
is outgoing it will have negative $p_a^0$. Our goal will be to analyze this amplitude
at one loop order, involving an internal photon connecting two matter lines. The relevant 
diagrams have been shown in Figs.~\ref{figqed} and \ref{figself}. 
We denote by $\Gamma^{(n,1)}$ the sum over
tree and one loop contribution to this amplitude. $\Gamma^{(n)}$ will denote the amplitude
without the external soft photon to one loop order. One loop contribution to $\Gamma^{(n)}$ has
been shown in Fig.~\ref{figqedtwo}. 

In our analysis we shall ignore graphs with self energy insertions on external legs and assume
that we follow on-shell renormalization so that the mass parameters appearing in the tree
level propagators are the physical masses. The wave-function renormalization of the external
scalars cancel between $\Gamma^{(n)}$ and $\Gamma^{(n,1)}$.

We shall use Feynman gauge and decompose  the photon propagator of momentum $\ell$,
connecting the leg $a$ to the leg $b$ for $b\ne a$, with $\ell$ flowing from the $a$-th
leg to the $b$-th leg, as\cite{grammer}
\be
-i {\eta^{\mu\nu}\over \ell^2 - i\eps} = -{i\over \ell^2-i\eps} 
\left\{ K_{(ab)}^{\mu\nu} + G_{(ab)}^{\mu\nu}\right\}
\ee
where,
\be \label{eGYqed}
K_{(ab)}^{\mu\nu} = \ell^\mu \ell^\nu {(2p_a-\ell).(2p_b+\ell) \over (2p_a.\ell -\ell^2+i\eps) 
(2p_b.\ell+\ell^2-i\eps)} ,
\quad G_{(ab)}^{\mu\nu} = \eta^{\mu\nu} - K_{(ab)}^{\mu\nu}\, .
\ee
Note that $p_a$ and $p_b$ refer to the external momenta flowing into the legs $a$ and $b$, and not
necessarily
the momenta of the lines to which the photon propagator attaches (which may have additional contribution
from external soft momentum, {\it e.g.} in Figs.~\ref{figqed}(a)). 
$\ell$ denotes the momentum flowing from leg $a$ to leg $b$.
For $a=b$ we do not carry out any decomposition.

\begin{figure}
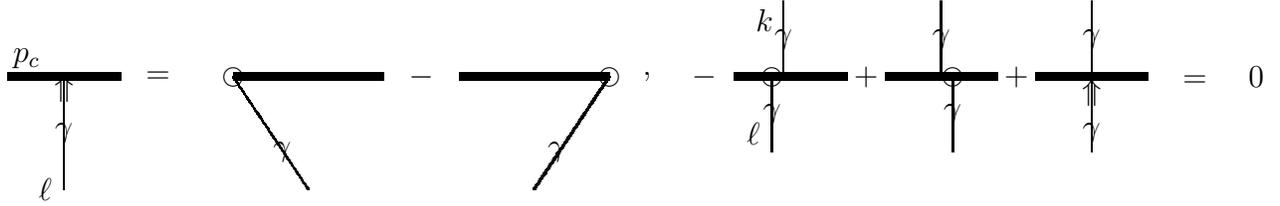


\begin{center}

\hbox{ \hskip .5in 
\figmiddle   
\figKgaugenew
}

\end{center}

\vskip -.5in

\caption{Diagrammatic representations of \refb{eward1} and \refb{eward2}. 
The arrow on the photon line represents that the polarization of the photon is taken
to be equal to the momentum entering the vertex.
The circle
denotes a simple vertex $-q_c$ with the polarization of the incoming photon 
stripped off.
\label{figward}}

\end{figure}

Since the K-photon polarization is proportional to $\ell^\mu \ell^\nu$, it is pure gauge.
This allows us to sum over K-photon insertions using Ward identities
\be \label{eward1}
{-i\over p_c^2+m_c^2}\, \ell^\mu \, i \, q_c 
\, (2p_{c\mu} +\ell_\mu) \, {-i\over (p_c+\ell)^2 + m_c^2}
= -q_c \left[{-i\over (p_c+\ell)^2 + m_c^2} - {-i\over p_c^2+m_c^2}\right]\, ,
\ee
and
\be \label{eward2}
q_c \left[ i\, q_c \, \ve.(2p_c+2\ell+k) - i\, q_c \, \ve.(2p_c+k)\right]-2\, i\, q_c^2 \, \ve.\ell=0\, ,
\ee
whose diagrammatic
representations have been shown in Fig.~\ref{figward}.
Sum over all insertions of the K-photons to either $\Gamma^{(n)}$ or $\Gamma^{(n,1)}$ produces
an exponential factor\cite{grammer}
\be \label{e4.5nn}
\exp\left[ i \, \sum_{a<b}
q_a \, q_b \int {d^4\ell\over (2\pi)^4} \, {1\over \ell^2-i\eps} {(2p_a-\ell).(2p_b+\ell) \over (2p_a.\ell -\ell^2
+i\eps) (2p_b.\ell+\ell^2-i\eps)}\right]\, .
\ee
Therefore we may write 
\ben \label{e3.6G}
\Gamma^{(n)} &=& \exp\left[ K_{\rm em} \right]  \left\{ \Gamma^{(n)}_{\rm tree}+ \Gamma^{(n)}_{\rm G} \right\}, 
\qquad \Gamma^{(n,1)} 
 = \exp\left[ K_{\rm em}
\right]  \left\{ \Gamma^{(n,1)}_{\rm tree}+ \Gamma^{(n,1)}_{\rm G} + \Gamma^{(n,1)}_{\rm self}\right\}\, , \non\\ 
K_{\rm em} &\equiv& {i\over 2} \, \sum_{a,b\atop b\ne a} q_a \, q_b \int {d^4\ell\over (2\pi)^4} \, {1\over \ell^2-i\epsilon}
 {(2p_a-\ell).(2p_b+\ell) \over (2p_a.\ell -\ell^2+i\epsilon) (2p_b.\ell+\ell^2-i\epsilon)}\, ,
\een
where 
$\Gamma^{(n)}_{\rm G}$ and $\Gamma^{(n,1)}_{\rm G}$ are computed by replacing the  internal 
photons by the
G-photons in Figs.~\ref{figqedtwo} and \ref{figqed} respectively and
$\Gamma^{(n,1)}_{\rm self}$ 
denotes the sum of diagrams in Fig.~\ref{figself} for which we use the full
photon propagator. 
Therefore a relation of the form
$\Gamma^{(n,1)} = S_{\rm em} \Gamma^{(n)}$ takes the form
\be\label{e3.7}
\Gamma^{(n,1)}_{\rm tree}+ \Gamma^{(n,1)}_{\rm G} + \Gamma^{(n,1)}_{\rm self}
= S_{\rm em} \left\{ \Gamma^{(n)}_{\rm tree}+ \Gamma^{(n)}_{\rm G} \right\}\, .
\ee

Now it is easy to see that Fig.~\ref{figqedtwo} vanishes when we replace the internal photon by
G-photon. Therefore $\Gamma^{(n)}_{\rm G}=0$, and we have:\footnote{Note that we are not explicitly 
writing the momentum conserving delta function, but are implicitly assuming that both sides of
\refb{e3.7} are
multiplied by the appropriate delta functions. We also implicitly assume that 
the delta function $\delta(\sum_a p_a+k)$ on the left hand side has been expanded in a power series in 
$k$.}
\be \label{e3.7aa}
\Gamma^{(n)}_{\rm tree}+ \Gamma^{(n)}_{\rm G}=\Gamma^{(n)}_{\rm tree} = i\, \lambda\, .
\ee
If we write $S_{\rm em}
= S^{(0)}_{\rm em}+S^{(1)}_{\rm em}$ where $S^{(0)}_{\rm em}$ is the leading soft factor
$\sum_{a=1}^n q_a \, {\ve .p_a / k.p_a}$ and $S^{(1)}_{\rm em}$ is the subleading multiplicative factor
containing logarithmic terms, then eq.\refb{e3.7} can be written as
\be \label{eGnG}
\Gamma^{(n,1)}_{\rm tree}+ \Gamma^{(n,1)}_{\rm G} + \Gamma^{(n,1)}_{\rm self}
=  i\lambda\, \sum_{a=1}^n q_a {\ve .p_a\over k.p_a} + i\lambda\, S^{(1)}_{\rm em} \, ,
\ee
to one loop order. Now $\Gamma^{(n,1)}_{\rm tree}$ is equal to the
first term on the right hand side up to terms involving Taylor series expansion of the momentum
conserving delta function in powers of $k$, but the latter are subleading contributions
without any logarithmic terms and can be
ignored in our analysis.
Therefore \refb{eGnG}
can be rewritten as:
\be\label{e3.11}
 \Gamma^{(n,1)}_{\rm self} + \Gamma^{(n,1)}_{\rm G}= i\lambda \, S^{(1)}_{\rm em} \, .
\ee
This is a simple algorithm for determination of $S^{(1)}_{\rm em}$.

Therefore we need to focus on the evaluation of the one loop contribution to
$\Gamma^{(n,1)}_{\rm G}$ and $\Gamma^{(n,1)}_{\rm self}$ by summing the diagrams in
Figs.~\ref{figqed} and \ref{figself}, 
with the internal photon replaced by G-photon in Fig.~\ref{figqed}. 
We first consider the diagrams in Fig.~\ref{figqed}.
It is easy to see that the G-photon
contribution to Fig.~\ref{figqed}(c) vanishes. 
Therefore we need to focus on Figs,~\ref{figqed}(a) and (b).
The contribution from Fig.~\ref{figqed}(a) is given by
\ben
\II_1 &=& \lambda \, q_a^2 \, q_b \, {\eps.p_a\over k.p_a} \,  \int {d^4\ell\over (2\pi)^4}\,
\left[ 2k. (2p_b+\ell) -
{2k.\ell \,  (2p_a-\ell).(2p_b+\ell)
\over (2 p_a.\ell-\ell^2+i\eps) } \right] \nonumber \\ &&\hskip 1in \times 
{1\over \ell^2-i\eps} \, {1\over 2p_a.(k-\ell) + (k-\ell)^2-i\eps} \, {1\over 2p_b.\ell+\ell^2-i\eps} \, ,
\een
and the contribution from Fig.~\ref{figqed}(b) is given by
\ben 
\II_2 &=& - \lambda \, q_a^2 \, q_b \,  \int {d^4\ell\over (2\pi)^4}\,
\left[2\eps . (2p_b+\ell) -
{2\eps.\ell \,  (2p_a-\ell).(2p_b+\ell)
\over (2 p_a.\ell-\ell^2+i\eps) } \right] 
\nonumber \\ &&\hskip 1in \times 
{1\over \ell^2-i\eps} \, {1\over 2p_a.(k-\ell) + (k-\ell)^2-i\eps} \, {1\over 2p_b.\ell+\ell^2-i\eps} \, .
\een
Both $\II_1$ and $\II_2$ are infrared finite since for small $\ell$ the integrands diverge as
$1/\ell^3$. The terms involving logarithm of $k$ come from the region of $\ell$ integration 
where the components $|\ell^\mu|$ are large compared to $\omega\equiv k_0$ but small compared to the 
$p_a$'s. In this range we can approximate $\II_1$ and $\II_2$ as
\ben\label{e3.16}
\II_{1} &\simeq& -\lambda \, q_a^2 \, q_b \, {\eps.p_a\over k.p_a} \,  
\int_{\rm reg} {d^4\ell\over (2\pi)^4}\,
\left[ k.p_b -
{k.\ell \, p_a.p_b
\over p_a.\ell +i\eps}\right] \, 
{1\over \ell^2-i\eps} \, {1\over p_a.\ell+i\eps} \, {1\over p_b.\ell-i\eps} \nonumber \\ 
&=&\ -\lambda\ q_{a}^{2}q_{b}\ \f{\epsilon.p_{a}}{k.p_{a}}\ 
\Big[\ k.p_{b}+p_{a}.p_{b}\ k^{\mu}\f{\p}{\p p_{a}^{\mu}}\Big]\ 
\int_{\rm reg}\ \f{d^{4}\ell}{(2\pi)^{4}}\ \f{1}{\ell^{2}-i\epsilon}\ \f{1}{p_{a}.\ell+i\epsilon}\ 
\f{1}{p_{b}.\ell-i\epsilon}\, ,\non\\
\een
and
\ben\label{e3.17pre}
\II_{2} &\simeq& \lambda \, q_a \, q_a q_b  \,  \int_{\rm reg} {d^4\ell\over (2\pi)^4}\,
\left[ \eps.p_b -
{\eps.\ell \, p_a.p_b
\over p_a.\ell+i\eps }\right] \, 
{1\over \ell^2-i\eps} \, {1\over p_a.\ell+i\eps} \, {1\over p_b.\ell-i\eps} \nonumber \\ &=& \ \lambda\ q_{a}^{2}q_{b}\ \Big[\epsilon. p_{b}\ 
+ \ p_{a}.p_{b}\ \epsilon^{\mu}\f{\p}{\p p_{a}^{\mu}}\Big]\ \int_{\rm reg}\ 
\f{d^{4}\ell}{(2\pi)^{4}}\ \f{1}{\ell^{2}-i\epsilon}\ \f{1}{p_{a}.\ell+i\epsilon}\ 
\f{1}{p_{b}.\ell
-i\epsilon}\, , 
\een
where the subscript reg indicates that the integration needs to be carried out over the
region where $|\ell^\mu|$ is large compared to $\omega$ but small compared to the
energies of the finite energy particles.
Adding $\II_1$ and $\II_2$  and summing over $a,b$ we get 
the total contribution to $\Gamma^{(n,1)}_{\rm G}$ to one loop order:
\ben \label{e3.19Ga}
\Gamma^{(n,1)}_{G} &=& -\lambda\, \sum_{a,b\atop b\ne a} (q_a)^2 q_b \left[ \f{\epsilon.p_{a}}{k.p_{a}}\ 
\ k.p_{b}+\f{\epsilon.p_{a}}{k.p_{a}}\ p_{a}.p_{b}\ k^{\mu}\f{\p}{\p p_{a}^{\mu}} - 
\epsilon. p_{b}\ 
- \ p_{a}.p_{b}\ \epsilon^{\mu}\f{\p}{\p p_{a}^{\mu}}\right] \nonumber \\ &&
\hskip 1in \int_{\rm reg}\ 
\f{d^{4}\ell}{(2\pi)^{4}}\ \f{1}{\ell^{2}-i\epsilon}\ \f{1}{p_{a}.\ell+i\epsilon}\ 
\f{1}{p_{b}.\ell
-i\epsilon} \nonumber \\
&=& -\lambda\sum_{a, b\atop b\ne a} (q_a)^2 \, 
q_b{\ve_\mu k_\nu \over p_a.k} \left\{p_a^\mu {\p\over \p p_{a\nu}} - p_a^\nu {\p\over \p p_{a\mu}}
\right\}
 \int_{\rm reg} {d^4\ell\over (2\pi)^4} \, 
 {1\over \ell^2-i\eps} {p_a.p_b\over (p_a.\ell+i\eps) \, (p_b.\ell-i\eps)}\, .
 \nonumber \\
 \een
 
 \begin{figure}
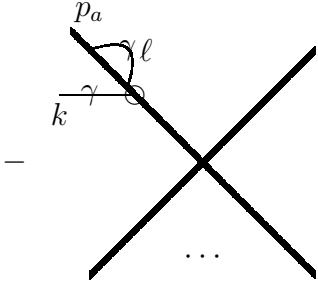


\begin{center}

\figselfsum

\end{center}

\vskip -.2in

\caption{Sum of the first four diagrams in Fig.~\ref{figself} with $\ve$ replaced by $k$.
\label{figres}}

\end{figure}
 
 The contribution to $\Gamma^{(n,1)}_{\rm self}$
 from Fig.~\ref{figself} can be analyzed using the following indirect method.
 First of all we note that the  net dependence on $\ve$ and $k$ from the first four diagrams must
 be of the form $\ve.p_a \, f(p_a.k)$ for some function $f$. To determine
 $f$, we can set $\ve=k$ and sum over all insertions of the external photon using the
 Ward identities shown in Fig.~\ref{figward}. The final result, given in Fig.~\ref{figres}, has
 the form:
 \be 
{C_1\over p_a.k}\, ,
 \ee
for some constant $C_1$. Therefore we get
\be
p_a.k \, f(p_a.k) = {C_1\over p_a.k} \quad \Rightarrow \quad f(p_a.k)={C_1\over (p_a.k)^2}\, .
\ee
The  fifth and sixth diagrams  also have the form 
\be 
{C_2\over (p_a.k)^2} \quad \hbox{and} \quad {C_3\over (p_a.k)^2}\, ,
\ee
 for appropriate constants $C_2$ and $C_3$. Now since we are using on-shell 
 renormalization the counterterm proportional to $C_3$ is to be adjusted precisely so that the net
 contribution proportional to $1/(p_a.k)^2$ vanishes. Therefore we must choose $C_3=-C_1-C_2$, and
 the total contribution to $\Gamma^{(n,1)}_{\rm self}$
 from all the diagrams in Fig.~\ref{figself} vanishes. We have verified this by
 explicitly computing the Feynman diagrams in Fig.~\ref{figself}.
 
From \refb{e3.11} we now see that the net contribution to the logarithmic terms in $S^{(1)}_{\rm em}$
is obtained by dividing $\Gamma^{(n,1)}_{\rm G}$ given in \refb{e3.19Ga} by $i\,\lambda$. 
This can be written as
\ben \label{e3.19G}
S^{(1)}_{\rm em}
&=& \sum_{c} q_c {\ve_\mu k_\nu \over p_c.k} \left\{p_c^\mu {\p\over \p p_{c\nu}} - 
 p_c^\nu {\p\over \p p_{c\mu}}\right\}
 \, K_{\rm em}^{\rm reg}\, ,
\een
where $K^{\rm reg}_{\rm em}$ is the factor $K_{\rm em}$ defined in \refb{e3.6G} with the understanding
that integration over the loop momentum $\ell$ will run over the range where $|\ell^\mu|$ is  larger
than $\omega$ but small compared to the momenta of the finite energy external states: 
\be \label{e3.6GKK}
K_{\rm em}^{\rm reg} \equiv {i\over 2} \, \sum_{a,b\atop b\ne a} q_a \, q_b \int_{\rm reg} 
{d^4\ell\over (2\pi)^4} \, {1\over \ell^2-i\epsilon}
 {(2p_a-\ell).(2p_b+\ell) \over (2p_a.\ell -\ell^2+i\epsilon) (2p_b.\ell+\ell^2-i\epsilon)}\, .
 \ee

So essentially we need to evaluate $K_{\rm em}^{\rm reg}$. 
For this we need to evaluate the  integral:\footnote{Since the $\ell^\mu$ integration runs over a 
limited range, one might wonder why we are choosing the $\ell^0$ integration range from $-\infty$
to $\infty$.
To this end, note that once we have imposed the range restriction on $|\vec\ell|$, we can let the $\ell^0$
integral in \refb{e3.17}
run over the entire real axis since the regions outside the allowed range do not generate any 
logarithmic contribution.}
\ben\label{e3.17}
\mathcal{I}_{ab}\ &\equiv&\ \int_{\rm reg}\ 
\f{d^{4}\ell}{(2\pi)^{4}}\ \f{1}{\ell^{2}-i\epsilon}\ \f{1}{p_{a}.\ell+i\epsilon}\ 
\f{1}{p_{b}.\ell
-i\epsilon} \\
&=&\ -\f{1}{E_{a}E_{b}}\ \int_{\rm reg}  \f{d^{3}{\ell}}{(2\pi)^{3}}\ 
\int_{-\infty}^{\infty} \f{d\ell^{0}}{2\pi}\ \f{1}{(\ell^{0}-|\vec{\ell}|
+i\epsilon)(\ell^{0}+|\vec{\ell}|-i\epsilon)}\ \f{1}{\ell^{0}-\vec{v}_{a}.\vec{\ell}
-i\epsilon}\ \f{1}{\ell^{0}-\vec{v}_{b}.\vec{\ell}+i\epsilon}\, , \non
\een
where $E_a=p_a^0$, $E_b=p_b^0$, 
$\vec{v}_{a}=\vec{p}_{a}/E_{a}$ and $\vec{v}_{b}=\vec{p}_{b}/ {E_{b}}$. 
In writing down the above equation we have assumed that $E_a$ and $E_b$ are
positive, i.e. both lines represent incoming states.
The integrand  has simple poles at,
\be
\ell^{0}\ =\ (|\vec{\ell}|-i\epsilon)\ ,\ -(|\vec{\ell}|-i\epsilon)\ ,\ (\vec{v}_{a}.\vec{\ell}+i\epsilon)\ ,\ (\vec{v}_{b}.\vec{\ell}-i\epsilon)\, .
\ee
So now if we close the contour in the lower half plane
 we have to take the 
pole contributions from $\ell^{0}\ =\ (|\vec{\ell}|-i\epsilon)$ and $\ell^{0}=(\vec{v}_{b}.\vec{\ell}
-i\epsilon)$. This gives
\ben\label{eIexp}
\mathcal{I}_{ab}\ &=&\ \f{i}{E_{a}E_{b}}\ \int_{\rm reg} \f{d^{3}\vec{\ell}}{(2\pi)^{3}}\ 
\f{1}{2|\vec{\ell}|}\ \f{1}{|\vec{\ell}|-\vec{v}_{a}.\vec{\ell}}\ 
\f{1}{|\vec{\ell}|-\vec{v}_{b}.\vec{\ell}} \non\\ &&\
+\ \f{i}{E_{a}E_{b}}\ \int_{\rm reg} \f{d^{3}\vec{\ell}}{(2\pi)^{3}}\  
\f{1}{(\vec{v}_{b}.\vec{\ell})^{2}-|\vec{\ell}|^{2}}\ \f{1}{(\vec{v}_{b}
-\vec{v}_{a}).\vec{\ell}-i\epsilon}\, .
\een
Note that we have removed the $i\epsilon$'s from the denominators that never vanish.

Let us first analyze the second term.
Since the result should be Lorentz invariant, it should not depend on the 
chosen frame. For simplicity choose a frame 
in which $\vec v_b$ and $\vec v_a$ are along the positive $z$-axis
with $|\vec v_b|>|\vec v_a|$. Denoting by $\theta$ the angle between
$\vec\ell$ and the $z$-axis, we can express the second term in \refb{eIexp} as
\ben
\II'_{ab}\ &=&\ \f{i}{E_{a}E_{b}(2\pi)^{2}}\ \f{1}{|\vec{v}_{a}-\vec{v}_{b}|} \int_{\rm reg}
\f{d|\vec{\ell}|}{|\vec{\ell}|}\ \int_{-1}^{1}\ d(cos\theta)\ \f{1}{|\vec{v}_{b}|^{2}cos^{2}\theta-1}
\ \f{1}{cos\theta-i\epsilon}\, .
\een
Without the $i\epsilon$ piece of the last term the integral vanishes since 
the integrand is an odd function of $\cos\theta$. 
However the imaginary part of the last term makes the integral non-vanishing. 
Using ${1}/{(x-i\epsilon)}=i\pi\delta(x) + P(1/x)$ in the integral, and using the fact that
the value of $|\vec\ell|$ for which our approximation of the integrand is valid
ranges from $\omega$ to some finite energy, we get,
\be\label{eipab}
\II'_{ab}\ \simeq\ \f{1}{4\pi \, E_{a}E_{b}}\ \ln \omega^{-1}\ \f{1}{|\vec{v}_{a}-\vec{v}_{b}|}
=\ \f{1}{4\pi}\ \ln \omega^{-1}\ \f{1}{\sqrt{(p_{a}.p_{b})^{2}-m_{a}^{2}m_{b}^{2}}}\, ,
\ee
where in the intermediate stage we used $|\vec{p}_{a}||\vec{p}_{b}|=|\vec{p}_{a}.\vec{p}_{b}|$,
since $\vec p_a$ and $\vec p_b$ are parallel.

If both the legs $a$ and $b$ are outgoing instead of ingoing, then $E_a$ and $E_b$ are negative and
the signs of the $i\epsilon$ in the last two terms in \refb{e3.17} are reversed. But this can be brought
back to the form given in \refb{e3.17} by making a change of variables $\ell^\mu\to -\ell^\mu$. Therefore
the net result for the residue at $\ell^0=\vec v_b.\vec\ell -i\epsilon$ will continue to be described by
\refb{eipab}.
Finally if one of the momenta is outgoing and the other is ingoing, then  both the $i\epsilon$'s in the
last two terms of \refb{e3.17} come with the same sign. By changing variables from $\ell^\mu$ to
$-\ell^\mu$ if necessary, we can ensure that both the poles are in the upper half plane and close the
contour to the lower half plane. In this case there will be no analog of the contribution given in \refb{eipab}.

We now turn to the contribution from the first term on the right hand side
of \refb{eIexp}, which we will call $\mathcal{I^{\prime\prime}}_{ab}$. 
We will again evaluate this integral in the frame in which $\vec v_a$ and $\vec v_b$ are
parallel to the $z$-axis with $|\vec v_b|>|\vec v_a|$. We get
\ben \label{e3.24G}
\mathcal{I^{\prime\prime}}_{ab}\ &=&\ \f{i}{E_{a}E_{b}}\ \int_{\rm reg} 
\f{d^{3}\vec{l}}{(2\pi)^{3}}\ \f{1}{2|\vec{l}|}\ \f{1}{|\vec{l}|-\vec{v}_{a}.\vec{l}}\ 
\f{1}{|\vec{l}|-\vec{v}_{b}.\vec{l}}\non\\
&=&\ \f{i}{8\pi^{2}E_aE_b}\ \ln \omega^{-1}\  \int _{-1}^{1}\ d(cos\theta)\ {1\over v_b-v_a}
\left[{v_b\over 1-v_b \cos\theta} - {v_a\over 1-v_a\cos\theta}\right]\non\\
&=&\ \f{i}{8\pi^{2}}\ \ln \omega^{-1}\  {1\over |\vec p_b| E_a - |\vec p_a| E_b}
\ln \left[ {(E_a- |\vec p_a|) (E_b +|\vec p_b|)\over 
(E_a+ |\vec p_a|) (E_b -|\vec p_b|)}\right]
\non\\
&=&\ -\f{i}{8\pi^{2}}\ \ln \omega^{-1}\  {1\over \sqrt{(p_a.p_b)^2 - p_a^2 p_b^2}}
\ln\left[{p_a.p_b +  \sqrt{(p_a.p_b)^2 - p_a^2 p_b^2}\over
p_a.p_b -  \sqrt{(p_a.p_b)^2 - p_a^2 p_b^2}}
\right]\, . 
\een
It is easy to check that the form of the contribution remains unchanged even when both
legs are outgoing or one leg is incoming and the other leg is outgoing.

Combining these results we get
\be \label{edefkemreg}
K_{\rm em}^{\rm reg} =
{i\over 2} \sum_{a, b\atop b\ne a} q_a\, q_b \, 
\f{1}{4\pi}\ \ln \omega^{-1}\ \f{p_a.p_b}{\sqrt{(p_{a}.p_{b})^{2}-p_{a}^{2}p_{b}^{2}}}
\left\{\delta_{\eta_a\eta_b,1} - {i\over 2\pi}\ln\left({p_a.p_b +  \sqrt{(p_a.p_b)^2 - p_a^2 p_b^2}\over
p_a.p_b -  \sqrt{(p_a.p_b)^2 - p_a^2 p_b^2}}
\right)
\right\} \, .
\ee
Using \refb{e3.19G} we can now write down the expression for the logarithmic term in the subleading 
soft factor $S^{(1)}_{\rm em}$
\ben \label{e3.23G}
&&-{i \over 4\pi}\ \ln \omega^{-1}\sum_{a=1}^n\sum_{b\ne a\atop \eta_a\eta_b=1} \ q_{a}^2\, q_{b}
\ \f{\epsilon_{\mu}\, k_{\rho}}{p_{a}.k}\ \f{m_{a}^{2}
m_{b}^{2}\big[p_{a}^{\mu}p_{b}^{\rho}-p_{b}^{\mu}p_{a}^{\rho}\big]}
{\big[(p_{a}.p_{b})^{2}-m_{a}^{2}m_{b}^{2}\big]^{3/2}}\nonumber \\
&& -\f{1}{8\pi^{2}}\ \ln \omega^{-1}\ \sum_{a,b\atop b\ne a} q_a^2 \, q_b\,
\ln\left[{p_a.p_b +  \sqrt{(p_a.p_b)^2 - p_a^2 p_b^2}\over
p_a.p_b -  \sqrt{(p_a.p_b)^2 - p_a^2 p_b^2}}
\right]  {p_a^2 p_b^2 \over \{(p_a.p_b)^2 - p_a^2 p_b^2\}^{3/2}}
\left\{ -\eps.p_b +{\eps.p_a\over k.p_a} k.p_b\right\} \non\\
&& + \f{1}{4\pi^{2}}\ \ln \omega^{-1}\    \sum_{a,b\atop b\ne a} q_{a}^{2}q_{b}\, 
{p_a.p_b\over (p_a.p_b)^2 - p_a^2 p_b^2} 
\left\{ -\eps.p_b +{\eps.p_a\over k.p_a} k.p_b\right\} \, .
\een
The term in the first line agrees with the classical expression for 
$S^{(1)}_{\rm em}$ given by the second term of \refb{e1.16}. The rest of the contribution is extra.

We have also checked that \refb{e3.23G} holds if instead of scalars we have interacting fermions. This
confirms that the logarithmic correction to the soft factor is independent of the spin of the particle.

We end this section by making some observation on the results derived above:
\begin{enumerate}
\item
Suppose we assume the validity of the naive version of the subleading soft photon 
theorem:\footnote{Since the presence of the logarithmic term makes the finite part 
ambiguous, we consider only the logarithmic terms in the subleading factor.}
\be
\Gamma^{(n,1)} = \{ S^{(0)}_{\rm em} + \wh S^{(1)}_{\rm em}\} \, \Gamma^{(n)}\, ,
\ee
where the `hat' on $S^{(1)}$ denotes that we are using the differential operator form that arises in
the quantum theory:
\be 
S^{(0)}_{\rm em} = \sum_a q_a \, {\ve.p_a\over p_a.k}, \qquad
\wh S^{(1)}_{\rm em} = \sum_{a} q_a\, {\ve_\mu k_\nu \over p_a.k} \left\{p_a^\mu {\p\over \p p_{a\nu}} - p_a^\nu {\p\over \p p_{a\mu}}
\right\} \, .
\ee
Then using \refb{e3.6G} and the fact that $\Gamma^{(n)}_{\rm G}$ vanishes at one loop order, we get
\be 
\Gamma^{(n,1)}_{\rm tree}+ \Gamma^{(n,1)}_{\rm self}
+\Gamma^{(n,1)}_{\rm G} = S^{(0)}_{\rm em} \Gamma^{(n)}_{\rm tree} + \{ \wh S^{(1)}_{\rm em} K_{\rm em} \}  
\, \Gamma^{(n)}_{\rm tree} + \wh S^{(1)}_{\rm em}  \Gamma^{(n)}_{\rm tree} \, .
\ee
Using $\Gamma^{(n)}_{\rm tree}=i\,\lambda$, using \refb{eGnG} to replace the left hand side,
and throwing away terms like $\wh S^{(1)}_{\rm em}  \Gamma^{(n)}_{\rm tree}$ 
which vanishes, we get
\be\label{e4.4}
S^{(1)}_{\rm em} = \wh S^{(1)}_{\rm em} \, K_{\rm em}
\, .
\ee
In the definition of $K_{\rm em}$ the integration over loop momentum runs over all range and we have an
infrared divergence from the region of small $\ell$. However if we make an ad hoc restriction that the
loop momentum integral will run in the range much larger than the energy $\omega$ of the external
soft photon, then $K_{\rm em}$ reduces to $K_{\rm em}^{\rm reg}$ defined 
in \refb{e3.6GKK} and we recover the
correct logarithmic terms in $S^{(1)}_{\rm em}$ as given in \refb{e3.19G}. 
This suggests an ad hoc rule for computing the logarithmic terms in the soft
expansion in quantum theory -- begin with the usual soft expansion and explicitly evaluate the action
of the differential operator on the amplitude, restricting the region of loop momentum integration to
lie in a range larger than the  soft momenta but smaller than the momenta of the finite energy particles.
With hindsight, this prescription can be justified by noting that the general arguments of 
\cite{1703.00024,1706.00759}, that
assumes existence of 1PI effective action with no powers of soft momenta coming from the vertices,
breaks down for the contribution where the loop momentum is smaller than the external soft momenta.
On the other hand we do not expect any large contribution from the region of integration where the
loop momentum is of the order of the external momenta or larger.

This argument also suggests that although we have carried out the 
explicit calculation only at one
loop order, the result may be valid to all orders in perturbation theory, 
since $K_{\rm em}$ is known to be
valid to all orders in perturbation theory\cite{grammer}.

\item The second observation concerns the relation between the classical and the quantum results.
As already noted, compared to the classical result that agrees with the first line of
\refb{e3.23G}, 
the quantum result found here has an extra term given in the second and third line of
 \refb{e3.23G}.
If however we replace in \refb{e3.17} the Feynman propagator for the photon by the retarded propagator, 
we get only the contribution from the first line of \refb{e3.23G}, 
since the contribution from the pole at $\ell^2=0$ can then be avoided
by appropriate choice of contour.  Therefore at least for the soft photon theorem in quantum 
electrodynamics, the rule for relating the quantum and the classical result seems to be to replace the
Feynman propagator of 
the photon in the loop  in the quantum result by retarded propagator.
\end{enumerate}
We shall now write down the results for the other cases and test if the generalization of
observation 1 works. We shall also explore if the results 
satisfy the generalization of observation 2.

\sectiono{Soft graviton theorem in gravitational scattering} \label{s5}

We now turn to the analysis of the soft graviton theorem in the scattering of scalar 
particles, interacting via gravity, to one loop order. 
The action is taken to be
\be \label{egraction}
\int d^4x \, \sqrt{-\det g} \, \Bigg[ {1\over 16\pi G} \, R 
- \sum_{a=1}^n \left\{ g^{\mu\nu}\, \p_\mu \phi_a^*\, 
\p_\nu \phi_a + m_a^2 \phi_a^* \phi_a\right\}  + \lambda \, \phi_1\cdots \phi_n
+ \lambda \, \phi_1^*\cdots \phi_n^* \Bigg]\, .
\ee
Even though in this case we could take the scalar fields to be real, we have kept them complex in order to
extend the analysis to the case where the scalars have both electromagnetic and gravitational interaction.
As in \S\ref{s3}, we shall postulate a relation of the form
\be\label{epost}
\Gamma^{(n,1)} = \left\{ S^{(0)}_{\rm gr} + S^{(1)}_{\rm gr} \right\} \Gamma^{(n)}\, ,
\ee
and try to determine the logarithmic terms in $S^{(1)}_{\rm gr}$ by comparing the two sides up to one loop order.

We shall carry out our computation in the de Donder gauge in which the propagator of a graviton of momentum $\ell$
is given by:
\be 
-\, {i\over \ell^2-i\eps} \, {1\over 2} \, \big(\eta^{\mu\rho}\eta^{\nu\sigma}+\eta^{\mu\sigma}\eta^{\nu\rho}-\eta^{\mu\nu}
\eta^{\rho\sigma}\big)\, .
\ee
For our analysis we also need the vertices involving the graviton. 
The scalar-scalar-graviton vertex, with the scalars carrying ingoing momenta 
$p_1$, $p_2$ and the graviton carrying ingoing momentum $-p_1-p_2$ and
Lorentz index $(\mu\nu)$, is given by
\be
-i\, \kappa\, \big[p_{1\mu}p_{2\nu}+p_{1\nu}p_{2\mu}-\eta_{\mu\nu}(p_{1}.p_{2}-m^{2})\big]\, ,
\ee
where $\kappa=\sqrt{8\pi G}=1$ in our convention. 
The vertex involving two scalars carrying ingoing momenta $p_1$, $p_2$, and two 
gravitons carrying ingoing momenta $k_1$, $k_2$ and Lorentz indices $(\alpha\beta)$ and
$(\mu\nu)$ is given by\footnote{In writing this and other vertices we already include the symmetry factor related
to exchange of identical particles. Therefore if we were to use this vertex to compute tree level two graviton, two
scalar amplitude, no further symmetry factor is necessary.}
\ben \label{equadver}
&& 2\, i\, \kappa^{2}\, 
 \Big[-\eta_{\alpha\mu}\ \eta_{\beta\nu}\ p_{1}.p_{2} \ + \ 
 \f{1}{2}\eta_{\alpha\beta}\, \eta_{\mu\nu}\, p_{1}.p_{2} \ - \ \eta_{\alpha\beta}\, 
 p_{1\mu}p_{2\nu}
\ - \ \eta_{\mu\nu} \, p_{1\alpha}p_{2\beta}\non\\
&&\hskip 1in + \ 2\, \eta_{\alpha\mu} \left\{ p_{1\beta}p_{2\nu}+p_{2\beta}p_{1\nu}\right\}
\  + \
m^2\ (\eta_{\mu\alpha}\eta_{\nu\beta}-\f{1}{2}\eta_{\mu\nu}\eta_{\alpha\beta})\Big]\, .\non\\
\een
If we label the ingoing graviton momenta by $k_1$, $k_2$ and $k_3=-k_1-k_2$ and the
Lorentz indices carried by them by $(\mu\alpha)$, $(\nu\beta)$ and $(\sigma\gamma)$ respectively, then the
3-graviton vertex takes the  form:
\ben\label{esoftvert}
 i\, \kappa\ &\Big[ & \big(k_{1}.k_{2}\eta_{\mu\alpha}\eta_{\nu\sigma}\eta_{\beta\gamma}+k_{2}.k_{1}
\eta_{\nu\beta}\eta_{\mu\sigma}\eta_{\alpha\gamma}+k_{1}.k_{3}\eta_{\mu\alpha}
\eta_{\nu\sigma}\eta_{\beta\gamma}\non\\
&+&\ k_{3}.k_{1}\eta_{\sigma\gamma}\eta_{\mu\nu}\eta_{\alpha\beta}+k_{2}.k_{3}\eta_{\nu\beta}
\eta_{\mu\sigma}\eta_{\alpha\gamma}+k_{3}.k_{2}\eta_{\sigma\gamma}\eta_{\mu\nu}\eta_{\alpha\beta}\big)\non\\
&-&\ 2\, \big(k_{1\sigma}k_{2\gamma}\eta_{\mu\nu}\eta_{\alpha\beta}+k_{2\mu}k_{3\alpha}\eta_{\nu\sigma}\eta_{\beta\gamma}+k_{3\nu}
k_{1\beta}\eta_{\mu\sigma}\eta_{\alpha\gamma}\big)\non\\
&-&\ 4\big(k_{1}.k_{2}+k_{2}.k_{3}+k_{3}.k_{1}\big)\eta_{\alpha\nu}\eta_{\beta\sigma}\eta_{\gamma\mu}\non\\
&+& \big(k_{1}.k_{2}\eta_{\mu\nu}\eta_{\alpha\beta}\eta_{\sigma\gamma}+k_{2}.k_{3}\eta_{\nu\sigma}\eta_{\beta\gamma}\eta_{\mu\alpha}
+k_{3}.k_{1}\eta_{\mu\sigma}\eta_{\alpha\gamma}\eta_{\nu\beta}\big)\non\\
&+&2\big(k_{1\sigma}k_{2\mu}\eta_{\alpha\nu}\eta_{\beta\gamma}+k_{2\mu}k_{3\nu}\eta_{\sigma\alpha}\eta_{\gamma\beta}+k_{3\nu}
k_{1\sigma}\eta_{\mu\beta}\eta_{\alpha\gamma}\non\\
&+&k_{2\sigma}k_{1\nu}\eta_{\mu\beta}\eta_{\alpha\gamma}+k_{3\mu}k_{2\sigma}\eta_{\nu\gamma}\eta_{\beta\alpha}+k_{1\nu}k_{3\mu}
\eta_{\sigma\beta}\eta_{\gamma\alpha}\big)\non\\
&-&\f{1}{2}\big(k_{1}.k_{2}+k_{2}.k_{3}+k_{3}.k_{1}\big)\eta_{\mu\alpha}\eta_{\nu\beta}\eta_{\sigma\gamma}\Big]\, .
\een
In \refb{equadver} and \refb{esoftvert} it is understood that
the vertices need to be symmetrized under the exchange of the pair of Lorentz
indices carried by each external graviton, {\it e.g.} $\mu\leftrightarrow \nu$ and
$\alpha\leftrightarrow\beta$ in \refb{equadver} and
$\mu\leftrightarrow \alpha$, $\nu\leftrightarrow \beta$ and $\sigma\leftrightarrow \gamma$
in \refb{esoftvert}.
Even though \refb{esoftvert} has a complicated form, we
shall need the form of the vertex when one of the external momenta (say $k_3$) is small
compared to the others. In this limit it simplifies.

\begin{figure}
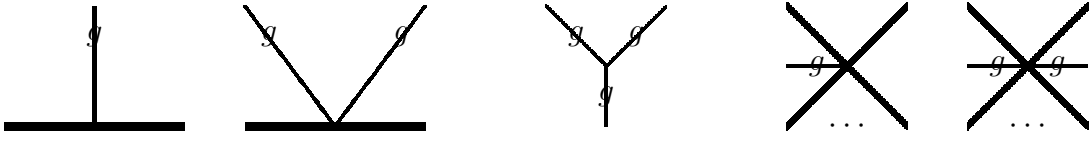


\begin{center}

\figgrvertex

\end{center}

\vskip -1.2in

\caption{This diagram shows various vertices induced from the action \refb{egraction} that are
needed for our computation. Here the \asnotex{thinner lines carrying the symbol $g$ denote
gravitons} and the thicker lines denote scalars. 
\label{figgrvertex}}

\end{figure}

\begin{figure}
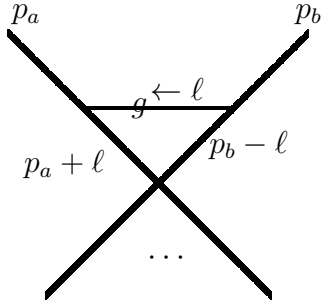


\begin{center}

\figqedtwoGR

\end{center}

\vskip -.4in

\caption{Diagram contributing to $\Gamma^{(n)}$.
\label{figqedtwoGR}}

\end{figure}

\begin{figure}
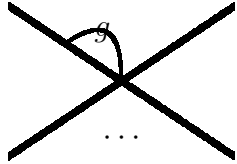


\begin{center}

\figgamman

\end{center}

\vskip -.6in

\caption{Another diagram contributing to $\Gamma^{(n)}$. We can also have a diagram where both ends of the
internal graviton are attached to the $n$-scalar vertex, but this vanishes in dimensional regularization and
so we have not displayed them.
\label{figadd}}

\end{figure}

\begin{figure}
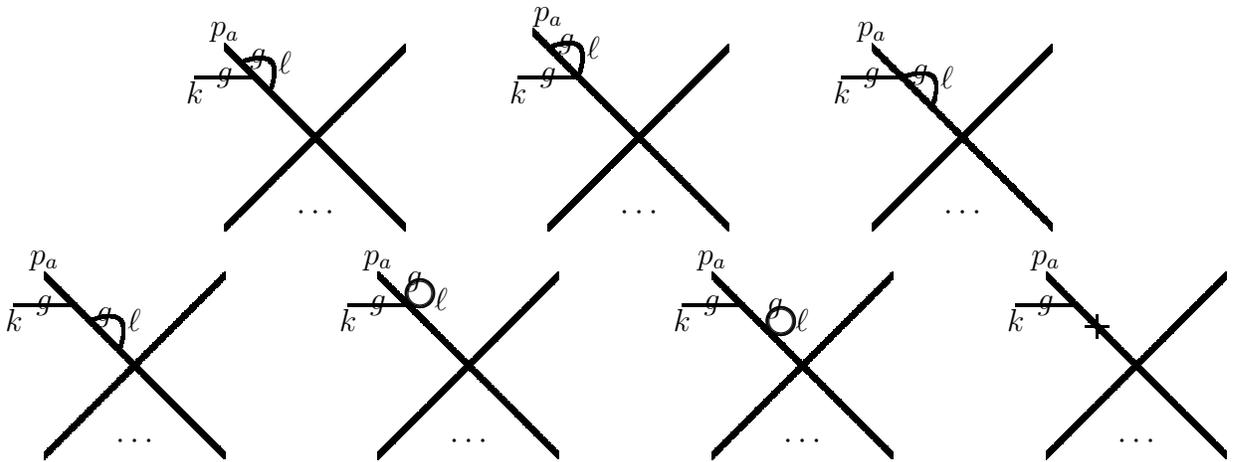


\begin{center}

\hbox{\hskip 1in \figselfaGR \qquad \figselfbGR \qquad \figselfcGR} 
\hbox{\figselfdGR \qquad \figselfgGR \qquad \figselfeGR \qquad \figselffGR}

\end{center}

\caption{One loop contribution to $\Gamma^{(n,1)}$ involving internal graviton line
connecting two different points on the same leg. \label{figselfGR}}

\end{figure}

The vertex where a graviton carrying Lorentz index $(\mu\nu)$
attaches to $n$ scalar fields  is given by:
\be \label{econt}
i\kappa \lambda \, \eta_{\mu\nu}\, .
\ee
The vertex where two gravitons carrying Lorentz index $(\mu\nu)$ and $(\rho\sigma)$
attach to $n$ scalar fields  is given by:
\be \label{econt1}
- \, {i}\, \kappa^2 \lambda \left( \eta_{\mu\nu} \eta_{\rho\sigma} - \eta_{\mu\rho}\eta_{\nu\sigma}
- \eta_{\mu\sigma}\eta_{\nu\rho}\right)\, .
\ee
We also need the vertex containing two scalars and three gravitons for evaluating the fifth diagram of
Fig.~\ref{figselfGR}.
However even without knowing the form of this
vertex one can see that this diagram does not generate contributions proportional to 
$\ln\omega^{-1}$.
Therefore we have not written down the expression for this vertex.

We can use these vertices to compute one loop contribution to the $n$ scalar amplitude
$\Gamma^{(n)}$ and $n$-scalar and one soft graviton amplitude $\Gamma^{(n,1)}$.
At one loop order $\Gamma^{(n)}$ receives contribution from diagrams shown in 
Fig.~\ref{figqedtwoGR} that are analogous to Fig.~\ref{figqedtwo} with
the internal  photon replaced by a graviton. There are also some 
additional diagrams shown in Fig.~\ref{figadd}.

\begin{figure}
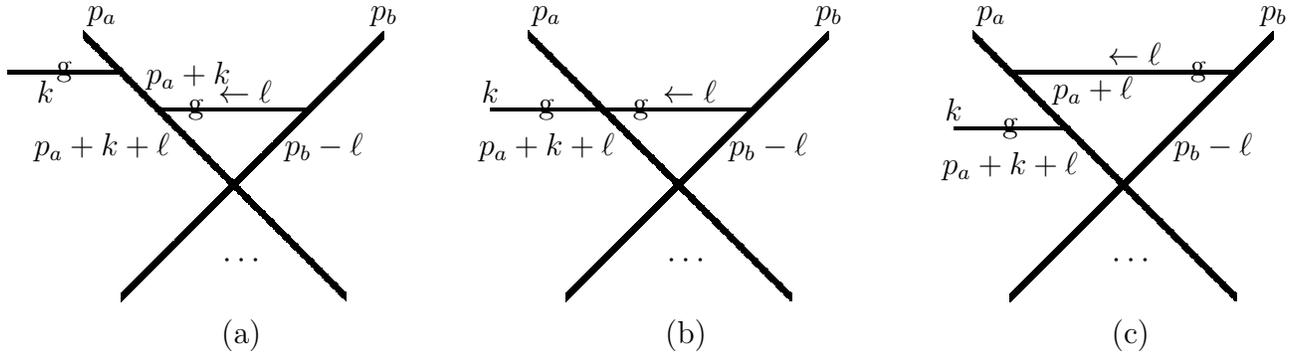


\begin{center}

\hbox{\hskip -.3in \figqedaGR \figqedbGR \figqedcGR}

\end{center}

\caption{One loop contribution to $\Gamma^{(n,1)}$ involving internal graviton line
connecting two different legs. The thicker lines represent scalar particles and the thinner
lines represent gravitons. 
\label{figqedGR}}

\end{figure}

\begin{figure}
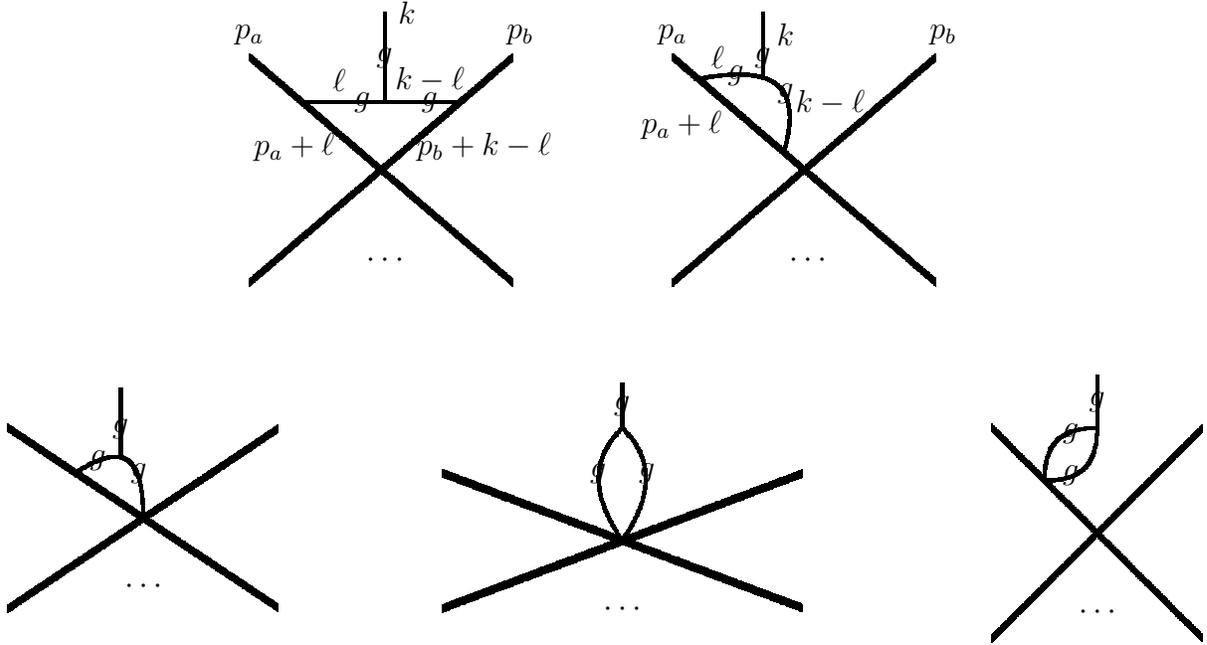


\begin{center}

\hbox{~\hskip 1in \figgravitya \quad \figgravityb} 
\hbox{~\hskip -.5in \figgrextrad \hskip -.2in \figtadpolec  \fignewfiggg}

\end{center}

\vskip -1in

\caption{Diagrams involving 3-graviton vertex. 
\label{figgravity}}

\end{figure}

\begin{figure}
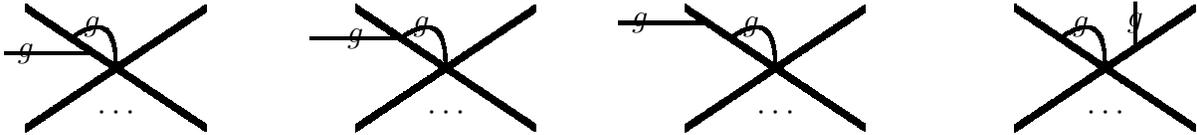


\begin{center}

\hbox{\hskip -.2in \figgrextraa \hskip .2in \figgrextrab   \hskip .2in \figgrextrac \hskip .2in \figgrextraf}


\end{center}

\vskip -.6in

\caption{Diagrams where the internal graviton attaches to the $n$-point vertex.
\label{figgrextra}}

\end{figure}

\begin{figure}
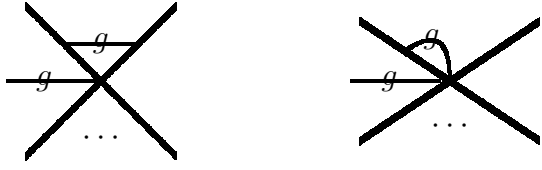


\begin{center}

\hbox{\hskip 1in \figextranew \qquad \figgrextrae}

\end{center}

\vskip -.6in

\caption{Diagrams where the external graviton attaches to the $n$-point vertex. The first diagram vanishes 
if we take the external graviton polarization to be traceless. The second diagram has no logarithmic terms.
\label{figgrtrivial}}

\end{figure}

\begin{figure}
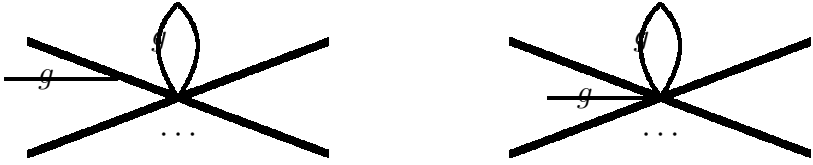


\begin{center}

\hbox{\figtadpolea  \figtadpoleb}


\end{center}

\vskip -.6in

\caption{Diagrams where both ends of the internal graviton attach to the $n$-point vertex.
In dimensional regularization these diagrams vanish. 
Even if we use momentum cut-off, these diagrams cannot have any contribution proportional
to $\ln\omega^{-1}$ since the soft momentum $k$ does not flow through any loop.
\label{figtadpole}}

\end{figure}

The relevant diagrams for $\Gamma^{(n,1)}$ include the analogs
Figs.~\ref{figqed} and \ref{figself} with all photons replaced by gravitons. This have been
shown in Figs.~\ref{figqedGR} and \ref{figselfGR}.
However there are also some extra
diagrams that we shall list below:
\begin{enumerate}
\item There are diagrams where the
external graviton couples to the internal graviton via the cubic coupling \refb{esoftvert}. 
Examples of these are 
shown in Fig.~\ref{figgravity}. 
\item There are diagrams where one end of the internal graviton attaches to the $n$-scalar vertex
via the coupling
\refb{econt}.
These have been shown in Figs.~\ref{figgrextra}.
\item There are diagrams where the external graviton attaches to the scalar $n$-point 
vertex via the coupling \refb{econt}  or \refb{econt1}. 
These have been shown in Fig.~\ref{figgrtrivial}.  The first diagram can be made to vanish by taking the external
graviton polarization to be traceless: $\ve_\rho^{~\rho}=0$. The second diagram has no logarithmic terms. Therefore
we shall ignore these diagrams in subsequent  discussions.
\item There are diagrams of the type shown in Fig.~\ref{figtadpole} where two ends of the
internal graviton attach to the $n$-scalar vertex.
In dimensional regularization these diagrams vanish. 
Therefore we shall ignore these diagrams in our analysis.
\end{enumerate}

Our analysis of these diagrams will proceed as
in \S\ref{s3}, but there will be some important differences that we shall point out below.
For an internal graviton of momentum $\ell$, whose two ends are attached to two scalar lines 
$a$ and $b$ with $\ell$
flowing from the leg $b$ towards the leg $a$, as in Figs.~\ref{figqedtwoGR},
\ref{figqedGR}, the analog of Grammer-Yennie decomposition of the 
graviton propagator will be taken to be
\be \label{eGYfirst}
G^{\mu\nu,\rho\sigma}_{(ab)}(\ell ,p_a,p_b)\ =\
 \big(\eta^{\mu\rho}\eta^{\nu\sigma}+\eta^{\mu\sigma}\eta^{\nu\rho}-\eta^{\mu\nu}
\eta^{\rho\sigma}\big)\ -\ K^{\mu\nu,\rho\sigma}_{(ab)}(\ell ,p_a,p_b)\, ,
\ee
\be
K^{\mu\nu,\rho\sigma}_{(ab)}(\ell ,p_{a},p_{b})\ =\ \mathcal{C}(\ell ,p_{a},p_{b})\ \big[(p_{a}+\ell )^{\mu}\ell ^{\nu}
+(p_{a}+\ell )^{\nu}\ell ^{\mu}\big]\ \big[(p_{b}-\ell )^{\rho}\ell ^{\sigma}+(p_{b}-\ell )^{\sigma}\ell ^{\rho}\big]\, ,\non\\
\ee
where
\ben
\mathcal{C}(\ell ,p_{a},p_{b})\  &&=\ \f{(-1)}{\{p_{a}.(p_{a}+\ell )-i\eps\}
\ \{p_{b}.(p_{b}-\ell )-i\eps\} \{\ \ell .(\ell +2p_{a})-i\eps\} \ \{\ell .(\ell -2p_{b})-i\eps\}}
\non\\
&& \Big[2(p_{a}.p_{b})^{2}-p_{a}^{2}p_{b}^{2}-\ell ^{2}(p_{a}.p_{b})-2(p_{a}.p_{b})(p_{a}.\ell )+2(p_{a}.p_{b})(p_{b}.\ell )
\Big]\, .
\een
If one end of an internal graviton is attached to the $n$-scalar vertex and the other end is attached
to the $a$'th scalar leg as in Figs.~\ref{figadd}, \ref{figgrextra}, 
with $\ell$ flowing from the vertex towards the $a$'th leg, 
we express the propagator as:
\be \label{eGK1}
-\, {i\over \ell^2-i\eps} 
\, {1\over 2}\, \left\{ G^{\mu\nu,\rho\sigma}_{(a)}(\ell ,p_a) + K^{\mu\nu,\rho\sigma}_{(a)}(\ell ,p_a)\right\}\, ,
\ee
where 
\be\label{eGK2}
G^{\mu\nu,\rho\sigma}_{(a)}(\ell ,p_a)\ =\
 \big(\eta^{\mu\rho}\eta^{\nu\sigma}+\eta^{\mu\sigma}\eta^{\nu\rho}-\eta^{\mu\nu}
\eta^{\rho\sigma}\big)\ -\ K^{\mu\nu,\rho\sigma}_{(a)}(\ell ,p_a)\, ,
\ee
\be\label{eGK3}
K^{\mu\nu,\rho\sigma}_{(a)}(\ell ,p_{a})\ =
\wt\CC(\ell ,p_{a})\ \big[(p_{a}+\ell )^{\mu}\ell ^{\nu}
+(p_{a}+\ell )^{\nu}\ell ^{\mu}\big]\ \eta^{\rho\sigma}
\, ,
\ee
and
\be\label{eGK4}
\wt\CC(\ell ,p_{a})\  =\ - \f{ 2\, p_a . (p_a-\ell)}{\{p_{a}.(p_{a}+\ell )-i\eps\}
\  \{\ \ell .(\ell +2p_{a})-i\eps\}}
\, .
\ee
For internal gravitons whose one end is attached to a 3-graviton vertex 
instead of a scalar, as in Fig~\ref{figgravity}, we do not carry out any 
Grammer-Yennie decomposition.

The decomposition into G and K-gravitons is not arbitrary but has been chosen to ensure two properties:
\begin{enumerate}
\item The K-graviton polarization, being proportional to $\ell$, is pure gauge and allows us to 
sum over K-graviton
insertions using Ward identities. The relevant Ward identities have been shown in Fig.~\ref{figKnon}, with 
the quantity $ A(p,k,\ell,\xi,\zeta)$
is given by
\ben \label{edefA}
A(p,k,\ell,\xi,\zeta) &=& 2\, i\, \xi.p\, \zeta^{\mu\nu} \bigg[ 2\, (2\, p_{\mu} + \ell_\mu) \, k_\nu  + 2\, k_\mu k_\nu
-\eta_{\mu\nu} \, \bigg\{ k.(2p+\ell) + k^2\bigg\}
\bigg] \nonumber \\
&& \hskip -1in + 2\, i\, \xi.(k+\ell) \, \zeta^{\mu\nu} \bigg[- 2\, p_\mu \, (p+\ell)_\nu +\eta_{\mu\nu}\,
\{ p.(p+\ell)+m^2\}
\bigg] \nonumber \\
&& \hskip -1in 
+ 2\, i\, (\xi^\alpha k^\beta+\xi^\beta k^\alpha) \, \zeta^{\mu\nu} \, \Bigg[\eta_{\alpha\mu}\,
\eta_{\beta\nu} \, p.(p+k+\ell)
-{1\over 2} \, \eta_{\alpha\beta}\, \eta_{\mu\nu} \, p.(p+k+\ell) \nonumber \\ && \hskip-1in
+ \eta_{\alpha\beta} \, p_\mu\, (p+k+\ell)_\nu
+ \eta_{\mu\nu} \, p_\alpha\, (p+k+\ell)_\beta 
- 2 \, \eta_{\alpha\mu} \, p_\beta (p+k+\ell)_\nu 
\nonumber \\ && \hskip -1in - 2 \, \eta_{\alpha\mu} \, p_\nu (p+k+\ell)_\beta
+m^2 \, \left( \eta_{\mu\alpha}\, \eta_{\nu\beta} -{1\over 2} \, \eta_{\mu\nu}\, \eta_{\alpha\beta}
\right)
\Bigg]\, .
\een
Due to this additional term, the sum over K-gravitons will leave behind some residual terms that will be discussed
below.
\item In any one loop diagram contributing to the amplitude $\Gamma^{(n)}$
without external soft graviton, the result vanishes if we
replace the internal graviton by G-graviton. 
\end{enumerate}

\begin{figure}

\begin{center}

\figmiddlegr

\hbox{~\hskip 0in \figKgaugenewgr}

\end{center}

\vskip -1in

\caption{Analog of Fig.~\ref{figward} for gravity.
The arrow on the graviton line represents that the polarization of the graviton carrying momentum
$k$ is taken
to be equal to $\xi_\mu k_\nu+\xi_\nu k_\mu$. The polarization of the graviton carrying momentum $k$ is
taken to be $\zeta_{\rho\sigma}$.
In the first diagram the circle on the left denotes a vertex $-2\, \xi.(p_c+k)$ while 
the circle on the  right 
denotes a vertex $-2\, \xi.p_c$. $A(p_c,k,\ell,\xi,\zeta)$ appearing 
on the right hand side of the second diagram  is
given in eq.\refb{edefA}.
\label{figKnon}}

\end{figure}

With this convention the  K-graviton contribution to
Fig.~\ref{figqedtwoGR} for gravity    
can be computed as in \S\ref{s3}, leading to a contribution of the form 
$i\lambda K_{\rm gr}$  to $\Gamma^{(n)}$,
where $K_{\rm gr}$ is the gravitational 
counterpart of $K_{\rm em}$. The 
relevant part of the expression for $K_{\rm gr}$ will be described later. 
The K-graviton contribution to Fig.~\ref{figadd} can be carried out similarly, leading to an
expression of the form $i\lambda \wt K_{\rm gr}$. $\wt K_{\rm gr}$  has no infrared divergence and
we shall not write down its expression explicitly although it is straightforward to do so. The G-graviton
contributions to Fig.~\ref{figqedtwoGR} and \ref{figadd} vanish by construction. 
Therefore the net contribution to $\Gamma^{(n)}$ to one loop order may be written as
$i\lambda\, \exp[K_{\rm gr} + \wt K_{\rm gr}]$.

\begin{figure}
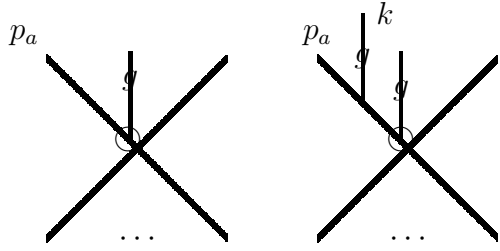


\begin{center}

\figcircle

\end{center}

\vskip -.5in

\caption{Figure illustrating the difference in the factorized K-graviton contribution to $\Gamma^{(n)}$ and
$\Gamma^{(n+1)}$.
\label{figcircle}}

\end{figure}

The K-graviton
contributions to Figs.~\ref{figqedGR} and \ref{figgrextra} may be evaluated similarly, with the factorized term giving
$i\lambda \, S^{(0)}_{\rm gr}\, \exp[K_{\rm gr} + \wt K_{\rm gr}]$. There are however some left-over terms arising as
follows:
\begin{enumerate}
\item As shown in Fig.~\ref{figKnon}, in the sum over K-graviton insertions in $\Gamma^{(n,1)}$
there is  a residual contribution $A$ that comes from
lack of complete cancellation among terms where a K-graviton is inserted to the two sides of a 
scalar-scalar-graviton
vertex and into the scalar-scalar-graviton
vertex. 
\item As explained in the caption of Fig.~\ref{figKnon},
the circled vertices 
 are momentum dependent. Therefore the two circled vertices shown in Fig.~\ref{figcircle} are not the same,
 one carries a factor of $\xi.p_a$ while the other carries a factor of $\xi.(p_a+k)$. The left hand figure is relevant for 
 $\Gamma^{(n)}$ while the right-hand figure is relevant for $\Gamma^{(n,1)}$. Therefore, even after factoring out
 $\exp[K_{\rm gr} + \wt K_{\rm em}]$ factor multiplying $\Gamma^{(n)}$, we are left with an additional 
 contribution to $\Gamma^{(n,1)}$ from sum over K-gravitons that must be accounted for.
 \end{enumerate}
We shall denote the sum of these two types of residual contributions as $\Gamma^{(n,1)}_{\rm residual}$.
The G-graviton contributions to Figs.~\ref{figqedGR} and \ref{figgrextra} will be denoted by $\Gamma^{(n,1)}_G$
and the net contribution from Fig.~\ref{figselfGR}  will be called 
$\Gamma^{(n,1)}_{\rm self}$. Finally the contribution to the 
diagrams in Fig.~\ref{figgravity} involving 3-graviton coupling will be denoted 
by $\Gamma^{(n,1)}_{\rm 3-graviton}$. 
In principle we should also include the contributions from Fig.~\ref{figgrtrivial} and
Fig.~\ref{figtadpole},
but we ignore them since they do not
generate logarithmic terms.
In this case the analog of \refb{e3.11} takes the 
form: 
\be\label{e3.11grav}
\Gamma^{(n,1)}_{\rm self}+\Gamma^{(n,1)}_{\rm G} + \Gamma^{(n,1)}_{\rm 3-graviton} +
\Gamma^{(n,1)}_{\rm residual} = i\lambda \, S^{(1)}_{\rm gr} \, .
\ee

We shall now briefly describe how we evaluate these contributions and then give the final result.
First let us consider $\Gamma^{(n,1)}_{\rm residual}$. This receives contribution
from Fig.~\ref{figqedGR}
and Fig.~\ref{figgrextra}.  
As explained above, there are two kinds of terms: one due to the right hand side of the second figure
of Fig.~\ref{figKnon}, and the other due to the momentum dependence of the circled vertices in 
Fig.~\ref{figKnon}.
It turns out that the residual part of the 
K-graviton contribution from Fig.~\ref{figgrextra} does not have any logarithmic term. On the
other hand the residual part of the 
K-graviton contribution from Fig.~\ref{figqedGR} receives logarithmic contribution only
from the region
where the loop momentum is large compared to $\omega$. The result
takes the form:
\be \label{eknonfac}
\Gamma^{(n,1)}_{\rm residual}=
-(i\lambda)\, \f{i}{2}\ \sum_{a=1}^n \sum_{b=1\atop b\ne a}^n \ 
\big[2(p_{a}.p_{b})^{2}-p_{a}^{2}p_{b}^{2}\big]\ \f{p_{a}.\ve.p_{a}}{p_{a}^{2}} \int_{\rm reg} 
\f{d^{4}l}{(2\pi)^{4}}\ \f{1}{\big[p_{a}.l-i\eps\big]\ \big[p_{b}.l+i\eps\big]\ \big[l^{2}-i\eps\big]}\, .
\ee
This contribution may be evaluated following a procedure similar
to the one used in \S\ref{s3}.

Contribution to $\Gamma^{(n,1)}_{\rm 3-graviton}$ arises from the five diagrams
in Fig.~\ref{figgravity}, but only the first two give terms proportional to $\ln\omega^{-1}$. 
Individually these diagrams suffer from collinear divergence from region of integration where the momenta of
the internal gravitons become parallel to that of the external graviton, but these divergences cancel in the sum over
such graphs after using momentum conservation. Therefore we always work with sum of these diagrams.
The net contribution from these diagrams receive logarithmic contribution from two regions  -- one where the
loop momentum is large compared to $\omega$ and the other where the loop momentum is small compared to
$\omega$. We shall analyze the contribution from the region of small loop momentum later. Contribution from the
region where the loop momentum is large compared to $\omega$  may be approximated as
\ben\label{e6.4}
&& - \, (i\lambda) \, \f{i}{4}\  \sum_{a=1}^{n}\sum_{b=1\atop b\ne a}^{n}\int_{\rm reg}
\f{d^{4}\ell }{(2\pi)^{4}}\ \f{1}{\big[p_{a}.\ell -i\eps\big]\ \big[p_{b}.\ell +
i\eps\big]\ \big[\ell ^{2}-i\eps\big]}\non\\
&&\left[-8\ (p_{a}.\ve.p_{b})\ (p_{a}.p_{b})\ +\ 2\ (p_{a}.\ve.p_{a})\ p_{b}^{2}\ +\ 2\ (p_{b}.\ve.p_{b})\ p_{a}^{2} - 2\, \big\{2(p_{a}.p_{b})^{2}-p_{a}^{2}p_{b}^{2}\big\} \,
{\ell.\ve.\ell\over \ell^2-i\eps}\right]\non\\
&& \ - (i\lambda) \, {i\over 2}\, \sum_{a=1}^n
\ \int_{\rm reg} \f{d^{4}\ell }{(2\pi)^{4}}\ \f{1}{\left[p_{a}.\ell -i\epsilon\right]^2}\ \f{1}{\big[\ell ^{2}-i\epsilon\big]}\ 
\Bigg[- 2\,  p_{a}^{2}\ (p_{a}.\ve.p_{a})\ +  \f{(p_{a}^{2})^2}
{p_a.\ell-i\epsilon}\ (p_a.\ve.\ell )\ \Bigg]\, . \non\\
\een
In arriving at this result we have used integration by parts and also conservation of total momentum
$\sum_{a=1}^n p_a=0$. We have also used the fact that in the expression for
the graviton propagator carrying momentum $(k-\ell)$  in the second diagram of Fig~.\ref{figgravity}, we 
can use the identity
\be \label{ebreak}
{1\over (k-\ell)^2-i\eps} = {2\ell.k\over \{(k-\ell)^2-i\eps\} \{\ell^2-i\eps\}} + {1\over \ell^2 -i\epsilon}\, ,
\ee
and ignore the contribution from the $(\ell^2-i\eps)^{-1}$ term, since the expression for the amplitude
involving this term has no $k$-dependent denominator and therefore cannot have a $\ln\omega^{-1}$
term.\footnote{This manipulation can be carried out only for terms containing at least two powers of $\ell$ in
the numerator so that each of the terms in \refb{ebreak} generates infrared finite integral.}
Similar manipulations will be used in other terms as well.

Contribution to $\Gamma^{(n,1)}_{\rm self}$ given in Fig.~\ref{figselfGR} 
may be analyzed following the argument given below \refb{e3.19Ga}.
We assume a general form $\ve_{\mu\nu} p_{a}^{\mu} p_{a}^{\nu}\, f(p_a.k)$ for this amplitude based on
Lorentz invariance and replace $\ve_{\mu\nu}$ by $\xi_\mu k_\nu + \xi_\nu k_\mu$ for an arbitrary vector $\xi$
satisfying $k.\xi=0$. Then the amplitude reduces to $2\, p_a.\xi \, p_a.k\, f(p_a.k)$. On the other hand the
diagrams in Fig.~\ref{figselfGR} for this choice of polarization may be evaluated using the Ward identity given in
Fig.~\ref{figKnon}. Due to the presence of the non-vanishing right-hand side in Fig.~\ref{figKnon},
the result does not vanish. Comparing this with the expected result
$2\, p_a.\xi \, p_a.k\, f(p_a.k)$, we can compute $f(p_a.k)$ and hence $\Gamma^{(n,1)}_{\rm self}$. It turns
out that it receives logarithmic contribution from region of integration where the loop momentum is large
compared to $\omega$. The result is:
\be \label{eGres}
\Gamma^{(n,1)}_{\rm self}= - (i\lambda)\, {i\over 2} \, \sum_{a=1}^{n} \,
p_a^2 \, p_a.\ve.p_a \, 
\int_{\rm reg}
\f{d^{4}\ell }{(2\pi)^{4}}\ \f{1}{\big[p_{a}.\ell -i\eps\big]^2\  \big[\ell ^{2}-i\eps\big]}\, .
\ee
This cancels the term in the last line of \refb{e6.4}.

One loop contribution from the diagrams involving 
G-gravitons in Figs.~\ref{figqedGR} and \ref{figgrextra}
may be evaluated following the procedure described in \S\ref{s3}. 
We find that 
the G-graviton contribution to Fig.~\ref{figgrextra} has no
logarithmic contribution. Therefore
we are left with the G-graviton contributions to Fig.~\ref{figqedGR}. 
These diagrams
have the same structure as in scalar QED and can be evaluated similarly. 
As in the case of scalar QED,
these diagrams receive significant contribution only
from the region where the loop momentum is large compared to $\omega$ and small compared to the momenta
of finite energy particles. The net logarithmic contributions from these diagrams is given by
\ben \label{egrG}
\Gamma^{(n,1)}_G &=& -(i\lambda) \, \f{i}{2}\ \sum_{a=1}^n \sum_{b=1\atop b\ne a}^n
 \int \f{d^{4}l}{(2\pi)^{4}}\ \f{1}{\big[p_{a}.l-i\epsilon\big]\ \big[p_{b}.l+i\epsilon\big]\ \big[l^{2}-i\epsilon\big]}\non\\
&&  \Bigg[8(p_{a}.p_{b})\ (p_{a}.\ve.p_{b})\ -\ 2p_{b}^{2}\ (p_{a}.\ve.p_{a})-\ \big[2(p_{a}.p_{b})^{2}-p_{a}^{2}p_{b}^{2}\big]\ \Bigg(\f{p_{a}.\ve.p_{a}}{p_{a}^{2}}\ +\ 2\ \f{p_{a}.\ve.l}{p_{a}.l}\Bigg)\ \Bigg] \non\\
&& + (i\lambda) \, \f{i}{2}\ \Big(\f{p_{a}.\ve.p_{a}}{p_{a}.k}\Big)\ 
\int \f{d^{4}l}{(2\pi)^{4}} \ \f{1}{\big[p_{a}.l-i\epsilon\big]\ \big[p_{b}.l+i\epsilon\big]\ 
\big[l^{2}-i\epsilon\big]}  \non\\ &&  \hskip 1in 
\Bigg[4(p_{a}.p_{b})\ (p_{b}.k)-\big[2(p_{a}.p_{b})^{2}-p_{a}^{2}p_{b}^{2}\big]\ 
\f{k.l}{p_{a}.l}\ \Bigg]\, .
\een

The total logarithmic terms in 
$\Gamma^{(n,1)}_{\rm G}$, $\Gamma^{(n,1)}_{\rm self}$, $\Gamma^{(n,1)}_{\rm residual}$ and  
$\Gamma^{(n,1)}_{\rm 3-graviton}$ from the region of integration where the loop momentum is large
compared to $\omega$,
can be expressed as\footnote{\asnotex{It is natural to conjecture that this pattern continues to hold also for
subsubleading soft graviton theorem, i.e.\ the universal part of the subsubleading contribution is given by
the action of the subsubleading soft graviton operator $\wh S^{(2)}_{\rm gr}$ acting on $\exp[K_{\rm gr}^{\rm reg}]$.
But we have not verified this by explicit computation.}}
\be \label{egr11}
(i\lambda)\, \widehat{S}^{(1)}_{\rm gr}\ K_{\rm gr}^{\rm reg}\, ,
\ee
where $\wh S^{(1)}_{\rm gr}$ is the quantum subleading soft graviton operator
\be \label{edefhs1gr}
\wh S^{(1)}_{\rm gr} = \sum_{a} 
{\ve_{\mu\rho} p_a^\rho 
k_\nu \over p_a.k} \left\{p_a^\mu {\p\over \p p_{a\nu}} - p_a^\nu {\p\over \p p_{a\mu}}
\right\}\, ,
\ee
and 
\be\label{e4.9}
K_{\rm gr}^{\rm reg}
\equiv {i\over 2} \sum_{a,b\atop b\ne a} \, \left\{ (p_a.p_b)^2 - {1\over 2} p_a^2 p_b^2\right\}\, 
\int_{\rm reg} 
{d^4\ell\over (2\pi)^4} \, {1\over \ell^2-i\eps} {1\over (p_a.\ell-i\eps) \, (p_b.\ell+i\eps)}\, .
\ee
$K_{\rm gr}^{\rm reg}$ is the analog of $K_{\rm em}^{\rm reg}$ for gravitational scattering, namely it is the factor
that appears in the exponent of the soft factor in the scattering of $n$ scalars, with the understanding that the
integration over loop momentum is restricted to the region larger than $\omega$. We note however that the full
expression for $K_{\rm gr}$ has more terms -- \refb{e4.9} already involves an approximation that the loop momentum
is small compared to the energies of external lines since this is the region that generates $\ln\omega^{-1}$ terms.
Explicit evaluation gives the following expression for the terms involving $\ln\omega^{-1}$:
\be \label{eevalkgr}
K_{\rm gr}^{\rm reg} =
{i\over 2} \sum_{a, b\atop b\ne a} 
\f{1}{4\pi}\ \ln \omega^{-1}\ \f{\left\{ (p_a.p_b)^2 - {1\over 2} p_a^2 p_b^2\right\}
}{\sqrt{(p_{a}.p_{b})^{2}-p_{a}^{2}p_{b}^{2}}}
\left\{\delta_{\eta_a\eta_b,1} - {i\over 2\pi}\ln\left({p_a.p_b +  \sqrt{(p_a.p_b)^2 - p_a^2 p_b^2}\over
p_a.p_b -  \sqrt{(p_a.p_b)^2 - p_a^2 p_b^2}}
\right)
\right\} \, .
\ee

At this stage the only remaining terms are 
the contributions to $\Gamma^{(n,1)}_{\rm 3-graviton}$  
from regions of loop
momentum integration where the loop momentum is small compared to $\omega$. 
These come from the first two diagrams in Fig.~\ref{figgravity}.
In the first diagram 
there are two relevant regions:
when $\ell$ is small and when $k-\ell$ is small, but they are related to each other by $\ell\to k-\ell$ and 
$a\leftrightarrow b$ symmetry. 
In the second diagram the relevant region is when $\ell$ is small.
The net contribution from these regions may be approximated by
\be
\lambda\,  \sum_{a=1}^{n}\sum_{b=1}^{n}\int\f{d^{4}\ell }{(2\pi)^{4}}\ \f{1}{\big[2 k.\ell -\ell ^2+ i\eps\big]\ \big[p_{a}.\ell -i\eps\big]\ \big[\ell ^2-i\eps\Big]}\Bigg[2(p_{a}.\ve.p_{b})\ (p_{a}.k)\ -\ 2(p_{b}.\ve.p_{b})\ \f{(p_{a}.k)^{2}}{p_{b}.k}\Bigg]\, ,
\ee
with the understanding that the integration over $\ell$ runs in the region where  the components of $\ell$ are
small compared to $\omega$. The result may be
expressed as
\be\label{egr12}
i {\lambda}
\, (\ln\omega^{-1} +\ln R^{-1})\, \left[ {i\over 4\pi} \, \sum_{b\atop \eta_b=-1} k.p_b \,  \sum_a {\ve_{\mu\nu} p_a^\mu p_a^\nu \over p_a.k}  
+{1\over 8\pi^2} \, 
\sum_a {\ve_{\mu\nu} p_a^\mu p_a^\nu \over p_a.k} \sum_b  p_b.k\, \ln{m_b^2 \over (p_b.\hat k)^2}\right]
\, ,
\ee
where $1/R$ is an infrared lower cut-off on momentum integration 
and $\hat k =-k/\omega =(1, \hat n)$.

Adding \refb{egr11} to \refb{egr12} and dividing by $i\lambda$ we get the terms involving $\ln\omega^{-1}$ and
$\ln R$ in $S^{(1)}_{\rm gr}$:
\ben \label{e5.14}
S^{(1)}_{\rm gr} &=& \widehat{S}^{(1)}_{gr}\ K_{\rm gr}^{\rm reg} \nonumber \\
&& \hskip -.5in + {1\over 4\pi} (\ln\omega^{-1} +\ln R^{-1})\, \left[ i\, \sum_{b\atop \eta_b=-1} k.p_b \,  \sum_a {\ve_{\mu\nu} p_a^\mu p_a^\nu \over p_a.k}  
+ {1\over 2\pi} \, 
\sum_a {\ve_{\mu\nu} p_a^\mu p_a^\nu \over p_a.k} \sum_b  p_b.k\, \ln{m_b^2 \over (p_b.\hat k)^2}\right]\, .
\nonumber \\
\een 


\sectiono{Generalizations} \label{s6}

In this section we shall consider the case where the scalars interact via both electromagnetic and gravitational
interaction via the action:
\ben
&& \int d^4x \, \sqrt{-\det g} \, \Bigg[ -{1\over 4} F_{\mu\nu} F^{\mu\nu} +{1\over 16\pi G} R 
-\sum_{a=1}^n \bigg\{ g^{\mu\nu} (\p_\mu \phi_a^* + i q_a A_\mu \phi_a^*) 
(\p_\nu \phi_a - i q_a A_\nu \phi_a)  \nonumber \\
&& \hskip 1in + m_a^2 \phi_a^*\phi_a\bigg\} 
+ \lambda \, \phi_1\cdots \phi_n + \lambda 
\, \phi_1^*\cdots
\phi_n^*
\Bigg]\, .
\een
For this analysis we need two new vertices, the graviton-photon-photon vertex
and the graviton-photon-scalar-scalar vertex.
If the graviton carries an ingoing momentum $q$ and Lorentz index $(\rho\sigma)$, and the
two photons carry ingoing momenta $k_1$ and $k_2$ and Lorentz indices $\mu$ and
$\nu$ respectively, then the graviton-photon-photon vertex is given by:
\ben
&& -i\, \kappa \ \Big[\eta_{\rho\sigma}\ 
\Big(-k_{1}.k_{2}\ \eta_{\mu\nu}+\ k_{1\nu}k_{2\mu}\Big) 
+\ \eta_{\mu\nu}\ \Big(k_{1\rho}k_{2\sigma}\ +\ k_{2\rho}k_{1\sigma}\Big)\ +\ k_{1}.k_{2}\Big( \eta_{\mu\rho}\eta_{\nu\sigma}+\eta_{\mu\sigma}\eta_{\nu\rho}\Big)\non\\
&&\ -\ \Big(k_{1\sigma}k_{2\mu}\eta_{\rho\nu}+k_{2\sigma}k_{1\nu}\eta_{\rho\mu}+k_{1\rho
}k_{2\mu}\eta_{\sigma\nu}+k_{2\rho}k_{1\nu}\eta_{\sigma\mu}\Big)\Big]\, .
\een
On the other hand the vertex with a pair of scalars carrying charges $q$, $-q$ and momenta $p_1$ and $p_2$,
a graviton carrying Lorentz indices $(\mu\nu)$ and momentum $k_1$ and a photon carrying Lorentz index $\rho$
and momentum $k_2$, all counted ingoing, is given by
\be
-i \, \kappa\, q\, \Big[\eta_{\mu\rho} (p_1-p_2)_\nu + \eta_{\nu\rho} (p_1-p_2)_\mu -
\eta_{\mu\nu} (p_1-p_2)_\rho\Big]\, .
\ee
In this theory we shall analyze  the extra terms in both the soft graviton theorem and the soft photon theorem.

There are two other vertices that are needed for our analysis. For example the sixth diagram of 
Fig.~\ref{figselfqedgr} needs
the vertex containing two scalars, two photons and one graviton, whereas the sixth diagram of Fig.~\ref{figfin2}
requires the two scalar, two graviton and one photon vertex. However even without knowing the form of these
vertices one can see that these diagrams do not generate contributions proportional to $\ln\omega^{-1}$.
Therefore we have not written down the expressions for these vertices.

\subsection{Soft graviton theorem}

\begin{figure}
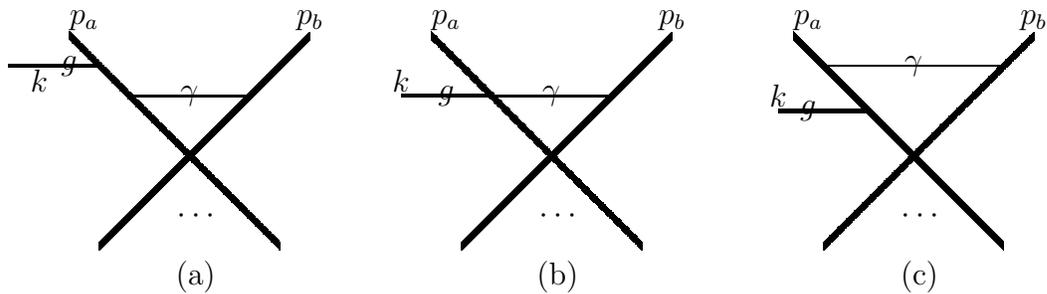


\begin{center}

\hbox{\figqedaqgr \figqedbqgr \figqedcqgr}

\end{center}

\caption{Contribution to soft graviton amplitude due to internal photon whose two ends are connected to
two different scalar lines.
Here the thickest lines denote scalars, \asnotex{lines of medium thickness carrying the symbol $g$
denote gravitons and the thin lines carrying the symbol $\gamma$
denote photons.  }
\label{figqedgr}}

\end{figure}

\begin{figure}
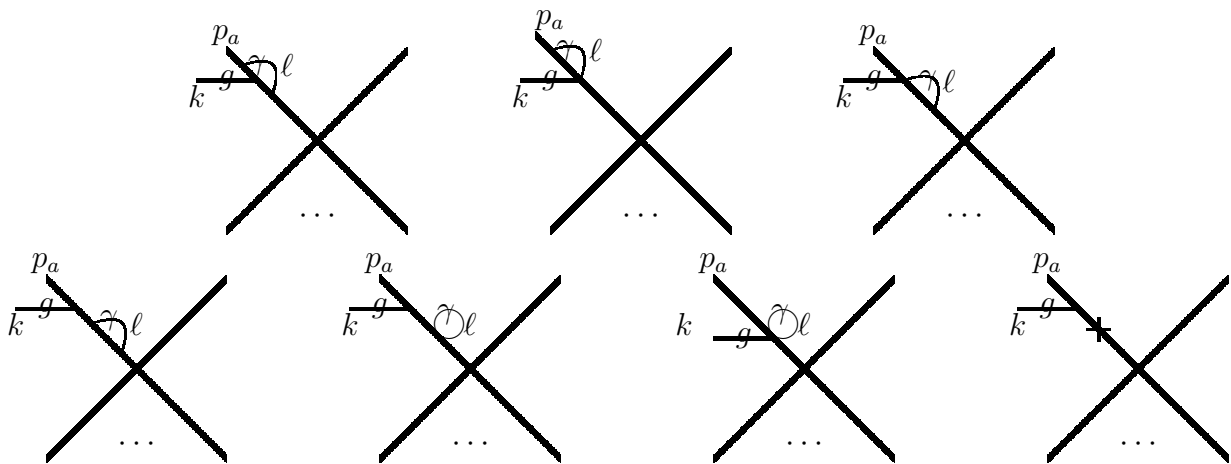


\begin{center}

\hbox{\hskip 1in \figselfaqgr \qquad \figselfbqgr \qquad \figselfcqgr} 
\hbox{\figselfdqgr \qquad \figselfeqgr
\qquad \figselfgqgr \qquad \figselffqgr}

\end{center}

\caption{One loop contribution to soft graviton amplitude involving internal photon line
connecting two points on the same leg. \label{figselfqedgr}}

\end{figure}

\begin{figure}
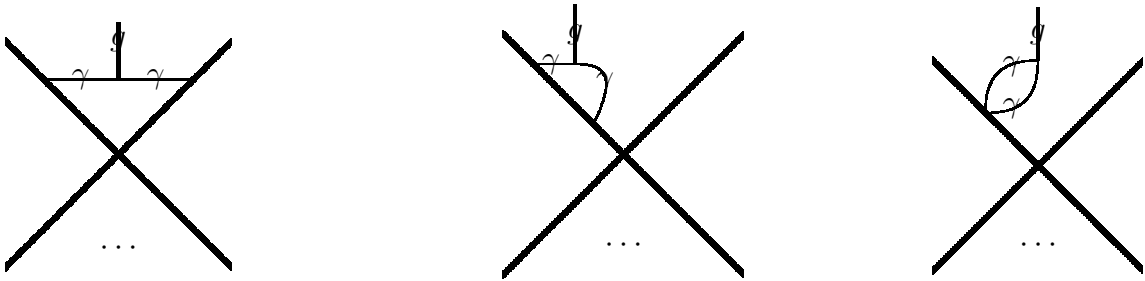


\begin{center}

\hbox{\hskip -.5in \figphgr \figgrselfem \fignewfigpg}

\end{center}

\vskip -1in

\caption{Diagrams containing graviton-photon-photon vertex 
that contribute to the soft photon contribution to the soft graviton 
theorem.
\label{figphgr}}

\end{figure}

\begin{figure}

\begin{center}



\figKgaugenewqedgr

\end{center}

\vskip -.6in

\caption{The Ward identity for the photon in the presence of a graviton-scalar-scalar vertex.
\label{fignewid}}

\end{figure}

We first consider the soft graviton theorem. In this case besides the 
contributions analyzed in \S\ref{s5}, we also
have the diagrams of Fig.~\ref{figqedgr} and Fig.~\ref{figselfqedgr},
obtained by replacing, in the diagrams in \S\ref{s3}, the external photon by a
graviton but keeping the internal line as a photon. We also have an additional set of
diagrams 
shown in Fig.~\ref{figphgr} where the external graviton connects to the internal
photon. Diagrams in which the external graviton attaches to the $n$-scalar vertex vanish for 
$\ve_\rho^{~\rho}=0$ and have not been displayed.
We carry out Grammer-Yennie decomposition for the internal photons
in Fig.~\ref{figqedgr} following \refb{eGYqed}, but not for diagrams of the form shown in Fig.~\ref{figselfqedgr} and
Fig.~\ref{figphgr}. The sum over K-photons factorize as in \S\ref{s3} and gives the factor of 
$\exp[K_{\rm em}]$ that
cancels between $\Gamma^{(n)}$ and $\Gamma^{(n,1)}$. In this case there is no residual contribution
since the analog of Fig.~\ref{figward} holds with the upper photon in the second identity replaced by a graviton
(see Fig.~\ref{fignewid}).
This leads to the analog of \refb{e3.11} with an additional
contribution to the left hand side given by diagrams of the form shown in Fig.~\ref{figphgr}. Denoting this contribution
by $\Gamma^{(n,1)}_{\gamma\gamma g}$ we arrive at the relation
\be
\Gamma^{(n,1)}_{\rm self} 
+ \Gamma^{(n,1)}_{\rm G} + \Gamma^{(n,1)}_{\gamma\gamma g} = i\, \lambda\, S^{(1)}_{\rm gr}
\, ,
\ee
with the understanding that both sides represent contributions in addition to what already appear in \refb{e3.11grav}. 
None of the terms have any infrared divergence, and therefore there are no logarithmic terms from the
region of integration in which the loop momentum is small compared to $\omega$. 
We shall describe below the organization of the various terms and then state the final result:
\begin{enumerate}
\item One can analyze $\Gamma^{(n,1)}_{\rm self}$ represented by 
the graphs in Fig.~\ref{figselfqedgr} by
following the procedure described below \refb{e3.19Ga}.
We
replace the external graviton polarization by a pure gauge form $(\xi_\mu k_\nu+\xi_\nu k_\mu)$ 
and apply Ward identity to evaluate the sum
over the graphs in Fig.~\ref{figselfqedgr}. In this case the Ward identity
has an additional contribution as shown in the right hand side of the second diagram in 
Fig.~\ref{fignonph}. It turns out however that its
contribution to the amplitude does not have any logarithmic term. Therefore $\Gamma^{(n,1)}_{\rm self}$ does not
generate any logarithmic contribution.
\item
$\Gamma^{(n,1)}_{\gamma\gamma g}$  receives
contribution proportional to $\ln\omega^{-1}$ from the first two diagrams of Fig.~\ref{figphgr},
from the region where the loop momentum is large compared 
to $\omega$.
\item
Finally, the G-photon contribution $\Gamma^{(n,1)}_{\rm G}$ from the
first two diagrams in Fig.~\ref{figqedgr} also has terms
proportional to $\ln\omega^{-1}$ from the region where the loop momentum is large compared to $\omega$. 
\end{enumerate}

\begin{figure}

\begin{center}

\figmiddlegr 
\hbox{\hskip 0in \figKgaugenewgraa}

\end{center}

\vskip -.5in

\caption{Analog of Fig.~\ref{figward} for graviton in the presence of a photon. 
The graviton carries a polarization $(\xi_\mu \ell_\nu+\xi_\nu \ell_\mu)$ and the photon carries
a polarization $\eps$. The circled vertex has been explained
in the caption of Fig.~\ref{figKnon}.
\label{fignonph}}

\end{figure}

The net logarithmic
contribution from $\Gamma^{(n,1)}_{\gamma\gamma g}$ and $\Gamma^{(n,1)}_{\rm G}$
is given by:
\be
(i\lambda)\, \widehat{S}^{(1)}_{\rm gr}\ K_{\rm em}^{\rm reg}\, .
\ee
After removing the $i\,\lambda$ factor, we have to add this 
to \refb{e5.14} to get the total logarithmic contribution to $S^{(1)}_{\rm gr}$:
\ben \label{e5.14tot}
S^{(1)}_{\rm gr} &=& \widehat{S}^{(1)}_{gr}\ \left(K_{\rm em}^{\rm reg}+ K_{\rm gr}^{\rm reg}\right) \nonumber \\
&& \hskip -.5in + {1\over 4\pi} (\ln\omega^{-1} +\ln R^{-1})\, \left[ i\, \sum_{b\atop \eta_b=-1} k.p_b \,  \sum_a {\ve_{\mu\nu} p_a^\mu p_a^\nu \over p_a.k}  
+ {1\over 2\pi} \, 
\sum_a {\ve_{\mu\nu} p_a^\mu p_a^\nu \over p_a.k} \sum_b  p_b.k\, \ln{m_b^2 \over (p_b.\hat k)^2}\right]\, .
\nonumber \\
\een 
This reproduces terms proportional to $\ln\omega^{-1}$ in the sum of \refb{e1.17int} and \refb{eqgrsoft} after using
\refb{edefkemreg} and \refb{eevalkgr}.

\subsection{Soft photon theorem}

\begin{figure}
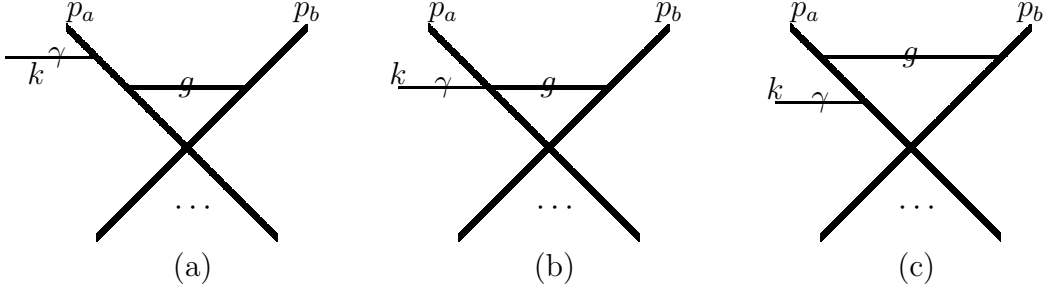


\begin{center}

~ 

\bigskip

\hbox{\figqedafin \figqedbfin \figqedcfin}

\end{center}

\caption{One loop contribution to soft photon amplitude involving internal graviton line
connecting two different legs. 
\label{figfin1}}

\end{figure}

\begin{figure}
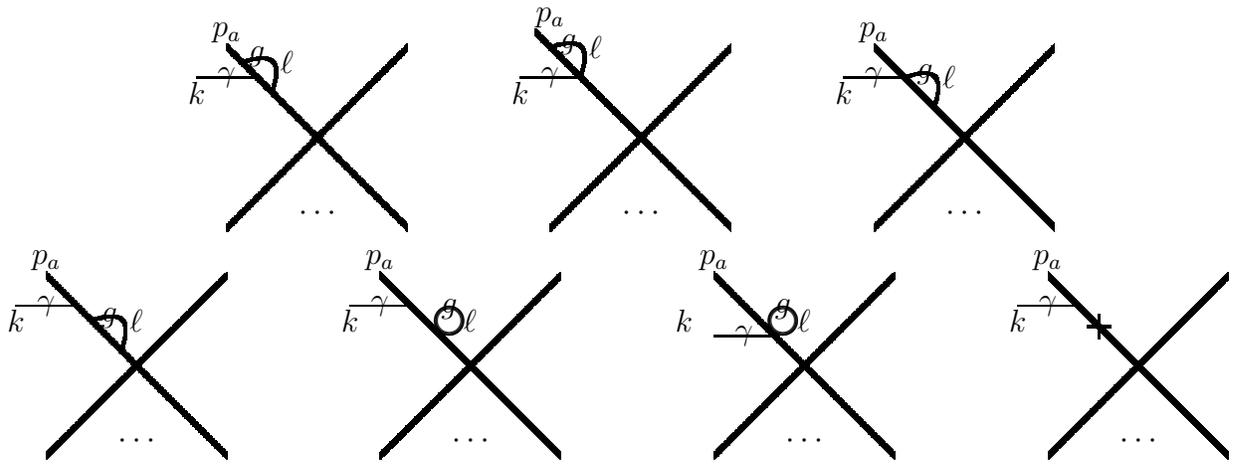


\begin{center}

\hbox{\hskip 1in \figselfafin \qquad \figselfbfin \qquad \figselfcfin} 
\hbox{\figselfdfin \qquad \figselfefin 
\qquad \figselfggrav \qquad \figselfffin}

\end{center}

\caption{Diagrams in which the external photon and both ends of the internal graviton attach to the same scalar
leg. 
\label{figfin2}}

\end{figure}

\begin{figure}
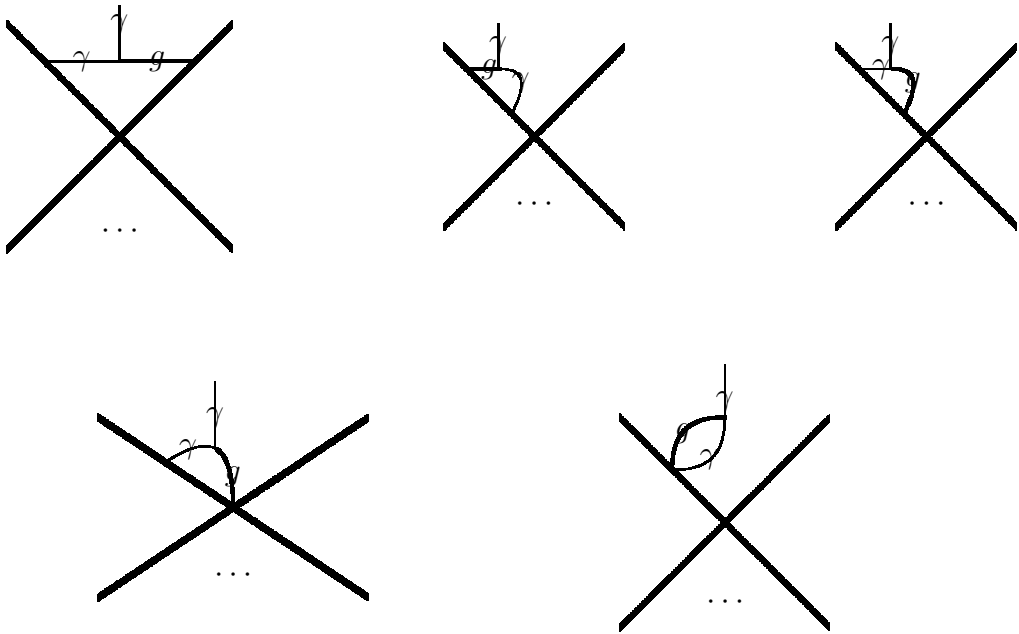


\begin{center}

\hbox{\hskip -.3in \figintgr \figemselfgrnew \figemselfgr}
\hbox{  \hskip .2in \figemextrad \qquad  \fignewfiggp}

\end{center}

\vskip -.9in

\caption{Diagrams involving graviton-photon-photon vertex
that need to be included in computing the soft graviton contribution to the soft photon 
theorem. 
\label{figintgr}}

\end{figure}

\begin{figure}
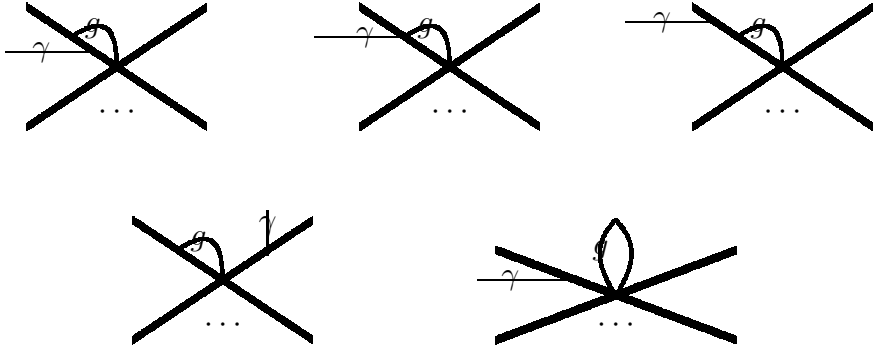


\begin{center}

~

\vskip .2in

\hbox{\hskip .5in \figemextraa \quad \figemextrab \quad \figemextrac
}
\hbox{
\hskip 1in  \figemextrae \quad \figtadpoleanew
}

\end{center}

\vskip -.5in

\caption{Diagrams with external soft photon and an internal
graviton where the internal graviton attaches to the $n$-point vertex.
\label{figemextra}}

\end{figure}

Next we shall consider the soft photon theorem. In this case we have all the diagrams considered in \S\ref{s3},
but also extra diagrams where the internal photon of Figs.~\ref{figqed} and \ref{figself} 
is replaced by an internal graviton, as shown in Figs.~\ref{figfin1} and
\ref{figfin2}, and two
additional sets of diagrams: one where one end of the internal graviton connects to the external photon as in 
Fig.~\ref{figintgr} and the other where one end of the internal graviton is attached
to the $n$-scalar vertex as in Fig.~\ref{figemextra}. 
There is also an additional diagram obtained by replacing in the first diagram of
Fig.~\ref{figtadpole} the external graviton by the external photon, but this vanishes
in dimensional regularization.

We shall analyze the diagrams in Figs.~\ref{figfin1} and \ref{figemextra} using 
Grammer-Yennie
decomposition for the internal graviton following the rules described in \refb{eGYfirst}-\refb{eGK4}. 
The result of summing over K-gravitons in 
$\Gamma^{(n,1)}$
will generate the factor of
$\exp[K_{\rm gr}+\wt K_{\rm gr}]$ which cancels a similar factor in the expression of 
$\Gamma^{(n)}$. However there will be 
residual part that will be left over due to non-cancellation of the sum over K-graviton insertions reflected in the
right-hand side of
Fig.~\ref{fignonph}. 
Another residual contribution arises due to the momentum dependence of the circled vertices;  as 
illustrated in Fig,~\ref{figcircle}, the factorized contribution of K-gravitons for $\Gamma^{(n)}$ and
$\Gamma^{(n,1)}$ differ. The only difference in the present case is that the external graviton carrying momentum
$k$ in Fig.~\ref{figcircle} is replaced by an external photon.
As in \S\ref{s5}, we shall denote these residual contributions in the sum over K-gravitons by 
$\Gamma^{(n,1)}_{\rm residual}$. The G-graviton contribution to Figs.~\ref{figfin1} and \ref{figemextra} will be
denoted
by $\Gamma^{(n,1)}_{\rm G}$. The contribution from diagrams involving the coupling of graviton to photon,
as shown in Fig.~\ref{figintgr}, will be denoted by  $\Gamma^{(n,1)}_{\gamma\gamma g}$, and the contributions 
from Fig.~\ref{figfin2} will be denoted by  
$\Gamma^{(n,1)}_{\rm self}$.
Then the generalization of \refb{e3.11}
takes the form:
\be \label{eleft}
\Gamma^{(n,1)}_{\rm self} 
+\Gamma^{(n,1)}_{\rm G} + \Gamma^{(n,1)}_{\gamma\gamma g}+ \Gamma^{(n,1)}_{\rm residual} 
= i\, \lambda\, S^{(1)}_{\rm gr}
\, ,
\ee
again with the understanding that both sides represent additional contribution besides those described
in \S\ref{s3}.

Analysis of various terms on the left hand side of \refb{eleft} goes as
follows:
\begin{enumerate}
\item $\Gamma^{(n,1)}_{\rm self}$ can be shown to vanish 
using the same
argument given below \refb{e3.19Ga}.
In this case the relevant Ward identities  given in Figs.~\ref{figward} and
\ref{fignewid} do not have any left-over extra contributions.
\item 
It turns out that $\Gamma^{(n,1)}_{\rm residual}$, given by the left-over contribution after
summing over K-graviton insertions in Figs.~\ref{figfin1} and \ref{figemextra},  does not receive any 
logarithmic terms either from the region of loop momentum integration small compared
to $\omega$ or from regions of loop momentum integration large compared to
$\omega$.
\item
$\Gamma^{(n,1)}_{\rm G}$ receives contributions proportional to
$\ln\omega^{-1}$ only from the G-graviton contribution to Fig.~\ref{figfin1}, 
from region of integration where the loop momentum is larger than $\omega$. 
\item The individual diagrams contributing to $\Gamma^{(n,1)}_{\gamma\gamma g}$  have collinear divergence
from the region where the momenta of the internal graviton and photon are parallel to the momentum
of the external photon. This cancels in the sum over all diagrams in Fig.~\ref{figintgr}. 
The second and third diagrams of Fig.~\ref{figintgr} each has contribution 
proportional to $\ln\omega^{-1}$ from the region of integration where
the loop momentum is large compared to $\omega$, but the sum of these 
contributions vanishes. Finally,
$\Gamma^{(n,1)}_{\gamma\gamma g}$ receives contributions 
proportional to $\ln\omega^{-1}$ from the first two diagrams in Fig.~\ref{figintgr},
from the region
where the  momentum of the internal graviton is  smaller than $\omega$. 
\end{enumerate}
The net logarithmic
contribution from the region of integration where the loop momentum is larger than $\omega$ is given by
\be \label{e6.5}
i\, \lambda \, \wh S^{(1)}_{\rm em} \, K^{\rm reg}_{\rm gr}\, .
\ee
On the other hand the contribution to $\Gamma^{(n,1)}_{\gamma\gamma g}$ from the small loop
momentum region
is given by:
\be \label{e6.6}
i\, \lambda\, (\ln\omega^{-1}+\ln R^{-1}) \left[{i\over 4\pi}\, 
\, \, \sum_{b\atop \eta_b=-1} k.p_b \, \sum_a {\ve_\mu p_a^\mu \over p_a.k} \, q_a 
+ \f{1}{8\pi^{2}}\, \sum_{a=1}^{n}\ \f{q_{a}\ve_\mu p_{a}^\mu}{p_{a}.k}\ \sum_{b=1}^{n}\ (p_{b}.k)\ \ln\Bigg(\f{-p_{b}^{2}}{(p_{b}.\hat{k})^{2}}\Bigg)
\right]
\, .
\ee
One difference from the previous diagrams of this type, {\it e.g.} the ones shown in 
Fig.~\ref{figgravity}, is
that the divergent contribution comes only from the region where the internal graviton momentum becomes small, and
not when the internal photon momentum becomes small. This reflects the fact that while photons feel the long range
gravitational force due to other particles, the graviton, being charge neutral, does not feel any long range Coulomb force. 
After removing the $i\,\lambda$ factors from \refb{e6.5} and \refb{e6.6}, we have to add them to \refb{e3.19G} 
to get the total
soft factor $S^{(1)}_{\rm em}$. This gives
\ben
S^{(1)}_{\rm em} &=& \wh S^{(1)}_{\rm em} \, \left(K^{\rm reg}_{\rm em}+ K^{\rm reg}_{\rm gr}\right)
\nonumber \\ && \hskip -.7in 
+ (\ln\omega^{-1}+\ln R^{-1}) \left[{i\over 4\pi}\, 
\, \, \sum_{b\atop \eta_b=-1} k.p_b \, \sum_a {\ve_\mu p_a^\mu \over p_a.k} \, q_a 
+ \f{1}{8\pi^{2}}\, \sum_{a=1}^{n}\ \f{q_{a}\ve_\mu p_{a}^\mu}{p_{a}.k}\ \sum_{b=1}^{n}\ (p_{b}.k)\ \ln\Bigg(\f{-p_{b}^{2}}{(p_{b}.\hat{k})^{2}}\Bigg)
\right]
\, . \nonumber \\
\een
This reproduces terms proportional to $\ln\omega^{-1}$ in the sum of \refb{e1.16int} and \refb{e3.23Gint} after 
using the explicit forms of $K^{\rm reg}_{\rm em}$ and $K^{\rm reg}_{\rm gr}$ given in
\refb{edefkemreg} and \refb{eevalkgr}.

\bigskip

{\bf Acknowledgement:}
We would like to thank Alok Laddha for many useful discussions and collaboration at
the early stages of this work. \asnotex{We would also like to thank Massimo Bianchi and
Gabriele Veneziano for discussions and the authors of \cite{1812.08137,1901.10986} for 
informing us of their results on the soft limit
of scattering amplitudes of massless states.}
The work of A.S. was
supported in part by the J. C. Bose fellowship of 
the Department of Science and Technology, India and also by the Infosys Chair Professorship.


\begin{thebibliography}{99}



\small

\baselineskip=14pt

\parskip=0pt

\bibitem{Gell-Mann}
M.~Gell-Mann and M.~L.~Goldberger, Phys.\ Rev.\ {\bf 96}, 1433 (1954).

\bibitem{low}
F.~E.~Low, Phys.\ Rev.\ {\bf 110}, 974 (1958).

\bibitem{saito}
S.~Saito, Phys.\ Rev.\ {\bf 184}, 1894 (1969).

\bibitem{burnett}
T.~H.~Burnett and N.~M.~Kroll, Phys.\ Rev.\ Lett.\ {\bf 20}, 86 (1968).

\bibitem{bell}
J.~S.~Bell and R. Van Royen, Nuovo Cim.\ {\bf A60}, 62 (1969).

\bibitem{duca}
V.~Del Duca, Nucl. Phys. {\bf B345}, 369 (1990).


\bibitem{weinberg1} 
  S.~Weinberg,
  ``Photons and Gravitons in s Matrix Theory: 
  Derivation of Charge Conservation and Equality of Gravitational and Inertial Mass,''
  Phys.\ Rev.\  {\bf 135}, B1049 (1964).
  doi:10.1103/PhysRev.135.B1049

\bibitem{weinberg2} 
  S.~Weinberg,
  ``Infrared photons and gravitons,''
  Phys.\ Rev.\  {\bf 140}, B516 (1965).
  doi:10.1103/PhysRev.140.B516

\bibitem{jackiw1} 
  D.~J.~Gross and R.~Jackiw,
  ``Low-Energy Theorem for Graviton Scattering,''
  Phys.\ Rev.\  {\bf 166}, 1287 (1968).
  doi:10.1103/PhysRev.166.1287
  
\bibitem{jackiw2} 
  R.~Jackiw,
  ``Low-Energy Theorems for Massless Bosons: Photons and Gravitons,''
  Phys.\ Rev.\  {\bf 168}, 1623 (1968).
  doi:10.1103/PhysRev.168.1623
  
  
\bibitem{ademollo} 
  M.~Ademollo, A.~D'Adda, R.~D'Auria, F.~Gliozzi, E.~Napolitano, S.~Sciuto and P.~Di Vecchia,
  ``Soft Dilations and Scale Renormalization in Dual Theories,''
  Nucl.\ Phys.\ B {\bf 94}, 221 (1975).
  doi:10.1016/0550-3213(75)90491-5

\bibitem{shapiro} 
  J.~A.~Shapiro,
  ``On the Renormalization of Dual Models,''
  Phys.\ Rev.\ D {\bf 11}, 2937 (1975).
  doi:10.1103/PhysRevD.11.2937

\bibitem{1312.2229} 
  A.~Strominger,
  ``On BMS Invariance of Gravitational Scattering,''
  JHEP {\bf 1407}, 152 (2014)
  doi:10.1007/JHEP07(2014)152
  [arXiv:1312.2229 [hep-th]].

\bibitem{1401.7026} 
  T.~He, V.~Lysov, P.~Mitra and A.~Strominger,
  ``BMS supertranslations and WeinbergÕs soft graviton theorem,''
  JHEP {\bf 1505}, 151 (2015)
  doi:10.1007/JHEP05(2015)151
  [arXiv:1401.7026 [hep-th]].

\bibitem{1408.2228} 
  M.~Campiglia and A.~Laddha,
  ``Asymptotic symmetries and subleading soft graviton theorem,''
  Phys.\ Rev.\ D {\bf 90}, no. 12, 124028 (2014)
  doi:10.1103/PhysRevD.90.124028
  [arXiv:1408.2228 [hep-th]].

\bibitem{1411.5745} 
  A.~Strominger and A.~Zhiboedov,
  ``Gravitational Memory, BMS Supertranslations and Soft Theorems,''
  JHEP {\bf 1601}, 086 (2016)
  doi:10.1007/JHEP01(2016)086
  [arXiv:1411.5745 [hep-th]].

\bibitem{1502.02318} 
  M.~Campiglia and A.~Laddha,
  ``New symmetries for the Gravitational S-matrix,''
  JHEP {\bf 1504}, 076 (2015)
  doi:10.1007/JHEP04(2015)076
  [arXiv:1502.02318 [hep-th]].

\bibitem{1502.06120} 
  S.~Pasterski, A.~Strominger and A.~Zhiboedov,
  ``New Gravitational Memories,''
  JHEP {\bf 1612}, 053 (2016)
  doi:10.1007/JHEP12(2016)053
  [arXiv:1502.06120 [hep-th]].

\bibitem{1502.07644} 
  D.~Kapec, V.~Lysov, S.~Pasterski and A.~Strominger,
  ``Higher-Dimensional Supertranslations and Weinberg's Soft Graviton Theorem,''
  Annals of Mathematical Sciences and Applications, Volume 2 (2017),
  pp 69-94
  doi:10.4310/AMSA.2017.v2.n1.a2
  [arXiv:1502.07644 [gr-qc]].
 
 \bibitem{1505.05346} 
  M.~Campiglia and A.~Laddha,
  ``Asymptotic symmetries of QED and Weinberg?s soft photon theorem,''
  JHEP {\bf 1507}, 115 (2015)
  doi:10.1007/JHEP07(2015)115
  [arXiv:1505.05346 [hep-th]].
  
\bibitem{1506.05789} 
  S.~G.~Avery and B.~U.~W.~Schwab,
  ``Burg-Metzner-Sachs symmetry, string theory, and soft theorems,''
  Phys.\ Rev.\ D {\bf 93}, 026003 (2016)
  doi:10.1103/PhysRevD.93.026003
  [arXiv:1506.05789 [hep-th]].

\bibitem{1509.01406} 
  M.~Campiglia and A.~Laddha,
  ``Asymptotic symmetries of gravity and soft theorems for massive particles,''
  JHEP {\bf 1512}, 094 (2015)
  doi:10.1007/JHEP12(2015)094
  [arXiv:1509.01406 [hep-th]].

\bibitem{1605.09094} 
  M.~Campiglia and A.~Laddha,
  ``Sub-subleading soft gravitons: New symmetries of quantum gravity?,''
  Phys.\ Lett.\ B {\bf 764}, 218 (2017)
  doi:10.1016/j.physletb.2016.11.046
  [arXiv:1605.09094 [gr-qc]].

\bibitem{1605.09677} 
  M.~Campiglia and A.~Laddha,
  ``Subleading soft photons and large gauge transformations,''
  JHEP {\bf 1611}, 012 (2016)
  doi:10.1007/JHEP11(2016)012
  [arXiv:1605.09677 [hep-th]].

\bibitem{1608.00685} 
  M.~Campiglia and A.~Laddha,
  ``Sub-subleading soft gravitons and large diffeomorphisms,''
  JHEP {\bf 1701}, 036 (2017)
  doi:10.1007/JHEP01(2017)036
  [arXiv:1608.00685 [gr-qc]].

\bibitem{1612.08294} 
  E.~Conde and P.~Mao,
  ``BMS Supertranslations and Not So Soft Gravitons,''
  arXiv:1612.08294 [hep-th].

\bibitem{1701.00496} 
  T.~He, D.~Kapec, A.~M.~Raclariu and A.~Strominger,
  ``Loop-Corrected Virasoro Symmetry of 4D Quantum Gravity,''
  arXiv:1701.00496 [hep-th].


\bibitem{1612.05886} 
  M.~Asorey, A.~P.~Balachandran, F.~Lizzi and G.~Marmo,
  ``Equations of Motion as Constraints: Superselection Rules, Ward Identities,''
  arXiv:1612.05886 [hep-th].

\bibitem{1703.01351} 
  A.~Campoleoni, D.~Francia and C.~Heissenberg,
  ``On higher-spin supertranslations and superrotations,''
  JHEP {\bf 1705}, 120 (2017)
  doi:10.1007/JHEP05(2017)120
  [arXiv:1703.01351 [hep-th]].

\bibitem{1703.05448} 
  A.~Strominger,
  ``Lectures on the Infrared Structure of Gravity and Gauge Theory,''
  arXiv:1703.05448 [hep-th].

\bibitem{1707.08016} 
  M.~Pate, A.~M.~Raclariu and A.~Strominger,
  ``Color Memory,''
  arXiv:1707.08016 [hep-th].

\bibitem{1709.03850} 
  A.~Laddha and P.~Mitra,
  ``Asymptotic Symmetries and Subleading Soft Photon Theorem in Effective Field Theories,''
  arXiv:1709.03850 [hep-th].
  
\bibitem{1711.04371} 
  D.~Kapec and P.~Mitra,
  ``A $d$-Dimensional Stress Tensor for Mink$_{d+2}$ Gravity,''
  arXiv:1711.04371 [hep-th].

\bibitem{1712.01204} 
  M.~Pate, A.~M.~Raclariu and A.~Strominger,
  ``Gravitational Memory in Higher Dimensions,''
  arXiv:1712.01204 [hep-th].

\bibitem{1712.09591} 
  A.~Campoleoni, D.~Francia and C.~Heissenberg,
  ``Asymptotic Charges at Null Infinity in Any Dimension,''
  Universe {\bf 4}, no. 3, 47 (2018)
  doi:10.3390/universe4030047
  [arXiv:1712.09591 [hep-th]].

\bibitem{1803.03023} 
  A.~H.~Anupam, A.~Kundu and K.~Ray,
  ``Double Soft Graviton Theorems and BMS Symmetries,''
  arXiv:1803.03023 [hep-th].


\bibitem{1103.2981} 
  C.~D.~White,
  ``Factorization Properties of Soft Graviton Amplitudes,''
  JHEP {\bf 1105}, 060 (2011)
  doi:10.1007/JHEP05(2011)060
  [arXiv:1103.2981 [hep-th]].

\bibitem{1404.4091} 
  F.~Cachazo and A.~Strominger,
  ``Evidence for a New Soft Graviton Theorem,''
  arXiv:1404.4091 [hep-th].

\bibitem{1404.5551} 
  E.~Casali,
  ``Soft sub-leading divergences in Yang-Mills amplitudes,''
  JHEP {\bf 1408}, 077 (2014)
  doi:10.1007/JHEP08(2014)077
  [arXiv:1404.5551 [hep-th]].

\bibitem{1404.7749} 
  B.~U.~W.~Schwab and A.~Volovich,
  ``Subleading Soft Theorem in Arbitrary Dimensions from Scattering Equations,''
  Phys.\ Rev.\ Lett.\  {\bf 113}, no. 10, 101601 (2014)
  doi:10.1103/PhysRevLett.113.101601
  [arXiv:1404.7749 [hep-th]].

\bibitem{1405.1015} 
  Z.~Bern, S.~Davies and J.~Nohle,
  ``On Loop Corrections to Subleading Soft Behavior of Gluons and Gravitons,''
  Phys.\ Rev.\ D {\bf 90}, no. 8, 085015 (2014)
  doi:10.1103/PhysRevD.90.085015
  [arXiv:1405.1015 [hep-th]].

\bibitem{1405.1410} 
  S.~He, Y.~t.~Huang and C.~Wen,
  ``Loop Corrections to Soft Theorems in Gauge Theories and Gravity,''
  JHEP {\bf 1412}, 115 (2014)
  doi:10.1007/JHEP12(2014)115
  [arXiv:1405.1410 [hep-th]].

\bibitem{1405.2346} 
  A.~J.~Larkoski,
  ``Conformal Invariance of the Subleading Soft Theorem in Gauge Theory,''
  Phys.\ Rev.\ D {\bf 90}, no. 8, 087701 (2014)
  doi:10.1103/PhysRevD.90.087701
  [arXiv:1405.2346 [hep-th]].


\bibitem{1405.3413} 
  F.~Cachazo and E.~Y.~Yuan,
  ``Are Soft Theorems Renormalized?,''
  arXiv:1405.3413 [hep-th].

\bibitem{1405.3533} 
  N.~Afkhami-Jeddi,
  ``Soft Graviton Theorem in Arbitrary Dimensions,''
  arXiv:1405.3533 [hep-th].

\bibitem{1406.4172} 
  B.~U.~W.~Schwab,
  ``Subleading Soft Factor for String Disk Amplitudes,''
  JHEP {\bf 1408}, 062 (2014)
  doi:10.1007/JHEP08(2014)062
  [arXiv:1406.4172 [hep-th]].

\bibitem{1406.5155} 
  M.~Bianchi, S.~He, Y.~t.~Huang and C.~Wen,
  ``More on Soft Theorems: Trees, Loops and Strings,''
  Phys.\ Rev.\ D {\bf 92}, no. 6, 065022 (2015)
  doi:10.1103/PhysRevD.92.065022
  [arXiv:1406.5155 [hep-th]].

\bibitem{1406.6574} 
  J.~Broedel, M.~de Leeuw, J.~Plefka and M.~Rosso,
  ``Constraining subleading soft gluon and graviton theorems,''
  Phys.\ Rev.\ D {\bf 90}, no. 6, 065024 (2014)
  doi:10.1103/PhysRevD.90.065024
  [arXiv:1406.6574 [hep-th]].

\bibitem{1406.6987} 
  Z.~Bern, S.~Davies, P.~Di Vecchia and J.~Nohle,
  ``Low-Energy Behavior of Gluons and Gravitons from Gauge Invariance,''
  Phys.\ Rev.\ D {\bf 90}, no. 8, 084035 (2014)
  doi:10.1103/PhysRevD.90.084035
  [arXiv:1406.6987 [hep-th]].
  
\bibitem{1406.7184} 
  C.~D.~White,
  ``Diagrammatic insights into next-to-soft corrections,''
  Phys.\ Lett.\ B {\bf 737}, 216 (2014)
  doi:10.1016/j.physletb.2014.08.041
  [arXiv:1406.7184 [hep-th]].
  
\bibitem{1407.5936} 
  M.~Zlotnikov,
  ``Sub-sub-leading soft-graviton theorem in arbitrary dimension,''
  JHEP {\bf 1410}, 148 (2014)
  doi:10.1007/JHEP10(2014)148
  [arXiv:1407.5936 [hep-th]].

\bibitem{1407.5982} 
  C.~Kalousios and F.~Rojas,
  ``Next to subleading soft-graviton theorem in arbitrary dimensions,''
  JHEP {\bf 1501}, 107 (2015)
  doi:10.1007/JHEP01(2015)107
  [arXiv:1407.5982 [hep-th]].

\bibitem{1408.4179} 
  Y.~J.~Du, B.~Feng, C.~H.~Fu and Y.~Wang,
  ``Note on Soft Graviton theorem by KLT Relation,''
  JHEP {\bf 1411}, 090 (2014)
  doi:10.1007/JHEP11(2014)090
  [arXiv:1408.4179 [hep-th]].

\bibitem{1410.6406} 
  D.~Bonocore, E.~Laenen, L.~Magnea, L.~Vernazza and C.~D.~White,
  ``The method of regions and next-to-soft corrections in Drell-Yan production,''
  Phys.\ Lett.\ B {\bf 742}, 375 (2015)
  doi:10.1016/j.physletb.2015.02.008
  [arXiv:1410.6406 [hep-ph]].

\bibitem{1411.6661} 
  B.~U.~W.~Schwab,
  ``A Note on Soft Factors for Closed String Scattering,''
  JHEP {\bf 1503}, 140 (2015)
  doi:10.1007/JHEP03(2015)140
  [arXiv:1411.6661 [hep-th]].

\bibitem{1412.3699} 
  A.~Sabio Vera and M.~A.~Vazquez-Mozo,
  ``The Double Copy Structure of Soft Gravitons,''
  JHEP {\bf 1503}, 070 (2015)
  doi:10.1007/JHEP03(2015)070
  [arXiv:1412.3699 [hep-th]].

\bibitem{1502.05258}
 P.~Di Vecchia, R.~Marotta and M.~Mojaza,
  ``Soft theorem for the graviton, dilaton and the Kalb-Ramond field in the bosonic string,''
  JHEP {\bf 1505}, 137 (2015)
  doi:10.1007/JHEP05(2015)137
  [arXiv:1502.05258 [hep-th]].




\bibitem{1503.04816} 
  F.~Cachazo, S.~He and E.~Y.~Yuan,
  ``New Double Soft Emission Theorems,''
  Phys.\ Rev.\ D {\bf 92}, no. 6, 065030 (2015)
  doi:10.1103/PhysRevD.92.065030
  [arXiv:1503.04816 [hep-th]].

\bibitem{1504.01364} 
  A.~E.~Lipstein,
  ``Soft Theorems from Conformal Field Theory,''
  JHEP {\bf 1506}, 166 (2015)
  doi:10.1007/JHEP06(2015)166
  [arXiv:1504.01364 [hep-th]].
  
  
\bibitem{1504.05558} 
  T.~Klose, T.~McLoughlin, D.~Nandan, J.~Plefka and G.~Travaglini,
  ``Double-Soft Limits of Gluons and Gravitons,''
  JHEP {\bf 1507}, 135 (2015)
  doi:10.1007/JHEP07(2015)135
  [arXiv:1504.05558 [hep-th]].

\bibitem{1504.05559} 
  A.~Volovich, C.~Wen and M.~Zlotnikov,
  ``Double Soft Theorems in Gauge and String Theories,''
  JHEP {\bf 1507}, 095 (2015)
  doi:10.1007/JHEP07(2015)095
  [arXiv:1504.05559 [hep-th]].

\bibitem{1505.05854} 
  M.~Bianchi and A.~L.~Guerrieri,
  ``On the soft limit of open string disk amplitudes with massive states,''
  JHEP {\bf 1509}, 164 (2015)
  doi:10.1007/JHEP09(2015)164
  [arXiv:1505.05854 [hep-th]].

\bibitem{1507.00938} 
  P.~Di Vecchia, R.~Marotta and M.~Mojaza,
  ``Double-soft behavior for scalars and gluons from string theory,''
  JHEP {\bf 1512}, 150 (2015)
  doi:10.1007/JHEP12(2015)150
  [arXiv:1507.00938 [hep-th]].

\bibitem{1507.08829} 
  A.~L.~Guerrieri,
  ``Soft behavior of string amplitudes with external massive states,''
  Nuovo Cim.\ C {\bf 39}, no. 1, 221 (2016)
  doi:10.1393/ncc/i2016-16221-2
  [arXiv:1507.08829 [hep-th]].

\bibitem{1507.08882} 
  S.~D.~Alston, D.~C.~Dunbar and W.~B.~Perkins,
  ``$n$-point amplitudes with a single negative-helicity graviton,''
  Phys.\ Rev.\ D {\bf 92}, no. 6, 065024 (2015)
  doi:10.1103/PhysRevD.92.065024
  [arXiv:1507.08882 [hep-th]].

\bibitem{1509.07840} 
  Y.~t.~Huang and C.~Wen,
  ``Soft theorems from anomalous symmetries,''
  JHEP {\bf 1512}, 143 (2015)
  doi:10.1007/JHEP12(2015)143
  [arXiv:1509.07840 [hep-th]].

\bibitem{1511.04921} 
  P.~Di Vecchia, R.~Marotta and M.~Mojaza,
  ``Soft Theorems from String Theory,''
  Fortsch.\ Phys.\  {\bf 64}, 389 (2016)
  doi:10.1002/prop.201500068
  [arXiv:1511.04921 [hep-th]].
  
  \bibitem{1512.00803} 
  M.~Bianchi and A.~L.~Guerrieri,
  ``On the soft limit of closed string amplitudes with massive states,''
  Nucl.\ Phys.\ B {\bf 905}, 188 (2016)
  doi:10.1016/j.nuclphysb.2016.02.005
  [arXiv:1512.00803 [hep-th]].

\bibitem{1601.03457} 
  M.~Bianchi and A.~L.~Guerrieri,
  ``On the soft limit of tree-level string amplitudes,''
  arXiv:1601.03457 [hep-th].


\bibitem{1604.00650} 
  J.~Rao and B.~Feng,
  ``Note on Identities Inspired by New Soft Theorems,''
  JHEP {\bf 1604}, 173 (2016)
  doi:10.1007/JHEP04(2016)173
  [arXiv:1604.00650 [hep-th]].

\bibitem{1604.03355} 
  P.~Di Vecchia, R.~Marotta and M.~Mojaza,
  ``Subsubleading soft theorems of gravitons and dilatons in the bosonic string,''
  JHEP {\bf 1606}, 054 (2016)
  doi:10.1007/JHEP06(2016)054
  [arXiv:1604.03355 [hep-th]].
  
 
  \bibitem{1604.02834} 
  S.~He, Z.~Liu and J.~B.~Wu,
  ``Scattering Equations, Twistor-string Formulas and Double-soft Limits 
  in Four Dimensions,''
  JHEP {\bf 1607}, 060 (2016)
  doi:10.1007/JHEP07(2016)060
  [arXiv:1604.02834 [hep-th]].

\bibitem{1604.03893} 
  F.~Cachazo, P.~Cha and S.~Mizera,
  ``Extensions of Theories from Soft Limits,''
  JHEP {\bf 1606}, 170 (2016)
  doi:10.1007/JHEP06(2016)170
  [arXiv:1604.03893 [hep-th]].

\bibitem{1607.02700} 
  A.~P.~Saha,
  ``Double Soft Theorem for Perturbative Gravity,''
  JHEP {\bf 1609}, 165 (2016)
  doi:10.1007/JHEP09(2016)165
  [arXiv:1607.02700 [hep-th]].

\bibitem{1610.03481} 
  P.~Di Vecchia, R.~Marotta and M.~Mojaza,
  ``Soft behavior of a closed massless state in superstring and universality in the soft behavior of the dilaton,''
  JHEP {\bf 1612}, 020 (2016)
  doi:10.1007/JHEP12(2016)020
  [arXiv:1610.03481 [hep-th]].

\bibitem{1611.02172} 
  A.~Luna, S.~Melville, S.~G.~Naculich and C.~D.~White,
  ``Next-to-soft corrections to high energy scattering in QCD and gravity,''
  JHEP {\bf 1701}, 052 (2017)
  doi:10.1007/JHEP01(2017)052
  [arXiv:1611.02172 [hep-th]].

\bibitem{1611.07534} 
  H.~Elvang, C.~R.~T.~Jones and S.~G.~Naculich,
  ``Soft Photon and Graviton Theorems in Effective Field Theory,''
  arXiv:1611.07534 [hep-th].

\bibitem{1611.03137} 
  C.~Cheung, K.~Kampf, J.~Novotny, C.~H.~Shen and J.~Trnka,
  ``A Periodic Table of Effective Field Theories,''
  arXiv:1611.03137 [hep-th].



\bibitem{1702.02350} 
  A.~P.~Saha,
  ``Double Soft Theorem for Perturbative Gravity II: Some Details on CHY Soft Limits,''
  arXiv:1702.02350 [hep-th].

\bibitem{1702.03934} 
  A.~Sen,
  ``Soft Theorems in Superstring Theory,''
  arXiv:1702.03934 [hep-th].

\bibitem{1703.00024} 
  A.~Sen,
  ``Subleading Soft Graviton Theorem for Loop Amplitudes,''
  arXiv:1703.00024 [hep-th].


\bibitem{1705.06175} 
  P.~Di Vecchia, R.~Marotta and M.~Mojaza,
  ``Double-soft behavior of the dilaton of spontaneously broken conformal invariance,''
  arXiv:1705.06175 [hep-th].

\bibitem{1706.00759} 
  A.~Laddha and A.~Sen,
  ``Sub-subleading Soft Graviton Theorem in Generic Theories of Quantum Gravity,''
  arXiv:1706.00759 [hep-th].
 
 \bibitem{1707.06803} 
  S.~Chakrabarti, S.~P.~Kashyap, B.~Sahoo, A.~Sen and M.~Verma,
  ``Subleading Soft Theorem for Multiple Soft Gravitons,''
  arXiv:1707.06803 [hep-th].

\bibitem{1801.05528} 
  Y.~Hamada and G.~Shiu,
  ``Infinite Set of Soft Theorems in Gauge-Gravity Theories as Ward-Takahashi Identities,''
  Phys.\ Rev.\ Lett.\  {\bf 120}, no. 20, 201601 (2018)
  doi:10.1103/PhysRevLett.120.201601
  [arXiv:1801.05528 [hep-th]].

\bibitem{1802.03148} 
  Z.~Z.~Li, H.~H.~Lin and S.~Q.~Zhang,
  ``Infinite Soft Theorems from Gauge Symmetry,''
  arXiv:1802.03148 [hep-th].

\bibitem{1805.11079} 
  S.~Higuchi and H.~Kawai,
  ``Universality of soft theorem from locality of soft vertex operators,''
  arXiv:1805.11079 [hep-th].

\bibitem{1801.07719} 
  A.~Laddha and A.~Sen,
  ``Gravity Waves from Soft Theorem in General Dimensions,''
  arXiv:1801.07719 [hep-th].

\bibitem{1804.09193} 
  A.~Laddha and A.~Sen,
  ``Logarithmic Terms in the Soft Expansion in Four Dimensions,''
  arXiv:1804.09193 [hep-th].

\bibitem{1806.01872} 
  A.~Laddha and A.~Sen,
  ``Observational Signature of the Logarithmic Terms in the Soft Graviton Theorem,''
  arXiv:1806.01872 [hep-th].
  
\bibitem{peter}
P.~C.~Peters, ``Relativistic Gravitational Bremsstrahlung.'', Phys.\ Rev.\ {\bf D1}, 1559 (1970).

\bibitem{blanchet} 
  L.~Blanchet and T.~Damour,
  ``Hereditary effects in gravitational radiation,''
  Phys.\ Rev.\ D {\bf 46}, 4304 (1992).
  doi:10.1103/PhysRevD.46.4304


\bibitem{0912.4254} 
  W.~D.~Goldberger and A.~Ross,
  ``Gravitational radiative corrections from effective field theory,''
  Phys.\ Rev.\ D {\bf 81}, 124015 (2010)
  doi:10.1103/PhysRevD.81.124015
  [arXiv:0912.4254 [gr-qc]].

\bibitem{1211.6095} 
  W.~D.~Goldberger, A.~Ross and I.~Z.~Rothstein,
  ``Black hole mass dynamics and renormalization group evolution,''
  Phys.\ Rev.\ D {\bf 89}, no. 12, 124033 (2014)
  doi:10.1103/PhysRevD.89.124033
  [arXiv:1211.6095 [hep-th]].

\bibitem{1409.4555} 
  A.~Gruzinov and G.~Veneziano,
  ``Gravitational Radiation from Massless Particle Collisions,''
  Class.\ Quant.\ Grav.\  {\bf 33}, no. 12, 125012 (2016)
  doi:10.1088/0264-9381/33/12/125012
  [arXiv:1409.4555 [gr-qc]].

\bibitem{1512.00281} 
  M.~Ciafaloni, D.~Colferai, F.~Coradeschi and G.~Veneziano,
  ``Unified limiting form of graviton radiation at extreme energies,''
  Phys.\ Rev.\ D {\bf 93}, no. 4, 044052 (2016)
  doi:10.1103/PhysRevD.93.044052
  [arXiv:1512.00281 [hep-th]].


\bibitem{1901.10986} 
  A.~Addazi, M.~Bianchi and G.~Veneziano,
  ``Soft gravitational radiation from ultra-relativistic collisions at sub- and sub-sub-leading order,''
  arXiv:1901.10986 [hep-th].
  
\bibitem{1812.08137} 
  M.~Ciafaloni, D.~Colferai and G.~Veneziano,
  ``Infrared features of gravitational scattering and radiation in the eikonal approach,''
  arXiv:1812.08137 [hep-th].
  
\bibitem{grammer} 
  G.~Grammer, Jr. and D.~R.~Yennie,
  ``Improved treatment for the infrared divergence problem in quantum electrodynamics,''
  Phys.\ Rev.\ D {\bf 8}, 4332 (1973).
  doi:10.1103/PhysRevD.8.4332

\end{thebibliography}
\end{document}